\pdfoutput=1

\documentclass[12pt]{article}
\usepackage{amsfonts}
\usepackage{amsmath}
\usepackage{amssymb}
\usepackage{bigints}
\usepackage{booktabs}
\usepackage[nosort]{cite}
\usepackage{color}
\usepackage{float}
\usepackage{framed}
\usepackage{graphicx}
\usepackage{indentfirst}
\usepackage{mathrsfs}
\usepackage{multirow}
\usepackage{pdflscape}
\usepackage{setspace}
\usepackage{subfig}
\usepackage{titlesec}
\usepackage[dotinlabels]{titletoc}
\usepackage{wrapfig}
\usepackage[all]{xy}
\usepackage{young}
\usepackage[vcentermath]{youngtab}
\usepackage{relsize}
\usepackage{datetime}
\usepackage{hyperref}
\usepackage{tikz}
\usetikzlibrary{trees}
\usetikzlibrary{decorations.pathmorphing}
\usetikzlibrary{decorations.markings}
\usetikzlibrary{shapes.misc}

\def\zb{\bar z}
\def\cA{\mathcal{A}}
\def\chib{\bar\chi}
\def\c{\cite}
\def\1{{\rm 1-loop}}
\def\C{\mathcal{C}}
\def\mO{\mathcal{O}}
\def\Ot{\widetilde{\mO}}

\def\vs{\vskip .1 in}
\def\Om{\Omega}
\def\G{\Gamma}
\def\p{\partial}
\def\o{\over}
\def\g{\gamma}
\def\D{\Delta}
\def\rar{\rightarrow}
\def\f{\phi}
\def\eqr{\eqref}
\def\dDisc{{\rm dDisc}}
\def\O{{\cal O}}
\def\ra{\rangle}
\def\la{\langle}
\def\ssec{\subsection}
\def\sssec{\subsubsection}

\def\i{\infty}
\def\foot{\footnote}
\newcommand{\es}[2] {\begin{equation} \label{#1} \begin{split} #2 \end{split} \end{equation}}
\newcommand{\e}[2] {\begin{equation} \label{#1} #2 \end{equation}}

\numberwithin{equation}{section}

\usepackage{verbatim}

\def\be#1\ee{\begin{align}#1\end{align}}
\newcommand{\Res}[1] {\underset{#1}{\text{Res}}\,}
\def\({\left(} \def\){\right)}
\def\[{\left[} \def\]{\right]}

\def\Re{\text{Re}}

\def\sgn{\text{sgn}}

\def\mF{\mathcal{F}}

\def\mJ{\mathcal{J}}

\def\mO{\mathcal{O}}
\def\mC{\mathcal{C}}

\def\mI{\mathcal{I}}

\def\eps{\epsilon}

\def\tl{\tilde}

\newcommand\cO{\mathcal{O}}
\newcommand\SO{\mathrm{SO}}
\newcommand\vol{\mathrm{vol}}
\newcommand\<\langle
\renewcommand\>\rangle
\newcommand\oo\infty
\newcommand\nn\nonumber
\renewcommand\.\cdot
\renewcommand\a\alpha
\renewcommand\g\gamma
\renewcommand\f\phi

\newcommand\x\times

\def\G{\Gamma}
\def\De{\Delta}

\newmuskip\pFqmuskip
\newcommand*\pFq[6][8]{%
  \begingroup % only local assignments
  \pFqmuskip=#1mu\relax
  \mathcode`\,=\string"8000
  \begingroup\lccode`\~=`\,
  \lowercase{\endgroup\let~}\pFqcomma
  {}_{#2}F_{#3}{\left(\genfrac..{0pt}{}{#4}{#5};#6\right)}%
  \endgroup
}
\newcommand{\pFqcomma}{\mskip\pFqmuskip}

\newmuskip\LpFqmuskip
\newcommand*\LpFq[6][8]{%
  \begingroup % only local assignments
  \pFqmuskip=#1mu\relax
  \mathcode`\,=\string"8000
  \begingroup\lccode`\~=`\,
  \lowercase{\endgroup\let~}\pFqcomma
  {}_{}F_{}{\left[\genfrac..{0pt}{}{#4}{#5};#6\right]}%
  \endgroup
}

\newmuskip\Ftmuskip
\newcommand*\Ft[6][8]{%
  \begingroup % only local assignments
  \pFqmuskip=#1mu\relax
  \mathcode`\,=\string"8000
  \begingroup\lccode`\~=`\,
  \lowercase{\endgroup\let~}\pFqcomma
  F_2{\left[\genfrac..{0pt}{}{#4}{#5};#6\right]}%
  \endgroup
}

\newcommand{\bea}{\begin{eqnarray}}
\newcommand{\eea}{\end{eqnarray}}

\newcommand{\bml}{\begin{multline}}
\newcommand{\emll}{\end{multline}}

\usepackage[left=2cm,right=2cm,top=2cm,bottom=2cm]{geometry}
\titleformat{\section}{\normalfont\bfseries}{\thesection.}{4pt}{}
\titlespacing{\section}{0pt}{22pt}{6pt}

\titleformat{\subsection}{\normalfont\itshape}{\thesubsection.}{4pt}{}
\titlespacing{\subsection}{0pt}{18pt}{6pt}

\titleformat{\subsubsection}{\normalfont\itshape}{\thesubsubsection.}{4pt}{}
\titlespacing{\subsubsection}{0pt}{16pt}{6pt}

\def\ie{\begin{equation}\begin{aligned}}
\def\fe{\end{aligned}\end{equation}}

\def\tilde{\widetilde}

\def\hat{\widehat}

\def\bar{\overline}
\def\b{\bar}

\def\Re{\mathop{\rm Re}}

\DeclareFontShape{OT1}{cmr}{mx}{n}%
    {<->cmr10}{}
\newcommand{\mytitlefont}{\fontseries{mx}\selectfont}
\DeclareMathAlphabet{\titlemath}{OT1}{cmr}{mx}{n}

\def\cJ{\mathtt{J}}

\begin{document}

\begin{titlepage}
\hbox{CALT-TH-2018-023}
\begin{center}

~\\[2cm]

{\fontsize{18pt}{0pt} \mytitlefont $d$-dimensional SYK, AdS Loops, and $6j$ Symbols}

~\\[0.5cm]

{\fontsize{14pt}{0pt} Junyu Liu$^{1,2}$, Eric Perlmutter$^1$, Vladimir Rosenhaus$^{3,4}$, and David Simmons-Duffin$^1$}

~\\[0.1cm]

\it{$^1$Walter Burke Institute for Theoretical Physics}\\{ Caltech, Pasadena, CA 91125}
~\\[.5cm]
\it{$^2$ Institute for Quantum Information and Matter}\\{ Caltech, Pasadena, CA 91125}
~\\[.5cm]
\it{$^3$ Kavli Institute for Theoretical Physics}\\ 
\it{University of California, Santa Barbara, CA 93106}\\
~\\[.5cm]
\it{$^4$ School of Natural Sciences, Institute for Advanced Study}\\ 
\it{Einstein Drive, Princeton, NJ 08540}

~\\[0.8cm]

\end{center}

\noindent 

We study the $6j$ symbol for the conformal group, and its appearance in three seemingly unrelated contexts:  the SYK model, conformal representation theory, and perturbative amplitudes in AdS. 
The contribution of the planar Feynman diagrams to the three-point function of the bilinear singlets in SYK is shown to be a $6j$ symbol. We  generalize the computation of these and other Feynman diagrams to $d$ dimensions. The $6j$ symbol can be viewed as the crossing kernel for conformal partial waves, which may be computed using the Lorentzian inversion formula. We provide closed-form expressions for $6j$ symbols in $d=1,2,4$. In AdS, we show that the $6j$ symbol is the Lorentzian inversion of a crossing-symmetric tree-level exchange amplitude, thus efficiently packaging the double-trace OPE data. Finally, we consider one-loop diagrams in AdS with internal scalars and external spinning operators, and show that the triangle diagram is a $6j$ symbol, while one-loop $n$-gon diagrams are built out of $6j$ symbols.

\vfill

\end{titlepage}

\tableofcontents

\section{Introduction}

Clebsch-Gordan coefficients are familiar quantities in quantum mechanics, encoding the addition of two angular momenta. When adding three angular momenta, one encounters the $6j$ symbol which, like the  Clebsch-Gordan coefficients, is of fundamental importance in the representation theory of $\text{SU}(2)$. In adding three angular momenta, one has a choice of combining the first two, and then the third, or the second and third, and then the first. The overlap of these two choices involves a  product of four Clebsch-Gordan coefficients, summed over the $m_i$ quantum numbers. As such, the $6j$ symbol can be represented by a tetrahedron, with each of the six sides labelled by a spin, and each of the four vertices representing a Clebsch-Gordan coefficient. 

These quantities, while most familiar for $\text{SU}(2)$, can be defined for any group. Our interest will be in the Euclidean conformal group, $\text{SO}(d+1, 1)$. In mapping from  $\text{SU}(2)$ to the conformal group, the angular momentum  is replaced by the dimension and spin of the operator, the $z$-component of angular momentum becomes the position $x$ of the operator, and the Clebsch-Gordan coefficient becomes a conformal three-point function. In this paper, we will compute $6j$ symbols for the conformal group, in one, two, and four dimensions.

\begin{figure}[t]
\centering
\includegraphics[width=5in]{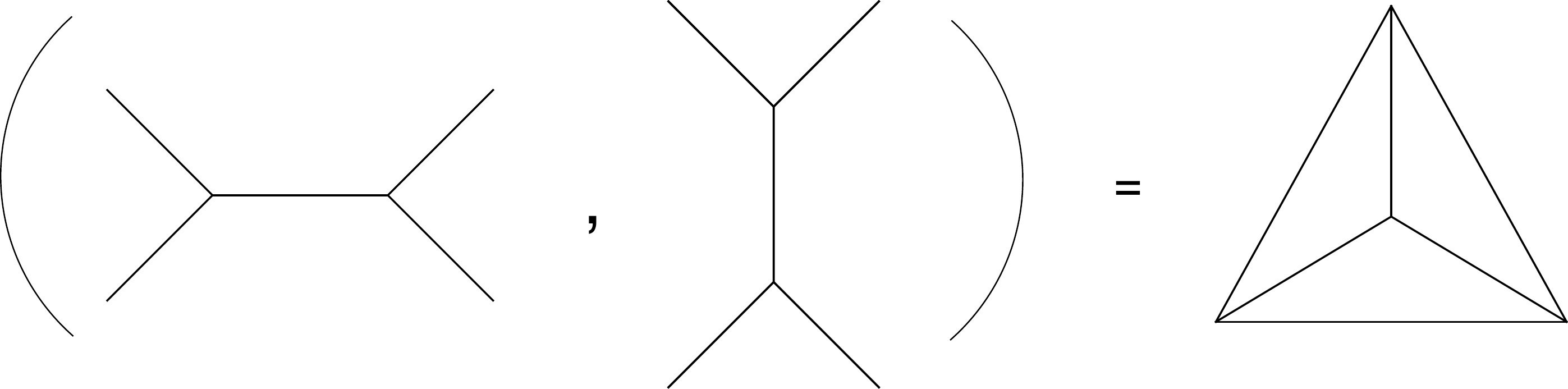}
\caption{The inner product of an $s$-channel conformal partial wave and a $t$-channel conformal partial wave is a $6j$ symbol, represented by a tetrahedron. Each line represents a coordinate, and each vertex is a conformal three-point function. } \label{tetraCut1}
\end{figure}

One context in which $6j$ symbols for the conformal group naturally appear is in the framework of the conformal bootstrap, as emphasized in \cite{Gadde:2017sjg}: a four-point function can be expanded in either $s$-channel or $t$-channel conformal blocks, and the equality of the two yields powerful constraints on CFT data. One can define single-valued conformal partial waves, as a sum of a conformal block and its ``shadow'', as we will recall in the main text. The overlap of an $s$-channel conformal partial wave with a $t$-partial conformal partial wave is the definition of a $6j$ symbol, as shown diagrammatically in Fig.~\ref{tetraCut1}. This is equivalent to a ``crossing kernel'' for the conformal group. This interpretation of the $6j$ symbol for the conformal group is a direct analogue of the one in quantum mechanics, as an overlap of the two different ways of combining three spins. 

In this work, we reveal two other contexts in which these $6j$ symbols appear: Feynman diagrams of the SYK model and their generalization to higher dimensions, and Witten diagrams for tree-level and one-loop scattering amplitudes in AdS, dual to large $N$ CFT correlators at leading and subleading orders in $1/N$.

\begin{figure}[t]
\centering
\subfloat[]{
\includegraphics[width=3in]{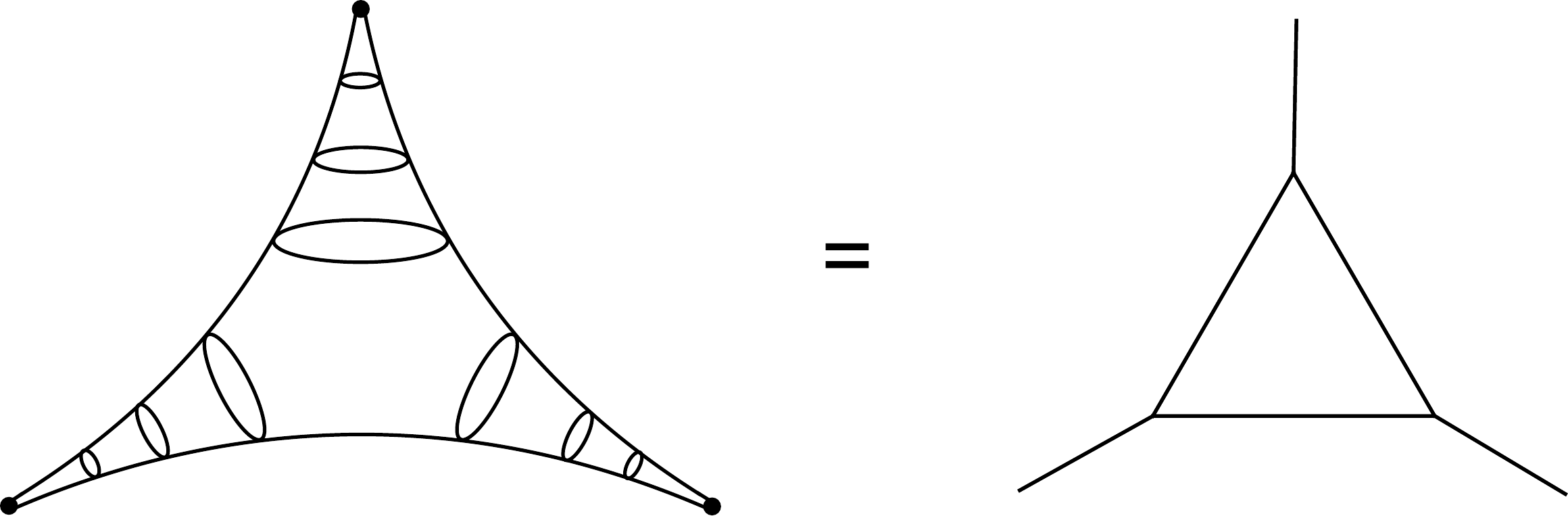}
}\\
\subfloat[]{
\includegraphics[width=5in]{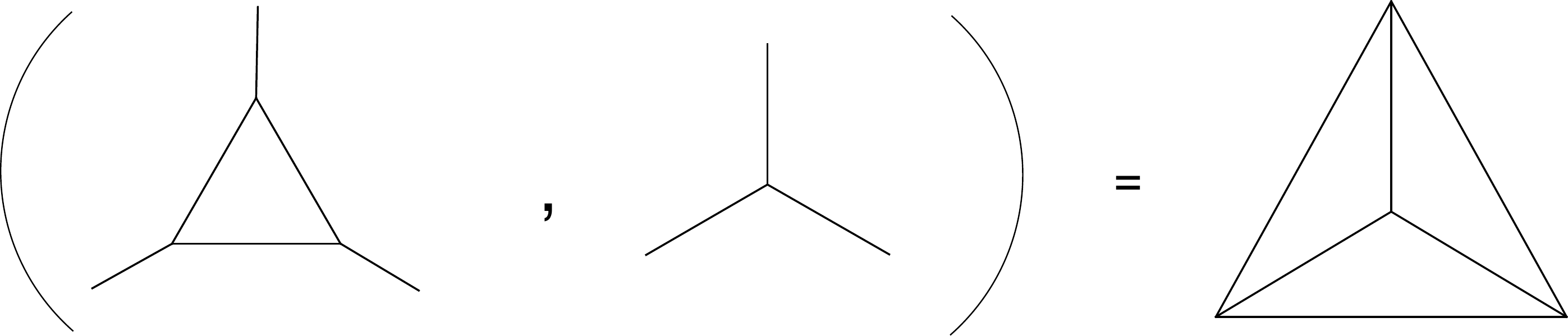}
}
\caption{ (a) The sum of the planar Feynman diagram contribution to the SYK three-point function of bilinears is given by three  conformal three-point functions glued together. The figure on the right is not a Feynman diagram; each vertex is a three-point function. (b) The inner product with a bare three-point function of  shadow operators extracts the structure constant, and is equal to a $6j$ symbol. } \label{tetraCut2}
\end{figure}

The one-dimensional SYK model \cite{SY, Kitaev, MS, Rosenhaus:2018dtp} is part of a new class of solvable large $N$ quantum field theories, which are dominated by  melonic Feynman diagrams at large $N$, and are conformally invariant in the infrared. The three-point function of the bilinear $\text{O}(N)$ singlets was recently computed in \cite{GR4}, by summing all Feynman diagrams. The contribution of the planar diagrams is shown in Fig.~\ref{tetraCut2}. There is a simple way to see that the planar three-point diagram is actually a $6j$ symbol: by taking the inner product with a three-point function of the shadow operators, one obtains a tetrahedron, i.e.\ a $6j$ symbol. This argument establishes an intriguing connection: the overlap of two conformal partial waves -- a group-theoretic quantity -- and the planar Feynman diagrams in an SYK correlation function -- a dynamical quantity -- are just  two different ways of splitting a tetrahedron. 

The SYK model is formulated in one dimension, and one would clearly like to have a higher-dimensional version. An obvious generalization is to take the SYK Lagrangian, replace the fermions by bosons, and put the theory in $d$ dimensions. Unfortunately, this $d$-dimensional bosonic SYK theory is not well defined because the potential has negative directions. Nevertheless, one can still study the theory formally, and this is what we will do in this paper: we compute the correlation functions of the $d$-dimensional bosonic SYK theory, in close analogy with the solution of the SYK model \cite{GR4}.  In fact, the extent to which a $d$-dimensional SYK model exists is irrelevant for our purposes:  we are simply taking the leading large $N$ Feynman diagrams that appear in the standard one-dimensional SYK model, and evaluating them in $d$ dimensions, using a conformal bosonic propagator. 
A two dimensional generalization of SYK that is well-defined is the supersymmetric version \cite{Murugan:2017eto}, and it should be straightforward to generalize our calculations to this context. There are also promising signs that more higher-dimensional SYK-like theories may still be found \cite{Klebanov:2016xxf, Giombi:2017dtl}.

In the AdS context, the appearance of the $6j$ symbol is quite natural, in the sense that the AdS isometry group is the conformal group. Where exactly it appears is less obvious; one answer, as we will show, is in the computation of tree- and loop-level scattering amplitudes. There has been recent progress on loops in AdS,\foot{For a long time, there were very few loop-level results. Progress was made via direct computation in Mellin space \c{Penedones:2010ue,Fitzpatrick:2011hu,Fitzpatrick:2011dm}. This inspired a prescription for an ``AdS unitarity method'' which uses ideas from CFT crossing symmetry \c{Aharony:2016dwx}. These developments led to further progress in the case of AdS$_5\times S^5$ \c{Alday:2017xua,Aprile:2017bgs,Aprile:2017qoy,Alday:2017vkk}, as well as applications of large spin perturbation theory/Lorentzian inversion \c{Alday:2017vkk} and, more recently, progress in direct computation of individual diagrams and their analytic structure \c{Cardona:2017tsw,Giombi:2017hpr,ellis1,ellis2,Bertan:2018khc}. The work of \c{ellis2}, in particular, performs an extensive analysis of the analytic structure of wide classes of individual diagrams.} though there are still relatively few explicit results for AdS diagrams beyond tree-level, especially compared to the extensive and sophisticated knowledge of S-matrices. Perhaps the clearest demonstration of this fact is the absence of a computation of the one-loop vertex correction in $\phi^3$ theory, a basic quantity. We will show that this diagram in AdS$_{d+1}$ is, in fact, given by a spectral integral over the $6j$ symbol for SO$(d+1,1)$. We also compute higher-point diagrams at one-loop with cubic vertices -- i.e.\ AdS ``$n$-gons'' -- and find that they may be written as gluings of $6j$ symbols. This includes the box diagram ($n=4$), which had not previously been computed.

The overall picture, then, is of a triangle of relations amongst SYK planar diagrams, the $6j$ symbol, and AdS one-loop diagrams, all of which can be expressed simply in terms of the others. 
 
In Sec.~\ref{sec:2} we study the  $d$-dimensional bosonic SYK theory. 
In Sec.~\ref{fourPoint} we  compute the four-point function of fundamentals, by summing ladder diagrams. Such calculations have become standard in the SYK literature. We are able to both generalize and simplify the computation,  by recognizing that the conformal three-point integrals that appear throughout are just shadow transforms, discussed in Appendix~\ref{shadow}. In Sec.~\ref{6pt} we turn to the three-point functions of bilinears, and demonstrate that the contribution of the planar Feynman diagrams is a $6j$ symbol. The contribution of the so called contact diagrams is calculated in Appendix~\ref{contact}. The analytically extended three-point function of bilinears plays an essential role in determining all higher-point correlation functions, as discussed in Sec.~\ref{allPoint}. 

 In Sec.~\ref{sec:6j} we compute the $6j$ symbols, which we view as the overlap of two conformal partial waves, as was shown in Fig.~\ref{tetraCut1}. This section can be read independently of the rest of the paper. In one dimension,  the conformal partial waves are the standard ${}_2 F_1$ hypergeometric functions of a conformally invariant cross ratio of four points. The overlap of two partial waves is an integral of a product of two ${}_2 F_1$'s, which is a generalized hypergeometric function ${}_4 F_3$. This straightforward computation is done in Appendix~\ref{app1d6j}. In two and four dimensions, there are two conformal cross-ratios, and the conformal partial waves are sums of products of two hypergeometric functions. The integral for the overlap of two partial waves, as formulated in Euclidean signature, does not appear to factorize. However, one can achieve factorization into one-dimensional integrals by an appropriate contour deformation into Lorentzian signature, i.e.\ by employing Caron-Huot's Lorentzian inversion formula  \cite{Caron-Huot:2017vep,SD17}; our integral is a special case of his more general formula, for a four-point function that is a $t$-channel partial wave. In Sec.~\ref{2d6j} we generalize Caron-Huot's formula in two dimensions to include external operators with spin (recovering the generalized inversion formula in \cite{Kravchuk:2018htv}), and compute the two-dimensional $6j$ symbol for arbitrary spin. In Sec.~\ref{4d6j} we compute the four-dimensional $6j$ symbol, for external scalars. In both two and four dimensions, the $6j$ symbol is expressed as a finite sum of products of two ${}_4 F_3$'s, as a result of the factorization into one-dimensional integrals. (See Eqs.~\eqr{6jd2}--\eqr{omegaexp} for $d=2$ and \eqr{6jd4} for $d=4$.)\footnote{The $6j$ symbol for scalar principal series representations of $\SO(d+1,1)$ was computed in \cite{Krasnov:2005fu} in terms of a four-fold Mellin-Barnes integral. The expressions we find in $d=2,4$ are much simpler.} We also explain how Lorentzian inversion allows one to formally invert a conformal block rather than a partial wave. 
 
In Sec.~\ref{6jcft}, we analyze the $6j$ symbol in order to address the following question: on what $s$-channel partial waves does the $t$-channel partial wave have support? The answer is encoded in the locations of the poles of the $6j$ symbol, whose residues encode the OPE data. That this is true can be seen from our method of using the Lorentzian inversion formula, which produces a function whose poles in $\D$ correspond to the physical operator spectrum. We show that there are poles when the twists $\tau_i=\D_i-J_i$ of any three operators that share a vertex obey $\tau_i=\tau_j+\tau_k+2n$ with $n\in\mathbb{Z}_{\geq0}$; in addition, for each of these poles, there is a pole at the shadow location. These locations correspond to the twists of infinite towers of ``double-twist operators'' in the $s$-channel, of the form $\cO_j(\partial^2)^n\partial_{\mu_1}\ldots \partial_{\mu_J}\cO_k$, whose existence is required by crossing in the lightcone limit \cite{Fitzpatrick:2012yx, Komargodski:2012ek}. The residues of these poles thus encode the OPE coefficients and anomalous dimensions of the double-twist operators, so the $6j$ symbol contains complete information about these quantities, even down to finite spin $J$. As an example, we present explicit results for leading-twist $(n=0)$ anomalous dimensions due to the Lorentzian inversion of a conformal block in Eqs. \eqr{eq:resulttwod} and \eqr{eq:resultfourd}, valid for finite spin $J$.\foot{There have been some recent works that compute anomalous dimensions from related perspectives \c{Alday:2017gde,Giombi:2018vtc,Cardona:2018dov,Sleight:2018epi,Sleight:2018ryu}. Comprehensive closed-form expressions for anomalous dimensions, valid to all orders in the $1/J$ expansion, may be found in \c{Cardona:2018dov,Sleight:2018epi,Sleight:2018ryu} for leading-twist $(n=0)$ double-twist operators, with an extension to subleading twists $(n>0)$ and external spins in \c{Sleight:2018epi,Sleight:2018ryu}. As a point of clarification, we emphasize that the $6j$ symbol is not actually computed in \c{Sleight:2018epi,Sleight:2018ryu}, but rather, parts of the residues of its poles. As such the use of the term ``crossing kernel'' in those papers is non-standard; in this work, we compute the complete crossing kernel in $d=1,2,4$, from which the finite $J$ OPE data may be extracted via residues. We explain in detail the difference between our results for leading-twist anomalous dimensions at finite $J$, and the asymptotic resummations of  \c{Cardona:2018dov,Sleight:2018epi,Sleight:2018ryu}, in footnote \ref{foot27}.}

In Sec.~\ref{AdS} we turn to the AdS story. In Sec.~\ref{tree} we start at tree-level. We show that Lorentzian inversion of a sum over channels of AdS exchange diagrams yields the $6j$ symbol; more precisely, it yields the Lorentzian inversion of a conformal block, which is a projection of the $6j$ symbol. (See Eq. \eqr{inj}.) Thus, the OPE function for leading-order connected correlators in large $N$ CFTs is a sum of $6j$ symbols, which in turn encode the OPE data for double-trace operators. In large $N$ CFTs with weakly curved, local AdS duals, there are a finite number of light operators, so the double-trace data is determined by a finite sum of $6j$ symbols. (There is a lower bound on the spin of the double-traces that the $6j$ symbol captures, as we recall later.) In Sec.~\ref{loop}, we turn to the one-loop diagrams, in general dimension $d$. We begin with the one-loop, three-point triangle diagram in AdS for arbitrary external spins and internal scalars, drawn in Fig. \ref{figtri}. This diagram --- indeed, all one-loop diagrams considered in this paper --- can be written as a spectral integral over the ``pre-amplitude'', which is defined as the same diagram but with the bulk-to-bulk propagators replaced by harmonic functions (i.e.\ the propagator minus the shadow propagator). There is some data on the analyticity properties of the triangle pre-amplitude \c{ellis2}, but it, too, has never been computed. We show that the triangle pre-amplitude equals a $6j$ symbol times a kinematic prefactor. (See Eqs.~\eqr{com456} and \eqr{com456j}.) This fact essentially follows immediately from the split representation of the harmonic function. This determines the full triangle amplitude by spectral integration of the $6j$ symbol, whose poles are easily read off from our explicit expressions for the $6j$ symbol. We then show that the pre-amplitude for $n$-gon diagrams (see Fig. \ref{figngon}) is equal to the $d$-dimensional SYK bilinear $n$-point planar diagram with no exchanged melons, times a kinematic prefactor. We give a more detailed exposition of the box diagram ($n=4$), which we write in a conformal block expansion. (See Eqs.~\eqr{a4pre}--\eqr{rho4pre}.) It takes the form of a sum of two $6j$ symbols times a conformal partial wave. Higher $n$-gon pre-amplitudes can also be written in terms of $6j$ symbols; in this sense, the $6j$ symbol is an atomic ingredient in the computation of AdS loop diagrams. We close with some open questions in Sec.~\ref{adsq}. 

In Appendix \ref{appa}, we give some simple analysis of the $d$-dimensional SYK bilinear two-point function. In Appendix \ref{app1d6j}, we present the one-dimensional $6j$ symbol. In Appendix \ref{contact}, we compute the contact diagram contributing to the $d$-dimensional bilinear three-point function. In Appendix \ref{shadow}, we present derivations of some shadow transforms of three-point functions.

\section{SYK and $d$-dimensional generalizations} \label{sec:2}
The SYK model  is a theory of $N \gg 1$ Majorana fermions $\chi_i$ with $q$-body all-to-all interactions with Gaussian-random coupling. The Lagrangian is
\be
\mathcal{L} =\frac{1}{2} \chi_i \partial_{\tau} \chi_i + \frac{i^{\frac{q}{2}}}{q!} J_{i_1 i_2 \ldots i_q} \chi_{i_1} \chi_{i_2} \cdots \chi_{i_q}~,
\ee
where the couplings have zero-mean and variance $\langle J_{i_1 i_2 \ldots i_q} J_{i_1 i_2 \ldots i_q} \rangle = N^{q-1} \cJ^2/ (q-1)!$ (no index summation).  After disorder averaging, the model has $\text{O}(N)$ symmetry. In the UV, the theory is free, and the fermions have a two-point function given by $\frac{1}{2} \sgn(\tau)$. At large $N$ the dominant Feynman diagrams are iterations of melons. In the infrared, for $\cJ |\tau| \gg 1$, the theory is nearly conformally  invariant, with a two-point function, at leading order in $1/N$,
\be \label{psiDelta}
G(\tau) = b\frac{\sgn(\tau)}{|\cJ\tau|^{2\Delta}}~, \ \ \ \ \ \ \ \Delta = 1/q~,\ \ \ \ \ \ \ \  b^{q} =  \frac{1}{2\pi}(1 - 2\Delta)\tan\pi \Delta~,
\ee
where $\Delta$ is the IR dimension of the fermions.

It is natural to consider a $d$-dimensional generalization of SYK.  It is simplest to take a bosonic theory, of $N$ scalars $\phi_i$, with Lagrangian
\be
\mathcal{L}  = \frac{1}{2}(\partial \phi_i)^2 + \frac{1}{q!}J_{i_1 i_2 \cdots i_q} \phi_{i_1} \phi_{i_2} \cdots \phi_{i_q}~.
\ee
It is clear that this theory will, at large $N$, be dominated by the same Feynman diagrams as those appearing in SYK. Assuming an infrared fixed point, the correlation functions would  then be found from a straightforward generalization, to $d$ dimensions,  of the SYK calculations. At the same time, it is clear that this $d$-dimensional bosonic theory is not well-defined: the couplings are random, so generically the potential will have negative directions.\footnote{Bosonic tensor models in $d$ dimensions suffer from similar problems \cite{Klebanov:2016xxf}. See \cite{Gurau:2009tw, Bonzom:2011zz, 2016LMaPh.106.1531C, Witten:2016iux} for some earlier studies of tensor models. } This will manifest itself in  some unphysical properties of the four-point function; however, as far as the diagrammatics are concerned, these will be irrelevant. Another difference with SYK is that a free scalar field $\phi$ in $d>2$ is not dimensionless. As a result, in $d>2$, one must take  $q\leq \frac{2 d}{d-2}$, in order for the interaction term to be marginal or relevant (equivalently, if one were to take $q$ larger than this and take an ansatz of an infrared fixed point, one would  find  $\phi$  to have a dimension below the unitarity bound).\footnote{One could try to let $q$ take any value, by modifying the kinetic term to be bilocal rather than local, as was done for the SYK model in \cite{GR3}. However with bosons, unlike with fermions, one will then encounter UV divergences.} 

Assuming an infrared fixed point, and performing the standard SYK analysis of dropping the kinetic term in the Schwinger-Dyson equation,~\footnote{As noted in \cite{Murugan:2017eto}, this may not be legitimate.} the large $N$  two-point function for the $d$-dimensional SYK model, in the infrared, is \footnote{The conventions in one dimension are that $\chi$ has dimension $\Delta$, and the dimension of other operators is denoted by $h$. In $d$ dimensions, we will use $\Delta_{\phi}$ to denote the dimension of $\phi$, and a general operator will have dimension $\Delta$ and spin $J$.}
\be  \label{2ptt}
G(x) = b \frac{1}{|x|^{2\Delta_{\phi}}}~, \ \ \ \ \ \Delta_{\phi} = \frac{ d}{q}~,\ \ \ \ \ \ {\cJ}^2 b^q = -\frac{1}{\pi^d} \frac{\Gamma(\Delta_{\phi})\Gamma( d- \Delta_{\phi})}{\Gamma(\frac{d}{2} - \Delta_{\phi}) \Gamma( \Delta_{\phi} - \frac{d}{2})}~.
\ee

In the rest of this section we will compute the correlation functions in the $d$-dimensional SYK model, assuming the above infrared fixed point. Equivalently, one can regard our calculation as simply taking the Feynman diagrams that appear in the standard one-dimensional SYK model, and evaluating them in $d$ dimensions, using the above conformal propagator. 

\subsection{Four-point function in $1d$ SYK} \label{fourPoint}
We start with the four-point function of the fundamentals $\phi_i$. 
First, recall that in SYK, the fermion four-point function is a sum of conformal blocks of the fermion bilinear $\text{O}(N)$ singlets, schematically of the form 
\be
\mO_h = \frac{1}{N}\chi_i \partial_{\tau}^{1+2n} \chi_i~.
\ee
If the anomalous dimensions were zero, the dimensions of the operators would be $h=2\Delta + 2n +1$.
In $d$-dimensional SYK, the result will be similar, except the operators now have spin: the four-point function will be a sum of conformal blocks of the scalar bilinear $\text{O}(N)$ singlets,  schematically of the form,\footnote{Instead of the bosonic SYK model, one could instead study the bosonic tensor model \cite{Klebanov:2016xxf}, with fields $\phi_{a b c}$ transforming in the tri-fundamental representation of $\text{O}(N)^3$, with an interaction $\phi_{a_1 b_1 c_1} \phi_{a_1 b_2 c_2} \phi_{a_2 b_1 c_2} \phi_{a_2 b_2 c_1}$. At leading nontrivial order in $1/N$, the correlators of the singlets $\frac{1}{N^3}  \phi_{a_1 b_1 c_1} (\partial^2)^n\, \partial_{\mu_1} \cdots \partial_{\mu_J} \phi_{a_1 b_1 c_1}$ will be identical to the correlators of the $\mO_{\Delta, J}$ in SYK. }
\e{mO}{
\mO_{\Delta, J} = \frac{1}{N}  \phi_i (\partial^2)^n\, \partial_{\mu_1} \cdots \partial_{\mu_J} \phi_i~.}
If the anomalous dimensions were zero, then the dimension of $\mO_{\Delta,J}$ would be $\Delta=2 \Delta_{\phi} + 2n + J$.

\subsubsection{SYK}
We first briefly recall the results for the SYK four-point function, following the conventions in \cite{GR4}.
The fermion four-point function is \cite{MS}
\be \label{4Rep}
\mF(\tau_1, \ldots, \tau_4) =  G(\tau_{12}) G(\tau_{34})\int_{\mC}\frac{d h}{2\pi i}\, \rho(h) \Psi_h(\tau_i)~,
\ee
where 
\be
  \rho (h) = \mu(h) \frac{\alpha_0 }{2} \frac{ k(h)}{1- k(h)}~, \ \ \ \ \ \mu(h) = \frac{2h - 1}{\pi \tan \frac{\pi h}{2}}~,\ \ \ \ \, \alpha_0 = \frac{2\pi \Delta}{(1- \Delta) (2- \Delta) \tan \pi \Delta}~,
  \ee
  and the contour $\mC$ consists of a piece parallel to the imaginary axis $h = \frac{1}{2} + i s$, as well as counter-clockwise circles around $h=2n$ for positive integer $n$. 
Here $\Psi_h$ is a conformal partial wave, defined as~\footnote{In this paper, unlike in \cite{GR4}, we use conventions in which correlation function denote just the conformal structure, unless otherwise noted.}  \be \label{PsiIntr}
2\,  \frac{\sgn(\tau_{12})\,  \sgn(\tau_{34})}{ |\cJ \tau_{12}|^{2 \Delta} |\cJ \tau_{34}|^{2\Delta}}  \Psi_h(\tau_i) = \int d\tau_0 \, \langle \chi(\tau_1) \chi(\tau_2) \mO_h(\tau_0)\rangle \langle  \chi(\tau_3) \chi(\tau_4) \mO_{\tilde{h}}(\tau_0)\rangle~,
\ee
and is given by the sum of a conformal block with dimension $h$ and its shadow block of dimension $\tilde{h}=1-h$,
\be \label{BS1d}
\frac{2}{|\tau_{12}|^{2\Delta} |\tau_{34}|^{2\Delta}} \Psi_h(\tau_i) = \beta(h, 0)\mF_{\Delta}^{h}(\tau_i) + \beta(\tilde{h},0) \mF_{\Delta}^{\tilde{h}}(\tau_i)~,
\ee
where $\mF_{\Delta}^h(\tau_i)$ denotes the conformal block
\be \label{FermionBlock}
\mF_{\Delta}^{h} (\tau_i) = \frac{\sgn(\tau_{12})\, \sgn(\tau_{34})}{|\tau_{12}|^{2\Delta}|\tau_{34}|^{2\Delta}}\, x^h\, {}_2 F_1(h, h, 2h, x)~, \ \ \ \ \ x \equiv \frac{\tau_{12} \tau_{34}}{\tau_{13}\tau_{24}}~.
\ee
while the prefactor $\beta(h, \Delta)$ is, 
\be \label{betaD}
\beta(h,\Delta) =\sqrt{\pi} \frac{\Gamma(\frac{h+\Delta}{2}) \Gamma(\frac{h-\Delta}{2})}{\Gamma(\frac{\tilde{h}+\Delta}{2})\Gamma(\frac{\tilde{h}-\Delta}{2})}\frac{\Gamma(\frac{1}{2}- h)}{\Gamma(h)}~.
\ee
After some manipulations involving trading $h$ for $\tilde{h}$, the four-point function can be written as
\be \label{F1d}
\mF(\tau_1, \ldots, \tau_4) = \frac{b^2}{\cJ^{4 \Delta}}\int_{\mC}\frac{d h}{2\pi i} \rho(h) \frac{\Gamma(h)^2}{\Gamma(2h)}\, \mF_{\Delta}^h(\tau_i)~.
\ee
With this form, one can close the contour to the right, and pick up the poles of $\rho(h)$, obtaining a sum of conformal blocks. 

\subsubsection{Higher dimensions}
We now consider the scalar four-point function in $d$-dimensional bosonic SYK. We may write it as (compare with the one-dimensional case, Eq.~\ref{4Rep}), 
\be \label{Fd}
\mF(x_1, \ldots, x_4) = \sum_{J=0}^{\infty} \int_{\frac{d}{2}}^{\frac{d}{2} + i \infty} \frac{ d \Delta}{2\pi i} \rho(\Delta, J)\, \Psi_{\Delta, J}^{\Delta_{\phi}}(x_i)~,
\ee
for some $\rho(\De, J)$, which is found by summing the ladder diagrams, and will be presented later in Sec.~\ref{LadderSum}.  Here $\Psi_{\Delta, J}^{\Delta_{\phi}}$ is the conformal partial wave for dimension-$\De$ and spin-$J$ exchange between external $\phi$'s, with propagators for the external operators included. The sum and integral is over principal series representations of $\SO(d+1,1)$, which have $\De\in \frac{d}{2} + i\mathbb{R}$. The significance of these representations is that the corresponding partial waves $\Psi_{\De,J}^{\D_{\phi}}$ form a complete set of eigenfunctions of the conformal Casimir in Euclidean space \cite{Dobrev:1977qv}.

We may write  $\Psi_{\Delta, J}^{\Delta_{\phi}}$ as a sum of a conformal block and its shadow. In the notation of \cite{SD17} (compare with Eq.~\ref{BS1d}),\footnote{Our normalization for a conformal block is the same as in \cite{SD17}. Specifically, for cross-ratios satisfying $\chi\ll \bar\chi \ll 1$, the conformal block is a product of standard dimensionful factors $x_{ij}^{\#}$, times $\chi^{\frac{\De-J}{2}}\bar\chi^{\frac{\De+J}{2}}$. Depending on context, we will use ``conformal block'' to refer to the function $G_{\D,J}^{\D_i}(x_i)$ with or without the power-law prefactors.}
\be \label{BSd}
\Psi_{\Delta, J}^{\Delta_i} = K_{\tilde{\Delta}, J}^{\Delta_3, \Delta_4}\, G_{\Delta, J}^{\Delta_i}(x_i) + K_{\Delta, J}^{\Delta_1, \Delta_2} G_{\tilde{\Delta}, J}^{\Delta_i}(x_i)~,
\ee
where $\tilde{\Delta} = d- \Delta$ and $K_{\Delta, J}^{\Delta_1, \Delta_2}  =  (-\frac{1}{2})^{J} S_{[\Delta, J]}^{\Delta_1, \Delta_2}$ where 
\be \label{S123}
S^{\De_1,\De_2}_{[\De_3,J]} = \frac{\pi^{\frac d 2} \G(\De_3-\frac d 2)\G(\De_3+J-1)\G(\frac{\tl\De_3+\De_1-\De_2+J}{2})\G(\frac{\tl\De_3+\De_2-\De_1+J}{2})}{\G(\De_3-1) \G(d-\De_3+J) \G(\frac{\De_3+\De_1-\De_2+J}{2})\G(\frac{\De_3+\De_2-\De_1+J}{2})}~.
\ee
In our current case, the external dimensions $\Delta_i$ are equal to $\Delta_{\phi}$. Notice that in $d=1$, for $J=0$, we recover the previous result in Eq.~\ref{betaD},
\be
K_{[h, 0]}^{\Delta_1, \Delta_2} = \beta(\tilde{h}, \Delta_{12})~, \ \ \ \ \ d=1~.
\ee
We now insert (\ref{BSd}) into (\ref{Fd}) and change variables $\Delta\rightarrow \tilde{\Delta}$ for the second term,
\be \label{Fd2}
\mF(x_1, \ldots, x_4) = \sum_{J=0}^{\infty}\left( \int_{\frac{d}{2}}^{\frac{d}{2} + i \infty} \frac{ d \Delta}{2\pi i} \rho(\Delta, J)\ K_{\tilde{\Delta}, J}^{\Delta_{\phi}, \Delta_{\phi}}\, G_{\Delta, J}^{\Delta_{\phi}}(x_i) + \int_{\frac{d}{2} - i\infty}^{\frac{d}{2}} \frac{ d\Delta }{2\pi i} \rho(\tilde{\Delta}, J)\, K_{\tilde{\Delta}, J}^{\Delta_{\phi}, \Delta_{\phi}} G_{ \Delta, J}^{\Delta_{\phi}}(x_i)\right).
\ee
Using symmetries of the partial wave, one can show  $\rho(\De,J) = \rho(\tilde{\Delta},J)$. Thus, we may combine the  two terms above,
\be \label{Fd3}
\mF(x_1, \ldots, x_4) = \sum_{J=0}^{\infty} \int_{\frac{d}{2} - i\infty}^{\frac{d}{2} + i \infty} \frac{ d \Delta}{2\pi i} \rho(\Delta, J)\ K_{\tl\Delta, J}^{\Delta_{\phi}, \Delta_{\phi}}\, G_{\Delta, J}^{\Delta_{\phi}}(x_i)~.
\ee
This expression should be compared with the one-dimensional equivalent, Eq.~\ref{F1d}. Closing the contour to the right, where the conformal block $G_{\De,J}^{\De_\f}$ is exponentially damped, we get a sum of conformal blocks,
\be \label{Fd4}
\mF(x_1, \ldots, x_4) = \sum_{J=0}^{\infty} \sum_{\Delta = \Delta_n} c_{\Delta,J}^2 G_{\Delta, J}^{\Delta_{\phi}}(x_i)~,
\ee
where the OPE coefficients $\phi_i \phi_i \sim c_{\Delta_n, J}^2 \mO_{\Delta_n, J}$ are, 
\be \label{221}
c_{\Delta_n, J}^2 =- \text{Res }  \rho(\Delta, J) K_{\tl\Delta, J}^{\Delta_{\phi}, \Delta_{\phi}} \Big|_{\Delta = \Delta_n}~.
\ee
These coefficients are what appear in a physical three-point function, 
\be \label{3pttt}
\langle \phi(x_1) \phi(x_2) \mO_{\Delta, J}(x_0)\rangle_{\mathrm{phys} } = c_{\Delta, J}\,b\, \langle \phi(x_1) \phi(x_2) \mO_{\Delta, J}(x_0)\rangle~,
\ee
where the right-hand side denotes just the conformal structure of the correlator.

\subsubsection{Summing ladders in $d$ dimensions} \label{LadderSum}
In this section, we perform the computation of the four-point function for the $d$-dimensional model by explicitly summing the ladder diagrams to determine $\rho(\D,J)$ in \eqr{Fd3}. This may be done via a conformally-invariant inner product between $\mathcal{F}$ and a partial wave $\Psi^{\tl \De_\f}_{\tl \De,J}$.

We first establish notation for conformal three-point integrals, which occur throughout the calculation. Given an operator with dimension $\De$ and spin $J$, the shadow representation $\tl \cO$ has dimension $\tl \De=d-\De$ and spin $J$. The shadow transform is a conformally invariant map between the representation of $\cO$ and the representation of $\tl \cO$. It is defined by\footnote{Here, if $\cO$ transforms in the $\SO(d)$ representation $\rho$, then $\cO^\dag$ transforms in the dual reflected representation $(\rho^R)^*$.
We will work almost exclusively with traceless symmetric tensors, so the representation of $ \cO^\dag$ is equivalent to the representation of $\cO$. Consequently we will often omit $\dagger$'s.}
\be \label{ShadowT}
\mathbf{S}[\mO](x) = \int d^d y\,  \langle \tl \mO(x) \tl \mO^\dag(y)\rangle \mO(y)~.
\ee
Here, $\<\tl \cO(x)\tl \cO^\dag(y)\>$ denotes the unique conformally invariant two-point structure for the given representation.
Shadow coefficients are defined by,
\be
\langle \mathbf{S}[\mO_1] \mO_2 \mO_3\rangle = S_{\mO_1}^{\mO_2 \mO_3}\langle \tl \mO_1 \mO_2 \mO_3\rangle~,
\ee
where $\<\tl \cO_1\cO_2\cO_3\>$ denotes a three-point structure for the given representations. (When the operators have spin, there can be multiple three-point structures, and the shadow coefficient becomes a matrix.)
The result for $S_{[\Delta_3, J]}^{\Delta_1, \Delta_2}$ was stated earlier (\ref{S123}), while, 
\be \label{Sfac}
S_{\Delta_2}^{\Delta_1, [\Delta_3, J]} = \frac{\pi^{\frac d 2}\G(\De_2 - \frac d 2)\G(\frac{\tl \De_2 + \De_1 - \De_3+J}{2}) \G(\frac {\tl \De_2 + \De_3 - \De_1+J}{2}) }{\G(d-\De_2)\G(\frac{\De_2+\De_1-\De_3+J}{2}) \G(\frac{\De_2+\De_3 - \De_1+J}{2})}~.
\ee
We give an elementary derivation of (\ref{S123}) and (\ref{Sfac}) in appendix~\ref{shadow}. They can also be computed efficiently using weight-shifting operators or Fourier space \cite{Karateev:2017jgd}. When $J=0$, the shadow transform of a three-point function becomes the famous star-triangle integral \cite{DEramo:1971hnd}. 

\begin{figure}[t]
\centering
\includegraphics[width=5in]{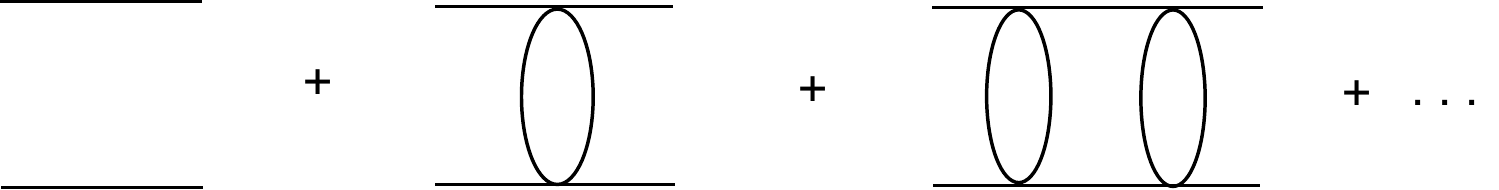}
\caption{The four-point function is a sum of ladder diagrams. } \label{ladder}
\end{figure} 
The four-point function is given by a sum of ladder diagrams, as shown in Fig.~\ref{ladder}.\footnote{Very similar ladder diagrams appear in  correlation functions of certain operators  in the fishnet theory \cite{Gurdogan:2015csr, Caetano:2016ydc,Gromov:2017cja,Grabner:2017pgm, Kazakov:2018qbr, Cavaglia:2018lxi,Basso:2018agi}.}
As is standard in SYK, one finds the spectrum of operators $\mO_{\Delta, J}$ by diagonalizing the kernel that adds a rung to the ladder \cite{Kitaev, PR, MS, GR1}.
The eigenvector equation is, 
\be \label{EigEqn}
\int d^d x_a d^d x_b\, K(x_1, x_2; x_a, x_b)  v(x_0; x_a, x_b) = k(\Delta, J)  v(x_0; x_1, x_2)~,
\ee
where the  kernel is, 
\be
K(x_1, x_2; x_a, x_b) = \cJ^2 (q-1) G(x_{1a}) G(x_{2b}) G(x_{ab})^{q-2}~.
\ee
We take the eigenvectors to be conformal three-point functions, 
\be
v(x_0; x_a, x_b) = \langle \phi(x_a) \phi(x_b) \mO_{\Delta, J}^{}(x_0)\rangle ~. 
\ee
Inserting these into the kernel, we find that the left-hand side of (\ref{EigEqn}) can be rewritten as, 
\bea \nonumber
&&\hspace{-1cm} \int\! d^d x_a d^d x_b\, K(x_1, x_2; x_a, x_b)  v(x_0; x_a, x_b) \\ \nonumber
\!\!\! &= &\!\!\!\!  (q\!-\!1) \cJ^2 b^q \int \! d^d x_a d^d x_b\, \langle \phi(x_2) \phi(x_b)\rangle\,  \langle \phi(x_1) \phi(x_a) \rangle\,  \langle \tl \phi (x_a) \tl \phi (x_b) \mO_{\Delta, J}(x_0) \rangle \\ 
 \!\!\! &= &\!\!\!\!  (q\!-\!1) \cJ^2 b^q S_{\tl \Delta_{\phi}}^{\tl \Delta_{\phi}, [\Delta, J]}\, S_{\tl \Delta_{\phi}}^{\Delta_{\phi}, [\Delta, J]} \langle \phi(x_1) \phi(x_2) \mO_{\Delta, J}(x_0)\rangle~.
\eea
Thus, the eigenvalues of the kernel are \cite{Giombi:2017dtl}
\be \label{kernelE}
k(\Delta, J) =(q-1)  \cJ^2 b^q  S_{\tl \Delta_{\phi}}^{\tl \Delta_{\phi}, [\Delta, J]}\, S_{\tl \Delta_{\phi}}^{\Delta_{\phi}, [\Delta, J]}~.
\ee
The $\Delta$ for which $k(\Delta, J) =1$ correspond to the dimensions of the bilinear operators $\mO_{\Delta, J}$ of (\ref{mO}). 

We can now proceed to sum the ladder diagrams. 
The zero-rung ladder is proportional to the four-point function of Mean Field Theory (MFT),
\be
\label{eq:zerorung}
b^2 \mathcal{F}^\mathrm{MFT}(x_i) = b^2 \<\f(x_1)\f(x_3)\>\<\f(x_2)\f(x_4)\> + (1\leftrightarrow 2)~.
\ee
We would like to express $\mathcal{F}^\mathrm{MFT}$ as a conformal partial wave expansion,
\be
\mathcal{F}^\mathrm{MFT}(x_i) = \sum_{J=0}^\oo \int_{\frac d 2}^{\frac d 2 + i\oo} \frac{d\De}{2\pi i} \rho^\mathrm{MFT}(\De,J) \Psi_{\De,J}^{\De_\f}(x_i)~.
\ee
This computation was originally done in \cite{Fitzpatrick:2011dm}. We will follow essentially the same method, though with slightly different language. This method is generalized to arbitrary spinning operators in \cite{MFTfuture}.
We first take a conformally invariant inner product of both sides with a partial wave,
\be
\left(\mathcal{F}^\mathrm{MFT},\Psi^{\tl \De_\f}_{\tl \De,J}\right) = \rho^\mathrm{MFT}(\De,J) n_{\De,J}~.
\ee
The conformally invariant inner product is defined by,
\be
\left(\mathcal{F},\mathcal{G}\right) = \int \frac{d^dx_1d^dx_2d^dx_3 d^dx_4}{\vol(\SO(d+1,1))} \mathcal{F}(x_i)\,\mathcal{G}(x_i)~,
\ee
and the normalization factor $n_{\De,J}$ is given by \cite{SD17}\footnote{Our expression for $n_{\De,J}$ differs by a factor of $2^{-d}$ from the one in \cite{SD17}, see below.}
\be
\left(\Psi^{ \De_\f}_{\frac d 2 + is,J}\,,\Psi^{\tl \De_\f}_{\frac d 2-is',J'}\right) &= 2\pi\delta(s-s')\delta_{J,J'} n_{\De,J}~, \nn\\
n_{\De,J} &= \frac{K^{\De_3,\De_4}_{\tl \De,J} K^{\tl \De_3,\tl \De_4}_{\De,J} \vol(S^{d-2})}{2^d\vol(\SO(d-1))} \frac{(2J+d-2)\pi \G(J+1)\G(J+d-2)}{2^{d-2} \G(J+\tfrac d 2)^2}~.
\ee
(Despite appearances, $n_{\De,J}$ is actually independent of $\De_3,\De_4$.)\footnote{Our definition of $\vol(\SO(n))$ is that $\vol(\SO(n))/\vol(\SO(n-1)) = \vol(S^{n-1})$. In all the physical quantities in this paper, the volumes of orthogonal groups will appear in ratios that give volumes of spheres.}

To compute the inner product, we use the shadow representation of the partial wave,
\be
\left(\mathcal{F}^\mathrm{MFT},\Psi^{\tl \De_\f}_{\tl \De,J}\right) &\supset
\int\frac{d^dx_1\cdots d^dx_5}{\vol(\SO(d+1,1))} \<\f(x_1)\f(x_3)\>\<\f(x_2)\f(x_4)\> \nn\\
&\qquad\<\tl \f(x_1)\tl \f(x_2) \cO_{\De,J}^{\mu_1\cdots\mu_J}(x_5)\>\<\tl \cO_{\De,J;\mu_1\cdots\mu_J}(x_5)\tl \f(x_3)\tl \f(x_4)\>~.
\ee
Here we have only written the contribution of the first term in (\ref{eq:zerorung}).
From the point of view of the $x_1,x_2$ integrals this is the exact same computation we did for the kernel. The result is a product of shadow factors
\be
&=S_{\tl \Delta_{\phi}}^{\tl \Delta_{\phi}, [\Delta, J]}\, S_{\tl \Delta_{\phi}}^{\Delta_{\phi}, [\Delta, J]} \int\frac{d^dx_3d^d x_4 d^d x_5}{\vol(\SO(d+1,1))} \<\f(x_3) \f(x_4) \cO_{\De,J}^{\mu_1\cdots\mu_J}(x_5)\>\<\tl \cO_{\De,J;\mu_1\cdots\mu_J}(x_5)\tl \f(x_3)\tl \f(x_4)\> \nn\\
&= S_{\tl \Delta_{\phi}}^{\tl \Delta_{\phi}, [\Delta, J]}\, S_{\tl \Delta_{\phi}}^{\Delta_{\phi}, [\Delta, J]}  
\left(\<\f\f\cO\>,\<\tl \f \tl \f\tl\cO\>\right)~.
\ee
The remaining factor is a conformal three-point integral. As explained in \cite{SD17}, it doesn't require any actual integration because one can fix the positions of the three points using conformal invariance. Setting $(x_3,x_4,x_5)=(0,e,\oo)$ for some unit vector $e$, the three-point pairing becomes,
\be
t_0\equiv \left(\<\f\f\cO\>,\<\tl \f \tl \f\tl\cO\>\right) &=
\frac{1}{2^d\vol(\SO(d-1))} \<\f(0) \f(1) \cO_{\De,J}^{\mu_1\cdots\mu_J}(\oo)\>\<\tl \cO_{\De,J;\mu_1\cdots\mu_J}(\oo)\tl \f(0)\tl \f(1)\> \nn\\
&=
\frac{\hat C_J(1)}{2^d\vol(\SO(d-1))}~. \label{t0}
\ee
Here, $\SO(d-1)$ is the stabilizer group of three points, $2^{-d}$ is the Faddeev-Popov determinant for our gauge-fixing,\footnote{Reference \cite{SD17} used a different convention for $\vol(\SO(d+1,1))$ which did not include the factor $2^{-d}$. We have chosen to include it because it leads to nicer expressions for several quantities, such as the generalized Lorentzian inversion formula \cite{Kravchuk:2018htv}.} and,
\be
\hat C_J(1) &= (e^{\mu_1}\cdots e^{\mu_J}-\textrm{traces})(e_{\mu_1}\cdots e_{\mu_J}-\textrm{traces}) = \frac{\G(\frac{d-2}{2})\G(J+d-2)}{2^J \G(d-2)\G(J+\frac{d-2}{2})}~.
\ee

Putting everything together, and including the $1\leftrightarrow 2$ term in (\ref{eq:zerorung}), which contributes the same thing times $(-1)^J$, we find
\be \label{rhomft}
 \rho^\mathrm{MFT}(\De,J) 
 &=
 \frac{(1+(-1)^J)}{n_{\De,J}}\,t_0\, S_{\tl \Delta_{\phi}}^{\tl \Delta_{\phi}, [\Delta, J]}\, S_{\tl \Delta_{\phi}}^{\Delta_{\phi}, [\Delta, J]}  \nn\\
 &= \frac{(1+(-1)^J)}{n_{\De,J}} \frac{t_0}{(q-1)\cJ^2 b^q}k(\De,J)~.
\ee
Our final result for the sum over ladders is thus
\be \label{rhogen}
\rho(\De,J) &= \frac{b^2 \rho^{\mathrm{MFT}}(\De,J)}{1-k(\De,J)} =  \frac{(1+(-1)^J)}{n_{\De,J}}  \frac{b^2\,t_0}{(q-1)\cJ^2 b^q}\frac{k(\De,J)}{1-k(\De,J)}~.
\ee
Plugging into \eqr{Fd3} gives the four-point function.

\subsection{Three-point function of bilinears} \label{6pt}
%{\color{red} Draw diagrams.}

In this section we consider the three-point function of the singlet,  $\text{O}(N)$-invariant, bilinears of the fundamental fields $\phi_i$. The analysis in $d$ dimensions closely resembles that of $d=1$ \cite{GR4, GR2}. There are two classes of Feynman diagrams that contribute at leading nontrivial order in $1/N$: planar diagrams, which we study in this section and show are related to $6j$ symbols, and nonplanar ``contact'' diagrams which are discussed in Appendix~ \ref{contact}. We correspondingly write the three-point function as,
\be\label{3phys}
\langle \mO_{\Delta_1, J_1}\mO_{\Delta_2, J_2}\mO_{\Delta_3, J_3}\rangle_\mathrm{phys} = \langle \mO_{\Delta_1, J_1}\mO_{\Delta_2, J_2}\mO_{\Delta_3, J_3}\rangle_1 + \langle \mO_{\Delta_1, J_1}\mO_{\Delta_2, J_2}\mO_{\Delta_3, J_3}\rangle_2~,
\ee
where the subscript $1$ denotes the contribution of the contact diagrams, and subscript $2$ denotes the contribution of the planar diagrams.\footnote{Intriguingly, similar diagrams appear in the 
calculation of the  interaction vertex of three BFKL pomerons \cite{Korchemsky:1997fy, Bialas:1997xp, Balitsky:2015tca}. We thank Evgeny Sobko for this observation. } In this work, correlators without subscripts $\<\cO \cO\>$ and $\<\cO_1\cO_2\cO_3\>$ represent conformally-invariant structures for the given representations --- i.e.\ they are  known functions and do not include OPE coefficients. We include the subscript ``phys" on the left-hand side of (\ref{3phys}) to emphasize that it represents a physical correlation function --- i.e.\ it {\it does\/} include an OPE coefficient.

The contribution of the planar diagrams is given by gluing together three partially amputated three-point functions \cite{GR2}, as depicted in Fig.~\ref{tetraCut2}(a), 
\begin{multline}\label{amp1}
 \langle \mO_{\Delta_1, J_1}(x_1)\mO_{\Delta_2, J_2}(x_2) \mO_{\Delta_3, J_3}(x_3)\rangle_2 = \int d^d x_a d^d x_b d^d x_c \langle \mO_{\Delta_1, J_1}(x_1) \phi(x_a) \phi(x_b)\rangle_{\text{amp}} \\ \langle \mO_{\Delta_2, J_2}(x_2)  \phi(x_c) \phi(x_a)\rangle_{\text{amp}}  \langle \mO_{\Delta_3, J_3}(x_3) \phi(x_b)  \phi(x_c) \rangle_{\text{amp}}~,
 \end{multline}
where the partially amputated three-point function involves stripping of the propagator on the last leg. Since the inverse of the propagator is proportional to  the two-point function of the shadow of $\phi$, we have that
\bea \nonumber  \label{ampeq}
 \langle \mO_{\Delta_1, J_1}(x_1) \phi(x_a) \phi(x_b)\rangle_{\text{amp}} &=&\cJ ^2 b^{q-1} \int d^d x_0 \langle \mO_{\Delta_1, J_1}(x_1) \phi(x_a) \phi(x_0)\rangle_\mathrm{phys} \langle \tl \phi(x_0) \tl \phi(x_b)\rangle\\
 & =& c_{\Delta_1, J_1}\, \cJ^2 b^q\,  S_{\Delta_{\phi}}^{\Delta_{\phi}, [\Delta_1, J_1]} \langle \mO_{\Delta_1, J_1}(x_1) \phi(x_a) \tl \phi(x_b)\rangle~,
 \eea
where in getting from the first line to the second we made use of (\ref{3pttt}), and recall that  $b^q$ is the normalization of the two-point function in Eq. \eqr{2ptt}, and $S_{\Delta_{\phi}}^{\Delta_{\phi}, [\Delta, J]}$ was given in Eq.~\ref{Sfac}. 
Thus, our three-point function is 
\e{OOO}{\langle \mO_{\Delta_1, J_1}(x_1)\mO_{\Delta_2, J_2}(x_2) \mO_{\Delta_3, J_3}(x_3)\rangle_2  = \left(\prod_{i=1}^3 c_{\Delta_i, J_i}\cJ^2 b^q\,   S_{\Delta_{\phi}}^{\Delta_{\phi}, [\Delta_i, J_i]}\right) \, I_{\Delta_i, J_i}^{(2)} (x_i)~,}
where
\be \label{I2v2}
  I_{\Delta_i, J_i}^{(2)} (x_i) = \int d^d x_a d^d x_b d^d x_c \langle \mO_{\Delta_1, J_1}(x_1) \phi(x_a)\tl \phi(x_b)\rangle  \langle \mO_{\Delta_2, J_2}(x_2)  \phi(x_c) \tl \phi(x_a)\rangle  \langle \mO_{\Delta_3, J_3}(x_3) \phi(x_b)\tl  \phi(x_c) \rangle~.
\ee
This is a straight-forward, yet formidable, integral. In one dimension, it was evaluated explicitly in \cite{GR4}. 
Here, we need only one additional step to notice that it is in fact a $6j$ symbol (see e.g.\ \cite{Gadde:2017sjg}). We note that, by conformal invariance, the functional form of the result is fixed, 
\be\label{246}
   I_{\Delta_i, J_i}^{(2)} (x_i)  = \sum_a \mathcal{I}_{\Delta_i, J_i,a}^{(2)}\langle \mO_{\Delta_1, J_1}(x_1)\mO_{\Delta_2, J_2}(x_2) \mO_{\Delta_3, J_3}(x_3)\rangle^a~,
   \ee
where the index $a$ runs over possible three-point structures in a three-point function of operators with spins $J_1,J_2,J_3$.\footnote{These structures were classified in~\cite{Kravchuk:2016qvl}.} To get the coefficient we are after, we contract both sides with a shadow three-point structure, 
 \be\label{247}
(t_0)^{ab}\,\mathcal{I}_{\Delta_i, J_i,a}^{(2)} = \int \frac{d^d x_1 d^d x_2 d^d x_3}{\text{vol}(\text{SO}(d+1,1))}\,   I_{\Delta_i, J_i}^{(2)} (x_i) \langle \tl \mO_{\Delta_1, J_1}(x_1) \tl \mO_{\Delta_2, J_2}(x_2) \tl \mO_{\Delta_3, J_3}(x_3)\rangle^b
\ee
where the constant $t_0$ is simply a three-point pairing,
\be\label{t0ab}
(t_0)^{ab} &= \int \frac{d^d x_1 d^d x_2 d^d x_3}{\text{vol}(\text{SO}(d+1,1))}\,   \langle \mO_1 \mO_2 \mO_3\rangle^a \langle \tl \mO_1 \tl \mO_2 \tl \mO_3\rangle^b \nn\\
&= \frac{1}{2^d\vol(\SO(d-1))}\langle \mO_1(0) \mO_2(e) \mO_3(\oo)\rangle^a \langle \tl \mO_1(0) \tl \mO_2(e) \tl \mO_3(\oo)\rangle^b~.
\ee
For the case that two of the three  operators are scalars, $t_0$ was given  previously in (\ref{t0}). Using (\ref{I2v2}), we can therefore write, 
\begin{multline}
\label{eq:sixjexpression}
\mathcal{I}_{\Delta_i, J_i;a}^{(2)} = (t_0^{-1})_{ab}  \int \frac{d^d x_1 d^d x_2 d^d x_3\, d^d x_a d^d x_b d^d x_c}{\text{vol}(\text{SO}(d+1,1))}\langle \mO_{\Delta_1, J_1}(x_1) \phi(x_a)\tl \phi(x_b)\rangle  \langle \mO_{\Delta_2, J_2}(x_2)  \phi(x_c) \tl \phi(x_a)\rangle \\
 \langle \mO_{\Delta_3, J_3}(x_3) \phi(x_b)\tl  \phi(x_c) \rangle \langle \tl \mO_{\Delta_1, J_1}(x_1)  \tl \mO_{\Delta_2, J_2}(x_2) \tl \mO_{\Delta_3, J_3}(x_3)\rangle^b~.
\end{multline}
We now have an integral of four three-point structures, glued into a ``tetrahedron" graph. Specifically, each pair of three-point structures shares exactly one coordinate. This is a particular example of a $6j$ symbol for the conformal group $\SO(d+1,1)$.

The $6j$ symbol can also be thought of as an overlap between conformal partial waves as follows. For simplicity, consider the case $J_1=J_2=0$. There is then a unique three-point structure for $I_{\De_i,J_i}^{(2)}$, and we can drop the structure label $a$. Performing the integrals over $x_a$ and $x_3$ in (\ref{eq:sixjexpression}), we obtain,
\be \label{3ptF}
\mathcal{I}_{\Delta_i, J_i}^{(2)}  = \frac{1}{t_0} \int\frac{ d^d x_1 d^d x_2 d^d x_b d^d x_c} {\text{vol}(\text{SO}(d+1,1))} \,  \Psi^{\tl \Delta_2, \tl \Delta_1, \Delta_{\phi}, \tl \Delta_{\phi}}_{\tl \Delta_3,J_3} (x_2, x_1, x_b, x_c) \Psi^{\tl \Delta_{\phi},\Delta_1, \Delta_2,  \Delta_{\phi}}_{\Delta_{\phi},0}(x_b, x_1, x_2, x_c) \,~.
\ee
To go further, we must actually compute some $6j$ symbols, which we will do in Sec.~\ref{sec:6j}. There we will denote a $6j$ symbol as $\mathcal{J}_d(\Delta,J;\Delta',J'|\Delta_1,\Delta_2,\Delta_3,\Delta_4)$ (see Eqs.~\eqr{6j} and \eqr{jdef}). In that notation,
\e{i6j}{\mathcal{I}_{\Delta_i, J_i}^{(2)}  =\frac{1}{t_0} \mathcal{J}_d(\Delta_3,J_3;\Delta_\phi,0|\Delta_2,\Delta_1,\tilde{\Delta}_\phi,\Delta_\phi)~.}

\subsection{All-point correlation functions} \label{allPoint}
\begin{figure}[t]
\centering
\includegraphics[width=3in]{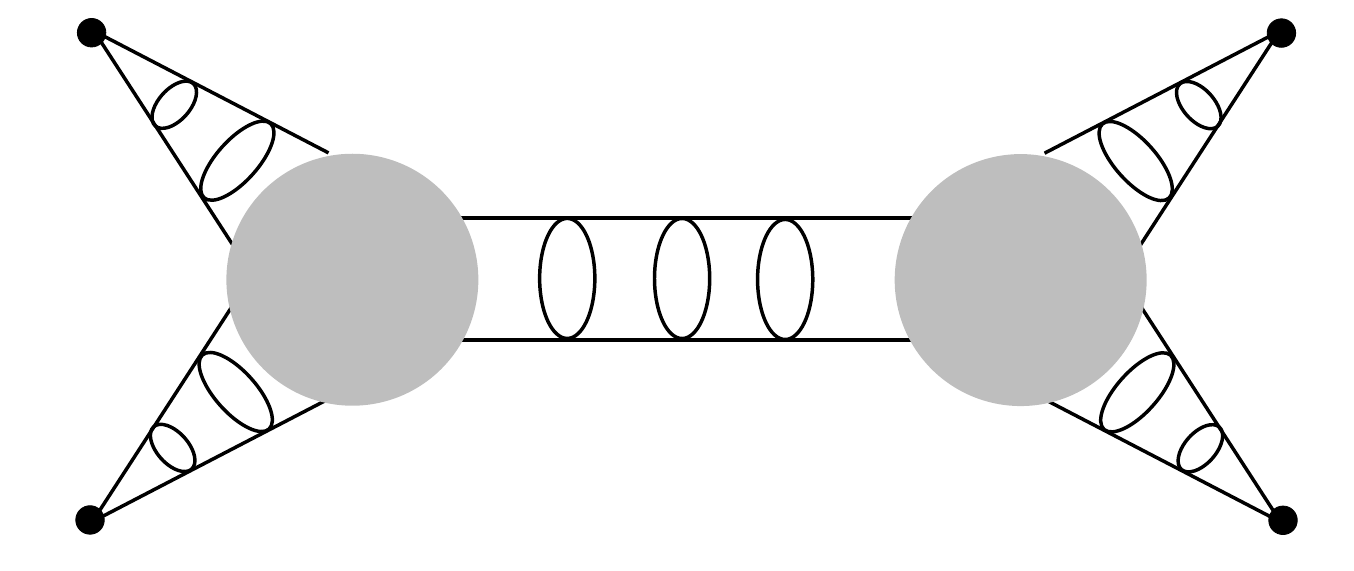}
\caption{A contribution to the bilinear four-point function. The shaded blob refers to the six-point function of fundamentals; for SYK, the six-point function is given by a sum of the planar and contact diagrams, though this is not relevant for our calculation, which relies only on the conformal properties of the ingredients.}\label{8split}
\end{figure} 
The three-point function of bilinears provides the basic building block for constructing higher-point correlation functions. In particular, the Feynman diagrams contributing to a four-point function of bilinears consist of ladder diagrams glued together, as in Fig.~\ref{8split}, summed over the three channels, with the diagram with no exchanged melons subtracted, and a contact diagram added. This is discussed in detail in one dimension in \cite{GR4}. The discussion in higher dimensions is nearly identical, the only difference being slightly more involved notation, to account for the spin of the operators.  Here we will evaluate the diagram in Fig.~\ref{8split}. One should note that the result is general, and holds for any four-point functions that are glued together in this way; it only relies on having a CFT, and expresses the sum of the diagrams of the type in Fig.~\ref{8split} in terms of the four-point function of fundamentals and the three-point function of bilinears.\footnote{The three-point function of bilinears that enters  is really an analytically extended three-point function: $\langle \mO_1 \mO_2 \mO_3\rangle$ viewed as an analytic function of the dimensions $\Delta_i$. We know this function, since in the computation in the previous section we were allowed to take the dimensions $\Delta_i$ to be arbitrary.}

An arbitrary three-point function of operators with spin may be written as
\e{ThreeCoeff}{\langle \mO_{\Delta_1, J_1}(x_1)\mO_{\Delta_2, J_2}(x_2) \mO_{\Delta_3, J_3}(x_3)\rangle_\mathrm{phys}  = \sum_a C_{\D_i,J_i; a}  \langle \mO_{\Delta_1, J_1}(x_1)\mO_{\Delta_2, J_2}(x_2) \mO_{\Delta_3, J_3}(x_3)\rangle^a~.}
 The index $a$ runs over possible three-point structures in a three-point function of operators with spins $J_1,J_2,J_3$.
 We will correspondingly also need a slightly more general form of the  conformal partial wave, in which the external operators have spin,
\be \nonumber %\label{PsiSpin}
 \Psi_{\Delta, J}^{\Delta_i, J_i,  a,b}(x_i)=\! \int d^d x_5\,  \langle \mO_{\Delta_1, J_1}(x_1) \mO_{\Delta_2, J_2}(x_2) \mO_{\Delta, J}(x_5)^{\mu_1 \cdots \mu_J}\rangle^a%\\
\langle \tilde \mO_{\Delta, J;\, \mu_1 \cdots \mu_J}(x_5)\mO_{\Delta_3, J_3}(x_3) \mO_{\Delta_4, J_4}(x_4)\rangle^b ~.
\ee
An equivalent form of this expression  that will be useful is in terms of the shadow transform,
%\begin{multline}
\es{}{\Psi _{\Delta ,J}^{{\Delta _i},{J_i},a,b}({x_i}) = [{(S_{[\Delta ,J]}^{[{\Delta _3},{J_3}],[{\Delta _4},{J_4}]})^{ - 1}}]{^b_c}\int {{d^d}} {x_5}\,& \langle {\mO_{{\Delta _1},{J_1}}}({x_1}){\mO_{{\Delta _2},{J_2}}}({x_2}){\mO_{\Delta ,J}}{({x_5})^{{\mu _1} \cdots {\mu _J}}}\rangle^a\\
&\langle {\bf{S}}[\mO_{\Delta ,J}]_{{\mu _1} \cdots {\mu _J}}({x_5}){\mO_{{\Delta _3},{J_3}}}({x_3}){\mO_{{\Delta _4},{J_4}}}({x_4})\rangle^c~.}
%\end{multline}
If one were to do the integral,  it would give a conformal block and its shadow,
\be \label{BSdG}
\Psi _{\Delta ,J}^{{\Delta _i},{J_i};a,b} = (K_{\tilde \Delta ,J}^{[{\Delta _3},{J_3}],[{\Delta _4},{J_4}]}\,){^b_c}G_{\Delta ,J}^{{\Delta _i},{J_i};a,c}({x_i}) + (K_{\Delta ,J}^{[{\Delta _1},{J_1}],[{\Delta _2},{J_2}]}\,){^a_c}G_{\tilde \Delta ,J}^{{\Delta _i},{J_i};c,b}({x_i})~.
\ee
We will not need to know the explicit form of  the $K$ prefactors that appear here. This partial wave is just a generalization of the partial wave (\ref{BSd}) to include external spins. 

Now, turning to the diagram in Fig.~\ref{8split}, to evaluate it we make use of the internal four-point function of fundamentals, in the form of (\ref{Fd}) with the internal partial wave written in terms of the split representation. We then recognize that the result involves two three-point functions, 
\bea \label{4s1}
&&\langle \mO_{\Delta_1, J_1}(x_1) \mO_{\Delta_2, J_2}(x_2) \mO_{\Delta_3, J_3}(x_3) \mO_{\Delta_4, J_4}(x_4)\rangle_s  = \sum_{J=0}^{\infty}\int_{\frac{d}{2}}^{\frac{d}{2} + i \infty} \frac{d\Delta}{2\pi i}\,\frac{ \rho(\Delta, J)}{c_{\Delta, J}\, c_{\tilde{\Delta}, J}}\,\\ \nonumber
&&\qquad\int d^d x_5 \langle \mO_{\Delta_1, J_1}(x_1) \mO_{\Delta_2, J_2}(x_2) \mO_{\Delta, J}(x_5)^{\mu_1 \cdots \mu_J}\rangle\langle \tilde \mO_{\Delta, J;\, \mu_1 \cdots \mu_J}(x_5)\mO_{\Delta_3, J_3}(x_3) \mO_{\Delta_4, J_4}(x_4)\rangle ~,
\eea
where the subscript refers to the $s$-channel. To go further, we use the general form of the three-point function (\ref{ThreeCoeff}), where we separate out the OPE coefficients of $ \phi \phi \sim  c_{\Delta_i, J_i}\mO_{\Delta_i J_i}$, 
  \be\label{255}
C_{\Delta_i,J_i; a} = (\prod_{i=1}^3 c_{\Delta_i, J_i})\, \,  \, \mathcal{I}_{\Delta_1, J_1, \Delta_2, J_2, \Delta_3, J_3; a}~.
\ee
This serves as the definition of $ \mathcal{I}_{\Delta_1, J_1, \Delta_2, J_2, \Delta_3, J_3; a}$. For SYK, the OPE coefficients $c_{\Delta_i, J_i}$ was given in (\ref{221}), while the three-point structure constant was a sum of the contributions from the contact and planar diagrams,
\be\label{257}
 \mathcal{I}_{\Delta_1, J_1, \Delta_2, J_2, \Delta_3, J_3; a} =(q-1)(q-2) {\cJ^2} b^q\,  \mathcal{I}_{\Delta_i, J_i;a}^{(1)}+ \left(\prod_{i=1}^3 \cJ^2 b^q\,   S_{\Delta_{\phi}}^{\Delta_{\phi}, [\Delta_i, J_i]}\right) \, \mathcal{I}_{\Delta_i, J_i;a}^{(2)}~.
 \ee 
where $\mathcal{I}^{(2)}_{\D_i,J_i;a}$ is a $6j$ symbol. However, our discussion will be completely general and  will not make use of this. 

With this we have that (\ref{4s1}) becomes,  
\be\nonumber
\langle \mO_{1}(x_1) \cdots \mO_{4}(x_4)\rangle_s 
= ( \prod_{i=1}^4 c_{\Delta_i, J_i})\, \sum_{J=0}^{\infty}\int_{\frac{d}{2}}^{\frac{d}{2} + i \infty} \frac{d\Delta}{2\pi i}\, \rho(\Delta, J) \,\mathcal{I}_{\Delta_1, J_1, \Delta_2, J_2, \Delta, J; a}\mathcal{I}_{\tilde{\Delta}, J, \Delta_3, J_3, \Delta_4, J_4; b}\, \Psi_{\Delta, J}^{\Delta_i, J_i,  a,b}(x_i)~.
\ee
This is almost our final answer. A more useful form will be one which just involves the conformal block, rather than the full partial wave. By applying the shadow transform, either to the $\mathcal{I}_{\tilde{\Delta}, J, \Delta_3, J_3, \Delta_4, J_4; b}$ above, or to the four-point function of the fundamentals we originally used, we can write this as
\es{}{&\langle \mO_{1}(x_1) \cdots \mO_{4}(x_4)\rangle_s= 
(\prod\limits_{i = 1}^4 {{c_{{\Delta _i},{J_i}}}} )\\&\times\sum\limits_{J = 0}^\infty  {\int_{\frac{d}{2}}^{\frac{d}{2} + i\infty } {\frac{{d\Delta }}{{2\pi i}}} } \,\,\frac{{(S_{[\Delta ,J]}^{[{\Delta _3},{J_3}],[{\Delta _4},{J_4}]})_b^c}}{{S_{[\Delta ,J]}^{{\Delta _\phi },{\Delta _\phi }}}}\rho (\Delta ,J)\,{{\cal I}_{{\Delta _1},{J_1},{\Delta _2},{J_2},\Delta ,J;a}}{{\cal I}_{\Delta ,J,{\Delta _3},{J_3},{\Delta _4},{J_4};c}}\,\Psi _{\Delta ,J}^{{\Delta _i},{J_i},a,b}({x_i})~.}
Notice that in this expression, in comparison to the one on the previous line, $\mathcal{I}_{\tilde{\Delta}, J, \Delta_3, J_3, \Delta_4, J_4; b}$ has been replaced by $\mathcal{I}_{{\Delta}, J, \Delta_3, J_3, \Delta_4, J_4; b}$.
We now insert the expression for the partial wave in terms of conformal blocks (\ref{BSdG}), and change integration variables on the second term, sending $\Delta \rightarrow d- \Delta$. After some shadow transforming of the second term, the integrands of both terms become the same, and we combine them to get
\begin{multline} \label{4Ans}
\langle \mO_{1}(x_1) \cdots \mO_{4}(x_4)\rangle_s =\\
 \hspace{-.5cm} ( \prod_{i=1}^4 c_{\Delta_i, J_i}) \sum_{J=0}^{\infty}\int_{\frac{d}{2}- i\infty}^{\frac{d}{2} + i \infty} \frac{d\Delta}{2\pi i}\, K_{[\tilde{\Delta}, J]}^{\Delta_{\phi}, \Delta_{\phi}}\, \rho(\Delta, J) \,\mathcal{I}_{\Delta_1, J_1, \Delta_2, J_2, \Delta, J; a} \,\mathcal{I}_{\Delta, J, \Delta_3, J_3, \Delta_4, J_4; b}\, \, G_{\Delta, J}^{\Delta_i, J_i; a,b }(x_i)~.
\end{multline}
In getting to this we used, 
\be \nonumber
&\frac{{(S_{[\Delta ,J]}^{[{\Delta _3},{J_3}],[{\Delta _4},{J_4}]})_b^c(K_{[\tilde \Delta ,J]}^{[{\Delta _3},{J_3}],[{\Delta _4},{J_4}]})_c^b}}{{S_{[\Delta ,J]}^{{\Delta _\phi },{\Delta _\phi }}}}\; = {( - \frac{1}{2})^J}\frac{{(S_{[\Delta ,J]}^{[{\Delta _3},{J_3}],[{\Delta _4},{J_4}]})_b^c(S_{[\tilde \Delta ,J]}^{[{\Delta _3},{J_3}],[{\Delta _4},{J_4}]})_c^b}}{{S_{[\Delta ,J]}^{{\Delta _\phi },{\Delta _\phi }}}} \nonumber\\
&= {( - \frac{1}{2})^J}\frac{{S_{[\Delta ,J]}^{{\Delta _\phi },{\Delta _\phi }}S_{[\tilde \Delta ,J]}^{{\Delta _\phi },{\Delta _\phi }}}}{{S_{[\Delta ,J]}^{{\Delta _\phi },{\Delta _\phi }}}} = K_{[\tilde \Delta ,J]}^{{\Delta _\phi },{\Delta _\phi }}~,
\ee
In the second equality we expressed $K$ in terms of $S$, to get in the numerator a term that is ${S}^2$, which must be proportional to the identity, and thus independent of $[\Delta_3, J_3]$ and $[\Delta_4, J_4]$ (see e.g.\ \eqr{spinshadow}). We can therefore replace it with the term in the numerator of the third equality.

The result (\ref{4Ans}) is our final answer. It is a $d$-dimensional generalization of the corresponding one-dimensional result, Eq. 4.16 of \cite{GR4}. 

Note that (\ref{4Ans}) is not the full four-point function, but rather, the sum of Feynman diagrams shown in Fig.~\ref{8split} which are a contribution to the four-point function. To obtain the full four-point function, one should add  to this the same diagrams but in the $t$ and $u$ channels, subtract a planar diagram with no exchanged melons (that will get double counted, among the three channels), and add an additional ``contact'' diagram; see \cite{GR4}.

\begin{figure}
\centering
\includegraphics[width=1.6in]{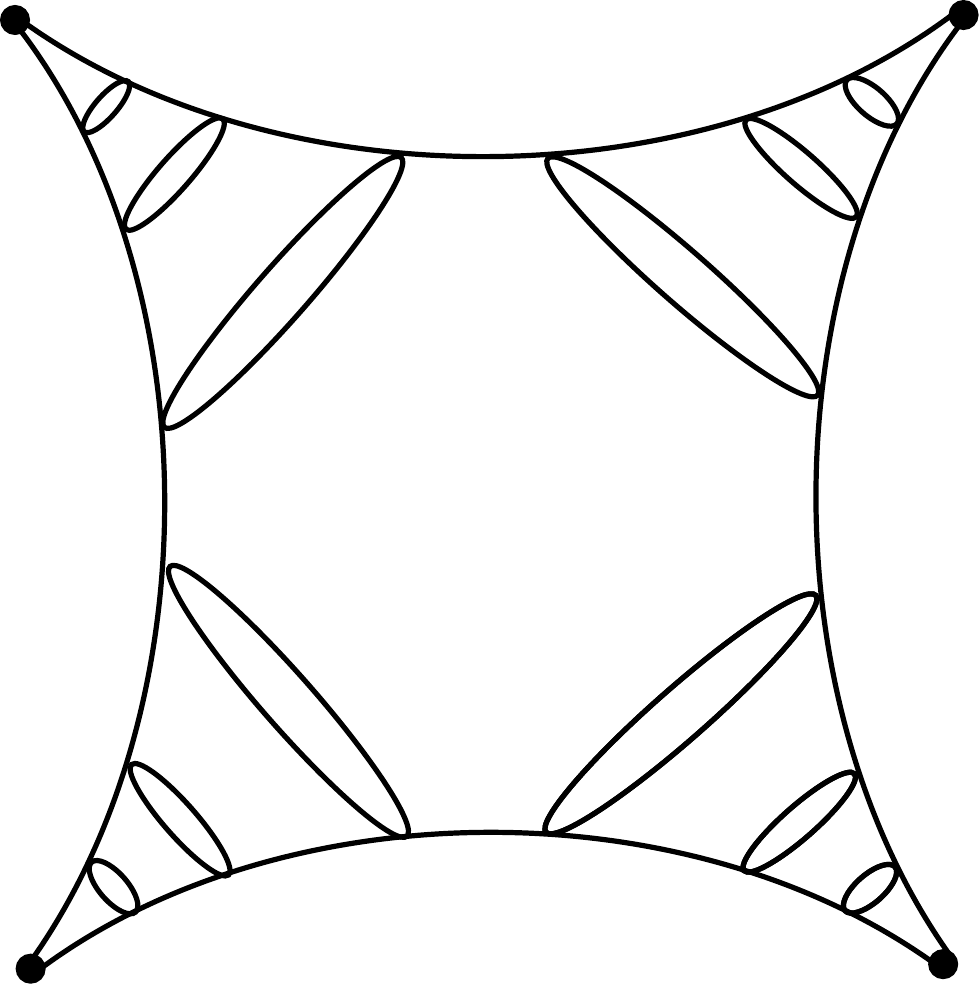}
\caption{The SYK bilinear four-point planar diagram with no exchanged melons. }
\label{figsykpl}
\end{figure} 

 The planar diagram with no exchanged melons is shown in Fig.~\ref{figsykpl}. It is similar to the planar diagram for the three-point function, shown previously in Fig.~\ref{tetraCut2}(a) and given in Eq. (\ref{amp1}) as three three-point functions glued together. We similarly write the diagram in  Fig.~\ref{figsykpl} as four three-point functions glued together, 
\es{4ptNoMelon}{
\langle \mO_{1}(x_1)& \cdots \mO_{4}(x_4)\rangle_s^0= \int d^d x_a d^d x_b d^d x_c d^d x_d \langle \mO_{\Delta_1, J_1}(x_1) \phi(x_a) \phi(x_b)\rangle_{\text{amp}}\\&\langle \mO_{\Delta_2, J_2}(x_2)  \phi(x_d) \phi(x_a)\rangle_{\text{amp}} \langle \mO_{\Delta_3, J_3}(x_3) \phi(x_c)  \phi(x_d) \rangle_{\text{amp}}  \langle \mO_{\Delta_4, J_4}(x_4) \phi(x_b)  \phi(x_c) \rangle_{\text{amp}}~.}
 In fact, there is a more appealing way of writing the answer, in terms of conformal blocks. One notices that the diagram in Fig.~\ref{figsykpl} is really a special case of the diagram in Fig.~\ref{8split} which we have been studying in this section, with the three-point function coefficients given by those for the planar diagram, and the intermediate four-point function of fundamentals given by that of free (mean) field theory. Thus, 
 \begin{multline} \label{4Ans2}
\langle \mO_{1}(x_1) \cdots \mO_{4}(x_4)\rangle_s^0 =\\
 \hspace{-.5cm} ( \prod_{i=1}^4 c_{\Delta_i, J_i}) \sum_{J=0}^{\infty}\int_{\frac{d}{2}- i\infty}^{\frac{d}{2} + i \infty} \frac{d\Delta}{2\pi i}\, K_{[\tilde{\Delta}, J]}^{\Delta_{\phi}, \Delta_{\phi}}\, b^2 \rho^{\rm MFT}(\Delta, J) \,\mathcal{I}^p_{\Delta_1, J_1, \Delta_2, J_2, \Delta, J; a} \,\mathcal{I}^p_{\Delta, J, \Delta_3, J_3, \Delta_4, J_4; b}\, \, G_{\Delta, J}^{\Delta_i, J_i; a,b }(x_i).
\end{multline}
 where $\rho^{\rm MFT}(\D,J)$ was given in \eqr{rhomft}, while the planar three-point structure constant was given in (\ref{OOO}), 
 \be
\mathcal{I}^p_{\Delta_1, J_1, \Delta_2, J_2, \Delta_3, J_3; a} =  \left(\prod_{i=1}^3 \cJ^2 b^q\,   S_{\Delta_{\phi}}^{\Delta_{\phi}, [\Delta_i, J_i]}\right) \, \mathcal{I}_{\Delta_1, J_1, \Delta_2, J_2, \Delta_3, J_3;a}^{(2)}~.
 \ee
The result Eq. (\ref{4Ans2}) is a $d$-dimensional generalization of the corresponding one-dimensional result, Eq. 4.19 of \cite{GR4}.

Finally, on general grounds, a connected four-point function at leading order in large $N$ is expressible as a sum of single-trace and double-trace conformal blocks. Closing the contour in (\ref{4Ans}) gives a sum of conformal blocks, arising from the poles of the integrand. The single-trace blocks are immediate, and come from the poles of $\rho(\Delta, J)$. The double-trace blocks should arise from the three-point function coefficients. A scalar four-point function, for instance, will contain an exchange of double-trace blocks of the schematic form 
\be
\mO_{1}( \partial^2)^n \partial_{\mu_1} \cdots \partial_{\mu_J} \mO_{2}~, \ \ \ \ \ \ \mO_{3}( \partial^2)^n \partial_{\mu_1} \cdots \partial_{\mu_J} \mO_{4}~.
\ee
where we take the $\O_i$ to be scalars for simplicity. This requires that $\mathcal{I}_{\Delta_1,0, \Delta_2,0, \Delta,J}$, regarded as a function of $\Delta$, have simple poles at
\be
\Delta = \Delta_1 + \Delta_2 + J + 2n~.
\ee
The explicit form of the $6j$ symbols that we will compute in the next section will  indeed have such singularities. Analogous poles exist for spinning operators, at twists $\tau_1+\tau_2+2n$, dressed by labels for the appropriate spinning structure and Lorentz representations.

\section{$6j$ symbols} \label{sec:6j}

\begin{figure}[t]
\centering
\includegraphics[width=1.7in]{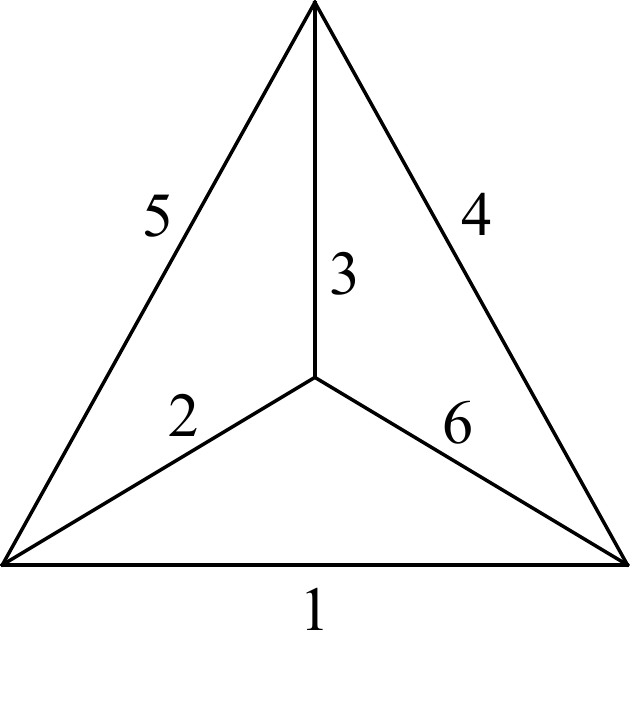}
\caption{The $6j$ symbol,  represented as a tetrahedron. } \label{tetraLabel}
\end{figure} 

In this section, we compute $6j$ symbols for principal series representations of $\SO(d+1,1)$ in two and four dimensions. 
We define the $6j$ symbol for these representations as a conformally-invariant integral of a product of four conformal three-point structures, 
\begin{align} \label{6jDef}
\left\{
\begin{matrix}
\cO_1 & \cO_2 & \cO_6 \\
\cO_3 & \cO_4 & \cO_5
\end{matrix}
\right\}^{abcd}
&=
 \int \frac{{{d^d}{x_1} \cdots {d^d}{x_6}}}{{{\rm{vol(SO(}}d + 1,1{\rm{))}}}}
\<\tl \cO_1 \tl \cO_2 \tl \cO_5\>^a\<\cO_5 \tl \cO_3 \tl \cO_4\>^b\<\cO_3 \cO_2\cO_6\>^c\<\tl \cO_6 \cO_1 \cO_4\>^d~,
\end{align}
where $\cO_i$ denotes an operator with dimension $\Delta_i\in \frac d 2 + i\mathbb{R}$, spin $J_i$, and position $x_i$. The shadow of an operator, $\tl \cO_i$,  has dimension $\tilde{\Delta}_i=d-\Delta_i$. In total, the $6j$ symbol depends on the six representations $\cO_1,\dots,\cO_6$, together with four indices $a,b,c,d$ that label conformally-invariant three-point structures. For most of this section, we consider the case where there is a unique three-point structure for the given representations, so we can drop the structure labels $a,b,c,d$.

It is useful to represent the $6j$ symbol graphically as a tetrahedron, shown in Fig.~\ref{tetraLabel}. Each of the six edges represent positions which are integrated over. Each of the four vertices is a three-point structure. Notice that  every one of the six operators appears once, and every one of the shadow operators appears once. Our notation for the $6j$ symbol is such that the columns of (\ref{6jDef}) and (\ref{eq:relatedobject}) contain edges that do not meet at a vertex. For example, $\cO_1$ and $\cO_3$ are not both present in any of the four three-point structures.

The $6j$ symbol (\ref{6jDef}) is invariant under symmetries of the tetrahedron $S_4$, which acts by permuting the vertices. The action of a permutation also swaps some representations with their shadows. For example, the symmetry group of the $6j$ symbol is generated by the relations
\be
\label{eq:tetrahedrasixj}
\left\{
\begin{matrix}
\cO_1 & \cO_2 & \cO_6 \\
\cO_3 & \cO_4 & \cO_5
\end{matrix}
\right\} = \left\{
\begin{matrix}
\tl \cO_5 & \cO_3 & \cO_2 \\
\cO_6 & \tl \cO_1 & \cO_4
\end{matrix}
\right\} 
= \left\{
\begin{matrix}
\cO_2 & \cO_5 & \tl \cO_3 \\
\tl \cO_4 & \cO_6 & \cO_1
\end{matrix}
\right\} =\left\{
\begin{matrix}
\cO_2 & \cO_1 & \tl\cO_6 \\
\cO_4 & \cO_3 & \cO_5
\end{matrix}
\right\},
\ee
all of which are manifest in the expression (\ref{6jDef}).

In defining the $6j$ symbol (\ref{6jDef}), we made an arbitrary choice of which of the four three-point structures contains the operator versus its shadow.  A related object which does not require these choices is
\be \nonumber
\! \!\begin{bmatrix}
\cO_1 &\!\! \cO_2 & \!\!\cO_6\!\! \\
\cO_3 & \cO_4 & \cO_5
\end{bmatrix}
\!=\!\! \int \! \frac{{{d^d}{x_1}\! \cdots\! {d^d}{x_6}{d^d}{x_{1'}}\! \cdots\! {d^d}{x_{6'}}}}{{{\rm{vol}}({\rm{SO}}(d + 1,1))}}\prod_{i = 1}^6 \<\tl \cO_i \tl \cO_{i'}\> \<\cO_{1'} \cO_{2'} \cO_{5'}\>\<\cO_{5}\cO_{3'}\cO_{4'}\>\<\cO_{3}\cO_{2} \cO_6\>\<\cO_{6'}\cO_{1}\cO_{4}\>~.
\label{eq:relatedobject}
\ee
This object is  manifestly symmetric under $S_4$, where permutations do not swap representations with their shadows.  It is trivial to evaluate six of the integrals appearing in (\ref{eq:relatedobject}) to relate it to (\ref{6jDef}), as these are simply the shadow transforms discussed in Appendix~\ref{shadow}, and the integrals give the shadow factors derived there. In particular, performing the integrals over $x_{i'}$ relates our two definitions,
\be
\begin{bmatrix}
\cO_1 & \cO_2 & \cO_6 \\
\cO_3 & \cO_4 & \cO_5
\end{bmatrix}
&=
S^{\cO_2 \cO_5}_{\cO_1} S^{\tl \cO_1 \cO_5}_{\cO_2} S^{\cO_5\cO_4}_{\cO_3} S^{\tl \cO_3 \cO_5}_{\cO_4} S^{\tl \cO_1 \tl \cO_2}_{\cO_5} S^{\cO_1 \cO_4}_{\cO_6} 
\left\{
\begin{matrix}
\cO_1 & \cO_2 & \cO_6 \\
\cO_3 & \cO_4 & \cO_5
\end{matrix}
\right\}~
.
\ee

It is clear that the manifest tetrahedral symmetry of the $6j$ symbol is broken by choices of which integrals to perform first. In calculating the $6j$ symbol, we will be forced to make such choices, and as a result, our answer will not be manifestly symmetric under $S_4$; the presence of this symmetry,  nevertheless, constitutes a strong check of our calculation. 

Continuing with \eqr{6jDef}, we see that the integrals over  $x_5$ and $x_6$ are simple to evaluate, each producing a conformal partial wave, and we recognize the $6j$ symbol to be the overlap of two partial waves, as previously shown in Fig.~\ref{tetraCut1}~,
\begin{align}
\label{6j}
\left\{
\begin{matrix}
\cO_1 & \cO_2 & \cO_6 \\
\cO_3 & \cO_4 & \cO_5
\end{matrix}
\right\}
&=
\left(
\Psi _{\tilde \Delta ,J}^{{{\tilde \Delta }_1},{{\tilde \Delta }_2},{{\tilde \Delta }_3},{{\tilde \Delta }_4}},
\Psi _{\Delta ',J'}^{{\Delta _3},{\Delta _2},{\Delta _1},{\Delta _4}}
\right)
\nn\\
&= \int {\frac{{{d^d}{x_1} \cdots {d^d}{x_4}}}{{{\rm{vol}}({\rm{SO}}(d + 1,1))}}} \Psi _{\tilde \Delta ,J}^{{{\tilde \Delta }_1},{{\tilde \Delta }_2},{{\tilde \Delta }_3},{{\tilde \Delta }_4}}({x_1},{x_2},{x_3},{x_4})\Psi _{\Delta ',J'}^{{\Delta _3},{\Delta _2},{\Delta _1},{\Delta _4}}({x_3},{x_2},{x_1},{x_4})~,
\end{align}
where we have changed notation from Fig. \ref{tetraCut1}, $(\Delta_5, J_5) \rightarrow (\Delta, J)$ and $(\Delta_6, J_6) \rightarrow  (\Delta', J')$, to reflect the distinction now present between external operators and internal operators. We have also dropped the spin indices for the external operators. We will refer to the first factor in (\ref{6j}) as an ``$s$-channel"  partial wave $(12\rar34)$, and the second factor as a ``$t$-channel" partial wave $(14\rar23)$. Instead of the somewhat cumbersome multi-line $\{\cdots\}$ notation, we will often use the following notation for the integral in \eqr{6j} in what follows:
\e{jdef}{{{\mathcal{ J}}_d}(\D,J;\D',J'|\D_1,\D_2,\D_3,\D_4) = \left(
\Psi _{\tilde \Delta ,J}^{{{\tilde \Delta }_1},{{\tilde \Delta }_2},{{\tilde \Delta }_3},{{\tilde \Delta }_4}},
\Psi _{\Delta ',J'}^{{\Delta _3},{\Delta _2},{\Delta _1},{\Delta _4}}
\right)
=\left\{
\begin{matrix}
\De_1 & \De_2 & [\De',J'] \\
\De_3 & \De_4 & [\De,J]
\end{matrix}
\right\}
~.
}

Our task is to evaluate the integral in (\ref{6j}). In fact,  (\ref{6j}) is  a special case of the more general quantity, 
\be\label{eucinv}
I_{\De,J} &= 
\left(
\Psi _{\tilde \Delta ,J}^{{{\tilde \Delta }_1},{{\tilde \Delta }_2},{{\tilde \Delta }_3},{{\tilde \Delta }_4}},
\<\cO_1\cO_2\cO_3\cO_4\>
\right)\nn\\
&= 
\int{\frac{{{d}^{d}}{{x}_{1}}\cdots {{d}^{d}}{{x}_{4}}}{\text{vol}(\text{SO}(d+1,1))}}\Psi _{\tl \Delta,J}^{{{{\tilde{\Delta }}}_{1}},{{{\tilde{\Delta }}}_{2}},{{{\tilde{\Delta }}}_{3}},{{{\tilde{\Delta }}}_{4}}}({{x}_{1}},{{x}_{2}},{{x}_{3}},{{x}_{4}}) 
\<{{\mathcal{O}_1}({x_1}){\mathcal{O}_2}({x_2}){\mathcal{O}_3}({x_3}){\mathcal{O}_4}({x_4})}\>~.
\ee
To get (\ref{6j}), we specialize to the case where the four-point function is a single $t$-channel conformal partial wave,
\be \label{Our4pt}
 \langle {{\mathcal{O}_1}({x_1}){\mathcal{O}_2}({x_2}){\mathcal{O}_3}({x_3}){\mathcal{O}_4}({x_4})}\rangle =\Psi _{\Delta' ,J'}^{{{\Delta }_{3}},{{\Delta }_{2}},{{\Delta }_{1}},{{\Delta }_{4}}}({{x}_{3}},{{x}_{2}},{{x}_{1}},{{x}_{4}})~.
 \ee
 \eqr{eucinv} is a Euclidean inversion formula. It may be obtained starting from a four-point function written in contour integral form,
\es{partialwaveexpansion}{\left\langle {{\mathcal{O}_1}({x_1}){\mathcal{O}_2}({x_2}){\mathcal{O}_3}({x_3}){\mathcal{O}_4}({x_4})} \right\rangle  &=\sum\limits_{J = 0}^{ \infty } {\int_{{d\o 2}}^{{d\o 2} + i\infty } {\frac{{d\Delta }}{{2\pi i}}{I_{\D,J}\o n_{\D,J}}\Psi _{\Delta,J} ^{{\Delta_1,\D_2,\D_3,\D_4}} }}({x_i})\\
& =\sum\limits_{J = 0}^{ \infty } {\int_{{d\o 2}-i\i}^{{d\o 2} + i\infty } {\frac{{d\Delta }}{{2\pi i}}{I_{\D,J}\o n_{\D,J}}K_{\tilde\D,J}^{\D_3,\D_4}G _{\Delta,J} ^{{\Delta_1,\D_2,\D_3,\D_4}}({x_i}) }}}
and applying orthogonality of partial waves. The function $I_{\D,J}$ has poles at physical operator locations with residues encoding the OPE coefficients,
\e{}{\Res{\D'=\D}I_{\D',J} \propto C_{12\O_{\D,J}}C_{34\O_{\D,J}} ~,}
after contour deformation away from the principal series $\D={d\o2}+i\nu$.\footnote{The coefficient function generally contains additional spurious poles that are cancelled by poles in the conformal blocks when deforming the contour from the principal series. Also, to write the partial wave expansion (\ref{partialwaveexpansion}), one must first subtract off ``non-normalizable" contributions coming from the unit operator and scalars with dimensions below $\frac d 2$, which we have suppressed above. See \cite{SD17} for details on both of these subtleties.}

Note that plugging (\ref{Our4pt}) into (\ref{partialwaveexpansion}) gives
\be
\Psi _{\Delta' ,J'}^{{{\Delta }_{3}},{{\Delta }_{2}},{{\Delta }_{1}},{{\Delta }_{4}}}({{x}_{3}},{{x}_{2}},{{x}_{1}},{{x}_{4}})
&=
\sum_{J=0}^\oo \int_{\frac d 2}^{\frac d 2 + i\oo} \frac{d\De}{2\pi i} \frac{\left(
\Psi _{\tilde \Delta ,J}^{{{\tilde \Delta }_1},{{\tilde \Delta }_2},{{\tilde \Delta }_3},{{\tilde \Delta }_4}},
\Psi _{\Delta ',J'}^{{\Delta _3},{\Delta _2},{\Delta _1},{\Delta _4}}
\right)}{n_{\De,J}} \Psi_{\De,J}^{\De_1,\De_2,\De_3,\De_4}(x_1,x_2,x_3,x_4)~.
\label{eq:crossingkernel}
\ee
Thus, the $6j$ symbol (\ref{jdef}), divided by an appropriate normalization factor $n_{\De,J}$, gives the coefficients for the expansion of a $t$-channel partial wave in $s$-channel partial waves.\footnote{Note that it is not possible to express a $t$-channel conformal block in terms of $s$-channel blocks in a canonical way because these two types of functions have different single-valuedness properties. In particular, the $s$-channel blocks are single-valued in Euclidean space near $z,\bar z = 0$, while $t$-channel blocks are not single-valued in this region. Conformal partial waves, by contrast are single-valued in Euclidean space.} For this reason, $6j$ symbols are sometimes referred to as ``crossing kernels."

Instead of integrating over Euclidean space, an appropriate contour deformation of \eqr{eucinv} allows one to  write it in terms of an integral of the double-commutator $\<[\cO_4,\cO_1][\cO_2,\cO_3]\>$ over a Lorentzian region. The result is Caron-Huot's Lorentzian inversion formula \cite{Caron-Huot:2017vep,SD17}. It may be expressed, in $d$ dimensions, as a double integral over cross-ratios,
\begin{align}\label{lorinv}
I_{\Delta,J} = \alpha_{\Delta,J} \Bigg[(-1)^J&\int_0^1\int_0^1 \frac{d\chi d\bar \chi}{(\chi \bar \chi)^d} |\chi-\b\chi|^{d-2} G^{\tl \De_i}_{J+d-1,\Delta-d+1}(\chi,\bar\chi)\frac{\<[\O_3,\O_2][\O_1,\O_4]\>}{T^{\De_i}}  \\
&+\int_{-\infty}^0\int_{-\infty}^0\frac{d\chi d\bar \chi}{(\chi \bar \chi)^d} |\chi-\b\chi|^{d-2} \hat G^{\tl \De_i}_{J+d-1,\Delta-d+1}(\chi,\bar\chi)\frac{\<[\O_4,\O_2][\O_1,\O_3]\>}{T^{\De_i}} \Bigg]~,\notag
\end{align}
where 
\begin{align} \label{Tdef}
{{T}^{{{\Delta }_{i}}}}=\frac{1}{|x_{12}|^{{{\Delta }_{1}}+{{\Delta }_{2}}}}\frac{1}{ |x_{34}|^{{{\Delta }_{3}}+{{\Delta }_{4}}}}{{\left( \frac{|x_{24}|}{|x_{14}|} \right)}^{{\Delta }_{12}}}{{\left( \frac{|x_{14}|}{|x_{13}|} \right)}^{ {{\Delta }_{34}}}}~,
\end{align}
and
\e{adj}{\alpha_{\D,J} = -{t_0\o 2^{d+1}}(2\pi)^{d-2}\frac{\Gamma(J+1)}{\Gamma(J+\frac{d}{2})}\frac{\Gamma(\Delta-\frac{d}{2})}{\Gamma(\Delta-1)}\frac{\Gamma(\frac{\De_{12}+J+\De}{2})\Gamma(\frac{\De_{21}+J+\De}{2})\Gamma(\frac{\De_{34}+J+\tl\De}{2})\Gamma(\frac{\De_{43}+J+\tl\De}{2})}{\Gamma(J+\Delta)\Gamma(J+d-\De)}~.}
Here $\De_{ij}\equiv \De_i-\De_j$. The function $\hat{G}_{\Delta,J}$ is defined in \cite{SD17} as the conformal block normalized as $(-\chi)^{\frac{\Delta-J}{2}}(-\bar{\chi})^{\frac{\Delta+J}{2}}$ for $|\chi|\ll |\bar{\chi}|\ll 1$ for negative cross ratios.

While in $d=1$ the Euclidean integral for the $6j$ symbol \eqr{6j} is straightforward (see Appendix~\ref{app1d6j}), in higher dimensions that integral appears more challenging. However, in $d=2$ and $d=4$, using the Lorentzian inversion formula \eqr{lorinv} instead will cause the integral to factorize into a product of one-dimensional integrals that can be explicitly performed. Carrying out this calculation is the goal of the next two subsections.

\subsection{Two dimensions} \label{2d6j}

In this section we compute the $6j$ symbol in two dimensions for arbitrary spinning operators.\footnote{
The two-dimensional $6j$ symbols have recently been discussed from a different, more mathematical viewpoint in \cite{ismo,Derkachov:2017mip}. } This requires that we generalize the inversion formula to arbitrary external spins, which we do first.\footnote{A general version of the Lorentzian inversion formula for arbitrary spinning operators in arbitrary spacetime dimension was derived in \cite{Kravchuk:2018htv}. The result here is a special case of the one in \cite{Kravchuk:2018htv}. However, we include the discussion here because the computation is very simple.}

It is convenient to use holomorphic and antiholomorphic coordinates.   We define principal series weights $h=(1+J+ir)/2$ and $\bar{h}=(1-J+ir)/2$ with $J\in \mathbb{Z}$ and $r\in \mathbb{R}$. In terms of the weights, $h+\bar{h}=1+ir=\Delta$ is dimension and $|h-\bar{h}|=|J|$ is spin. The integral for the $6j$ symbol is convergent when all weights lie on the principal series; for more general weights, we define it by analytic continuation. In  two dimensions,
\be\nonumber
\Psi _{h,\bar h}^{{{h}_{i}},\bar h_i}({{z}_{1}},{{z}_{2}},{{z}_{3}},{{z}_{4}})
=\!
\int{{{d}^{2}}{{z}_{5}}}\frac{1}{{{\left[ {{z}_{51}} \right]}^{{{h}_{1}}+h-{{h}_{2}}}}{{\left[ {{z}_{52}} \right]}^{{{h}_{2}}+h-{{h}_{1}}}}{{\left[ {{z}_{12}} \right]}^{{{h}_{1}}+{{h}_{2}}-h}}}\frac{1}{{{\left[ {{z}_{53}} \right]}^{{{h}_{3}}+\tilde{h}-{{h}_{4}}}}{{\left[ {{z}_{54}} \right]}^{{{h}_{4}}+\tilde{h}-{{h}_{3}}}}{{\left[ {{z}_{34}} \right]}^{{{h}_{3}}+{{h}_{4}}-\tilde{h}}}}
\ee
Performing the integral yields \cite{Osborn:2012vt}
\be
\Psi _{h,\bar h}^{{{h}_{i}},\bar h_i}({{z}_{1}},{{z}_{2}},{{z}_{3}},{{z}_{4}}) =K_{\tl h,\tl {\bar h}}^{{h_3},{h_4},{{\bar h}_3},{{\bar h}_4}}\,  T_s k^{h_i}_{2h}(\chi)\, k^{\bar h_i}_{2\bar h}(\bar \chi)+
(1\leftrightarrow 3,2\leftrightarrow 4,h\leftrightarrow \tl h,\bar h \leftrightarrow \tl {\bar h})~,
\ee
where $k_{2h}^{h_i}(\chi) $ are one-dimensional conformal blocks, 
\be\label{eq:sltwoblock}
k_{2h}^{h_i}(\chi) &\equiv \chi^{h}{}_2F_1(h-h_{12},h+h_{34},2h,\chi)~, 
\ee
the ${{T}_{s}}$ are leg factors, 
\be \label{eq:standardfactors}
{{T}_{s}}
&\equiv
\frac{1}{[z_{12}]^{h_1+h_2} [z_{34}]^{h_3+h_4}} \frac{[z_{24}]^{h_1-h_2}[z_{14}]^{h_3-h_4}}{[z_{14}]^{h_1-h_2}[z_{13}]^{h_3-h_4}}~,
\ee
and the prefactor is a ratio of gamma functions,
\begin{align}  \label{Kfactor2d}
K_{h,\bar h}^{{h_1},{h_2},{{\bar h}_1},{{\bar h}_2}} &\equiv \frac{\pi\G(1-2\tl h)}{\G(2\tl {\bar h})}
\frac{\G(\tl h-h_{12})\G(\tl{\bar h}+\bar h_{12})}{\G(h-h_{12})\G(\bar h + \bar h_{12})}~.
\end{align}
%\ee
Also, we use the notation ${{\left[ z \right]}^{h}}={{z}^{h}}{{\bar{z}}^{{\bar{h}}}}$ and $\tilde{h}=1-h$, $\tilde{\bar h} = 1-\bar h$. The leg factors $T_s$ contain the standard dimensionful factors in a four-point function. After factoring them out, the remaining quantities are functions of conformal cross-ratios,
\be
\chi = \frac{z_{12}z_{34}}{z_{13}z_{24}},\quad \bar \chi=\frac{\bar z_{12}\bar z_{34}}{\bar z_{13}\bar z_{24}}~.
\ee

\subsubsection{Lorentzian inversion for general spins in two dimensions}

We begin by deriving the two-dimensional Lorentzian inversion formula,  for general spinning operators, following the derivation in \cite{SD17}. This generalization could also be obtained by applying the formula of \cite{Kravchuk:2018htv} in two dimensions. The result can be found in Eqns. \eqr{eq:caronhuottwod}, \eqr{eq:ddisct} and \eqr{eq:ddiscu}.

We would like to compute the inner product,
\begin{align}\label{312}
\mathcal{P}=
\left(
\Psi _{\tilde{h},\tl{\bar h}}^{{{{\tilde{h}}}_{i}},\tl{\bar h_i}},
\<\cO_1\cO_2\cO_3\cO_4\>
\right)
=
\int{\frac{{{d}^{2}}{{z}_{1}}\cdots {{d}^{2}}{{z}_{4}}}{\text{vol}(\text{SO}(3,1))}} \Psi _{\tilde{h},\tl{\bar h}}^{{{{\tilde{h}}}_{i}},\tl{\bar h_i}}({{z}_{i}})\<\cO_1\cO_2\cO_3\cO_4\>~.
\end{align}
We will often use shorthand notation $\<\cO_1\cO_2\cO_3\cO_4\>$, where it is implicit that $\cO_i$ is at position $z_i$. Plugging in the shadow representation for the partial wave, this becomes
\begin{align} \nonumber
\mathcal{P}=\int{\frac{{{d}^{2}}{{z}_{1}}\cdots {{d}^{2}}{{z}_{5}}}{\text{vol}(\text{SO}(3,1))}}\frac{\left\langle {{\cO}_{1}}{{\cO}_{2}}{{\cO}_{3}}{{\cO}_{4}} \right\rangle }{{{\left[ {{z}_{51}} \right]}^{{{{\tilde{h}}}_{1}}+\tilde{h}-{{{\tilde{h}}}_{2}}}}{{\left[ {{z}_{52}} \right]}^{{{{\tilde{h}}}_{2}}+\tilde{h}-{{{\tilde{h}}}_{1}}}}{{\left[ {{z}_{12}} \right]}^{{{{\tilde{h}}}_{1}}+{{{\tilde{h}}}_{2}}-\tilde{h}}}{{\left[ {{z}_{53}} \right]}^{{{{\tilde{h}}}_{3}}+h-{{{\tilde{h}}}_{4}}}}{{\left[ {{z}_{54}} \right]}^{{{{\tilde{h}}}_{4}}+h-{{{\tilde{h}}}_{3}}}}{{\left[ {{z}_{34}} \right]}^{{{{\tilde{h}}}_{3}}+{{{\tilde{h}}}_{4}}-h}}}~,
\end{align}
As a first step, we choose the gauge $(z_1,z_2,z_3,z_4,z_5)=(1,0,z_3,z_4,\infty)$, giving 
\begin{align}
\mathcal{P}=\frac{1}{2^2}\int{{{d}^{2}}{{z}_{3}}{{d}^{2}}{{z}_{4}}}\frac{\left\langle {{\cO}_{1}}{{\cO}_{2}}{{\cO}_{3}}{{\cO}_{4}} \right\rangle }{{{\left[ {{z}_{34}} \right]}^{{{{\tilde{h}}}_{3}}+{{{\tilde{h}}}_{4}}-h}}}~
\end{align}
(the $2^{-2}$ is the Faddeev-Popov determinant).
We now Wick rotate the integrals over $z_3,z_4$ to Lorentzian signature, obtaining lightcone coordinates 
$u=x-t$ and $v=x+t$. The integral becomes,
\begin{align}
\mathcal{P}=-\frac{1}{2^4}\int{d{{u}_{3}}d{{u}_{4}}d{{v}_{3}}d{{v}_{4}}}\left\langle {{\cO}_{1}}{{\cO}_{2}}{{\cO}_{3}}{{\cO}_{4}} \right\rangle u_{34}^{-{{{\tilde{h}}}_{3}}-{{{\tilde{h}}}_{4}}+h}v_{34}^{-{{{\bar{\tilde{h}}}}_{3}}-{{{\bar{\tilde{h}}}}_{4}}+\bar{h}}~.
\end{align}
In this expression, $\<\cO_1\cdots\cO_4\>$ now represents a time-ordered correlator in Lorentzian signature.
Now, the key observation is that the integral of $\cO_3$ or $\cO_4$ over the $v$ direction kills the vacuum if $\bar h$ is sufficiently negative (since one can deform the $v$ contour away from the real axis to obtain zero if $\cO_3$ or $\cO_4$ act on the vacuum). This lets us replace the four-point function with a double-commutator\footnote{See Section 2 of \cite{SD17} for more details.},
\begin{align} 
  \mathcal{P}&=-\frac{{{(-1)}^{{{j}_{H}}}}}{2^4}\int_{{{R}_{1}}}{d{{u}_{3}}d{{u}_{4}}d{{v}_{3}}d{{v}_{4}}}\left\langle \left[ {\cO_{3}},{\cO_{2}} \right]\left[ {\cO_{1}},{\cO_{4}} \right] \right\rangle u_{43}^{H}v_{43}^{{\bar{H}}} \nonumber\\ \label{eq:jH}
 &\quad -\frac{1}{2^4}\int_{{{R}_{2}}}{d{{u}_{3}}d{{u}_{4}}d{{v}_{3}}d{{v}_{4}}}\left\langle \left[ {\cO_{4}},{\cO_{2}} \right]\left[ {\cO_{1}},{\cO_{3}} \right] \right\rangle u_{34}^{H}v_{34}^{{\bar{H}}}~,
\end{align}
where $H=h-\tilde{h}_3-\tilde{h}_4$, $j_H=h-\bar{h}-\tilde{h}_3+\bar{\tilde{h}}_3-\tilde{h}_4+\bar{\tilde{h}}_4$, and the regions $R_1,R_2$ are given by,
\begin{align}
  & {{R}_{1}}:{{v}_{3}}<0,\quad {{v}_{4}}>1,\quad 0<{{u}_{3}}<{{u}_{4}}<1~, \nonumber\\
 & {{R}_{2}}:{{v}_{3}}>1,\quad {{v}_{4}}<0,\quad 0<{{u}_{4}}<{{u}_{3}}<1~.
\end{align}

Let us write $|T_s|$ to indicate $T_s$ in (\ref{eq:standardfactors}), where each $z_{ij},\bar z_{ij}$ has been replaced with its absolute value $|z_{ij}|,|\bar z_{ij}|$.\footnote{This is a slight abuse of notation, as $h_i,\bar h_i$ can be complex.} Factoring out $|T_s|$, we  use,
\begin{align}
 \chi =\frac{{{u}_{34}}}{({{u}_{3}}-1){{u}_{4}}}~, 
\qquad \bar{\chi }=\frac{{{v}_{34}}}{({{v}_{3}}-1){{v}_{4}}}~,
\end{align}
to solve for $u_3,v_3$, and then integrate over $u_4,v_4$ to obtain,
\be
\label{eq:caronhuottwod}
\mathcal{P} &= - \frac{1}{2^4} \frac{\G(h+h_{12})\G(h+h_{21})\G(1-\bar h+\bar h_{43})\G(1-\bar h+\bar h_{34})}{\G(2h)\G(2-2\bar h)} \nn\\
&\quad \x \left[
(-1)^{j_H} \int_0^1 \int_0^1 \frac{d\chi}{\chi^2} \frac{d\bar\chi}{\bar\chi^2} 
\frac{\<[\cO_3,\cO_2][\cO_1,\cO_4]\>}{|T_s|}
k_{2h}^{\tl h_i}(\chi) k_{2\tl{\bar h}}^{\tl {\bar h}_i}(\bar \chi)
\right. \nn\\
&\left.\quad\quad\quad+ \int_{-\oo}^0 \int_{-\oo}^0 \frac{d\chi}{\chi^2} \frac{d\bar\chi}{\bar\chi^2} \frac{\<[\cO_4,\cO_2][\cO_1,\cO_3]\>}{|T_s|}
\hat k_{2h}^{\tl h_i}(\chi) \hat k_{2\tl{\bar h}}^{\tl {\bar h}_i}(\bar \chi)
 \right]~.
\ee
where we have defined
\be
\hat k_{2h}^{h_i}(\chi) &\equiv (-\chi)^h {}_2F_1(h-h_{12},h+h_{34},2h,\chi).
\ee

To apply the Lorentzian inversion formula, we  need to compute the double commutators $\<[\cO_3,\cO_2][\cO_1,\cO_4]\>$ and $\<[\cO_4,\cO_2][\cO_1,\cO_3]\>$ for the kinematic regions $R_1,R_2$. We take the four-point function to have the form,
\be
\<\cO_1\cO_2\cO_3\cO_4\> &= T_s \, g(\chi,\bar\chi),
\ee
and applying the appropriate $i\epsilon$ prescriptions we find,
\be
\frac{\<[\cO_3,\cO_2][\cO_1,\cO_4]\>}{|T_s|} &= -2 \cos(\pi(\bar h_2-\bar h_1+\bar h_3 - \bar h_4)) g(\chi,\bar\chi)\nn\\
&\quad + e^{i\pi(\bar h_2 - \bar h_1 +\bar h_3 - \bar h_4)} g^\circlearrowleft(\chi,\bar\chi) + e^{-i\pi(\bar h_2 - \bar h_1 + \bar h_3 - \bar h_4)}g^\circlearrowright(\chi,\bar\chi)\nn\\
&\equiv -2 \mathrm{dDisc}_t[g(\chi,\bar\chi)],
\label{eq:ddisct}
\ee
where $g^\circlearrowleft$ or $g^\circlearrowright$ indicates that we should take $\bar\chi$ around $1$ in the direction shown, while $\chi$ is held fixed.  Similarly,
\be
\frac{\<[\cO_4,\cO_2][\cO_1,\cO_3]\>}{|T_s|} &= -2\cos(\pi(\bar h_2 - \bar h_1 + \bar h_4 - \bar h_3))g(\chi,\bar\chi)\nn\\
&\quad + e^{i\pi(\bar h_3 - \bar h_4+\bar h_2 - \bar h_1)} g^\circlearrowright(\chi,\bar\chi) + e^{-i\pi(\bar h_3 - \bar h_4+\bar h_2 - \bar h_1)}g^\circlearrowleft(\chi,\bar\chi) \nn\\
&\equiv -2\mathrm{dDisc}_u[g(\chi,\bar\chi)],
\label{eq:ddiscu}
\ee
where now $g^\circlearrowleft$ or $g^\circlearrowright$ indicated that we should take $\bar\chi$ around $-\oo$ in the direction shown, while leaving $\chi$ fixed. The subscripts $\mathrm{dDisc}_{t,u}$ indicate the OPE limit around which we take discontinuities. The $t$-channel is $1\to 4$ and $2\to 3$, while the $u$-channel is $1\to 3$ and $2\to 4$.

To derive (\ref{eq:caronhuottwod}), we had to assume that $\bar h=\frac{\De-J}{2}$ was sufficiently negative, so that we could deform the $v$ contour to obtain a double-commutator. In general dimensions, Caron-Huot's formula for $(\Psi_{\De,J},\<\cdots\>)$ is valid if $J\geq J_0$, where $J_0$ is the rate of growth of the correlator $\<\cdots\>$ in the so-called Regge limit. The important fact for us is that a $t$-channel partial wave has bounded growth in the Regge limit \cite{Cornalba:2006xm}. In particular, the conformal Casimir equation can be used to show that a $t$-channel partial wave behaves as $\Psi^{\De_i}_{\De,J}(1-\chi,1-\bar \chi)\sim (\chi \bar\chi)^{\frac{\De_{12}-\De_{34}}{2}}+\mathrm{const}.$ as $\chi\bar\chi\to 0$ in general dimensions. Taking the double-discontinuity simply introduces some phases.
The factor $1/T^{\De_i}$ in the inversion formula contributes additional factors of $(\chi\bar\chi)^{\frac{\De_1+\De_2}{2}}$, so that altogether the $t$-channel partial wave behaves like a four-point function that vanishes as $(\chi \bar\chi)^{\frac{2\De_1-\De_{34}}{2}}$ for small $\chi\bar\chi$. For principal series representations, we have $\Re 2\De_1-\De_{34}=d$, so the Lorentzian inversion formula is analytic down to spin $J_0=1-d$ when acting on a $t$-channel partial wave. This implies that $6j$ symbols are analytic in $J$ for all nonnegative integer $J$.

\subsubsection{Inverting a $t$-channel partial wave}

To compute the general $6j$ symbol in two dimensions, we apply (\ref{eq:caronhuottwod}) to the case where the four-point function $\<\cO_1\cO_2\cO_3\cO_4\>$ is a $t$-channel conformal partial wave, with exchanged operator $(h',\bar{h}')$,
\be\label{323}
\<\cO_1\cO_2\cO_3\cO_4\> \rar \left.\Psi^{h_i,\bar h_i}_{h',\bar h'}(z_i)\right|_{1\leftrightarrow 3} 
= 
K^{h_1,h_4,\bar h_1,\bar h_4}_{\tl {h'},\tl {\bar h'}} T_t\,  g_t(\chi,\bar\chi) +
(1\leftrightarrow 3,2\leftrightarrow 4,h'\leftrightarrow \tl h',\bar h' \leftrightarrow \tl {\bar h'}),
\ee
where
\be
g_t(\chi,\bar\chi) \equiv 
k^{h_3,h_2,h_1,h_4}_{2h'}(1-\chi)\, k^{\bar h_3,\bar h_2,\bar h_1,\bar h_4}_{2\bar h'}(1-\bar \chi),
\ee
and $T_t=T_s|_{1\leftrightarrow 3}$.

To compute the double-commutators, note that,
\be
\frac{T_t}{T_s} g_t(\chi,\bar\chi) &\sim %\chi^{h_1+h_2}
(1-\chi)^{h'-h_2-h_3}
%\bar \chi^{\bar h_1+\bar h_2}
(1-\bar \chi)^{\bar h'-\bar h_2-\bar h_3} \x (1 + O(1-\chi,1-\bar\chi)).
\ee
Thus, $\mathrm{dDisc}_t$ simply introduces a constant factor:
\be
-2 \mathrm{dDisc}_t\left[\frac{T_t}{T_s} g_t(\chi,\bar\chi)\right]
&=
-4 \sin(\pi(\bar h'-\bar h_1-\bar h_4))\sin(\pi(\bar h'- \bar h_2 - \bar h_3)) \frac{T_t}{T_s} g_t(\chi,\bar\chi).
\ee
To compute $\mathrm{dDisc}_u$, we must expand the hypergeometric function around $\bar \chi=\oo$:
\be \nonumber
\frac{T_t}{T_s} k^{\bar h_3,\bar h_2,\bar h_1,\bar h_4}_{2\bar h'}(1-\bar \chi) &\sim
\#\,\bar{\chi }^{\bar h_1 - \bar h_2}+ \#\bar{\chi }^{\bar h_4-\bar h_3}~.
\ee
Applying (\ref{eq:ddiscu}), we find that $\mathrm{dDisc}_u$ vanishes. Since the $t$-channel partial wave is a linear combination of $t$-channel blocks, its $\mathrm{dDisc}_u$ vanishes as well. This is a general result: the double-discontinuity in one channel of a partial in another channel is zero. Let us briefly summarize the explanation from \cite{SD17}. The $t$-channel partial wave has a shadow representation, of schematic form
\be
\Psi_t &\sim \int d^dx \<\cO_1\cO_4\cO(x)\>\<\tl\cO(x) \cO_2\cO_3\>.
\ee
The $u$-channel double-discontinuity is proportional to the commutator $\<[\cO_2,\cO_4][\cO_1,\cO_3]\>$. In order for this to be nonzero, $\Psi_t$ must have a singularity when $2$ and $4$ become light-like separated and also when $1$ and $3$ become light-like separated. (The discontinuities of these singularities compute the commutators.) The integrand above has no singularity when $2$ and $4$ are light-like. However, the integral over $x$ can generate such a singularity if $x$ is near a light-like line between $2$ and $4$. Similar statements apply to $1$ and $3$. However, generically $x$ cannot be simultaneously light-like from all four points $1,2,3,4$. Thus, $\mathrm{dDisc}_u \Psi_t$ vanishes.

In summary, we only need to include the integral over the region $R_1$ in (\ref{eq:caronhuottwod}), i.e.\ the integral over $\chi,\bar\chi\in [0,1]$.  The final result for the $6j$ symbol is,
\begin{align}
\mathcal{J}_2 &= K_{\tilde h',\tilde{\bar {h'}}}^{{h_1},{h_4},{{\bar h}_1},{{\bar h}_4}} (\mathcal{B}_2)^{h_i,\bar h_i}_{[h,\bar h], [h',\bar h']} + K^{h_3,h_2,\bar h_3, \bar h_2}_{h',\bar h'} (\mathcal{B}_2)^{h_i,\bar h_i}_{[h,\bar h], [\tl h',\tl {\bar h'}]},\label{6jd2} 
\end{align}
where
\es{eq:inverseofblock}{(\mathcal{B}_2)^{h_i,\bar h_i}_{[h,\bar h],[ h',\bar h']} &= 
\frac{(-1)^{j_{H}}}{4}\frac{\G(h+h_{12})\G(h+h_{21})\G(\tl{\bar h}+\bar h_{43})\G(\tl{\bar h}+\bar h_{34})}{\G(2h)\G(2-2\bar h)} \\
&\quad \x
\sin(\pi(\bar h'-\bar h_1-\bar h_4))\sin(\pi(\bar h'- \bar h_2 - \bar h_3)) \, \Omega^{h_i}_{h,h',h_2 + h_3}\, \Omega^{\bar h_i}_{\tl{\bar h},\bar h',\bar h_2 + \bar h_3}.}
%\nn\\
%\mathcal{J}_2 &= \frac{(-1)^{j_{H}}}{4}\frac{\G(h+h_{12})\G(h+h_{21})\G(\tl{\bar h}+\bar h_{43})\G(\tl{\bar h}+\bar h_{34})}{\G(2h)\G(2-2\bar h)}
%\nn\\
%&\quad\quad \x
%\Bigg(
%K_{\tilde h',\tilde{\bar {h'}}}^{{h_1},{h_4},{{\bar h}_1},{{\bar h}_4}} \sin(\pi(\bar h'-\bar h_1-\bar h_4))\sin(\pi(\bar h'- \bar h_2 - \bar h_3)) \, \Omega^{h_i}_{h,h',h_2 + h_3}\, \Omega^{\bar h_i}_{\tl{\bar h},\bar h',\bar h_2 + \bar h_3}
%\nn\\
%& \quad\quad\quad + 
%K^{h_3,h_2,\bar h_3, \bar h_2}_{h',\bar h'}\sin(\pi(\tl{\bar h'}-\bar h_1-\bar h_4))\sin(\pi(\tl{\bar h'}- \bar h_2 - \bar h_3)) \, \Omega^{h_i}_{h,\tl h',h_2+h_3} \Omega^{\bar h_i}_{\tl {\bar h} \tl {\bar h'},\bar h_2+\bar h_3}\Bigg),
%\end{align}
$K_{h,\bar h}^{{h_1},{h_2},{{\bar h}_1},{{\bar h}_2}} $ was given in (\ref{Kfactor2d}), $j_H$ was given below (\ref{eq:jH}), and
\begin{align} \label{eq:Omega}
\Omega^{h_i}_{h,h',p} &\equiv \int_0^1 \frac{d\chi }{\chi^2} {\left( {\frac{\chi }{{1 - \chi }}} \right)^p}{\chi ^{{h_{13}}}}k_{2h}^{\tl h_1,\tl h_2,\tl h_3,\tl h_4}(\chi )k_{2h'}^{{h_3},{h_2},{h_1},{h_4}}(1 - \chi )~.
\end{align}
The quantity $\mathcal{B}_2$ represents the Lorentzian inversion of a $t$-channel conformal block (as opposed to a $t$-channel partial wave).
The $\Omega^{h_i}_{h,h',p}$ integral can be computed in closed form using the Mellin-Barnes representation for the hypergeometric functions in the variable $\tfrac{\chi}{1-\chi}$.\footnote{A similar integral  was  recently computed in \cite{Hogervorst:2017sfd}. }
 Performing the integral over $\chi$ gives a $\delta$-function in Mellin space, which trivializes one of the Mellin-Barnes integrals. To compute the remaining Mellin-Barnes integral, we must sum two series of poles, leading to a sum of two ${}_4F_3$ hypergeometric functions:
\es{omegaexp}{\Omega^{h_i}_{h,h',p} &= 
\frac{{\Gamma (2h)\Gamma (h' - p + 1)\Gamma \left( {h' - {h_{12}} + {h_{34}} - p + 1} \right)\Gamma \left( { - h' + {h_{12}} + h + p - 1} \right)}}{{\Gamma \left( {{h_{12}} + h} \right)\Gamma \left( {{h_{34}} + h} \right)\Gamma \left( {h' - {h_{12}} + h - p + 1} \right)}}\\
&~~\quad \times{}_4F_3\left( {\begin{array}{*{20}{c}}
{h' + {h_{23}},h' - {h_{14}},h' - {h_{12}} + {h_{34}} - p + 1,h' - p + 1}\\
{2h',h' - {h_{12}} + h - p + 1,h' - {h_{12}} - h - p + 2}
\end{array};1} \right)\\
&+ \frac{{\Gamma (2h')\Gamma \left( {h' - {h_{12}} - h - p + 1} \right)\Gamma \left( {{h_{13}} + h + p - 1} \right)\Gamma \left( {{h_{42}} + h + p - 1} \right)}}{{\Gamma \left( {h' + {h_{23}}} \right)\Gamma \left( {h' - {h_{14}}} \right)\Gamma \left( {h' + {h_{12}} + h + p - 1} \right)}}\\
&~~\quad \x{}_4F_3\left( {\begin{array}{*{20}{c}}
{{h_{13}} + h + p - 1,{h_{42}} + h + p - 1,{h_{34}} + h,{h_{12}} + h}\\
{h' + {h_{12}} + h + p - 1,2h, - h' + {h_{12}} + h + p}
\end{array};1} \right)~.}
The ordering of the parameters of the $6j$ symbol may be read off from \eqr{312} and \eqr{323}. In the notation of \eqr{jdef}, the ordering in \eqr{6jd2} is $\mathcal{J}_2(\D,J;\D',J'|\D_1,\D_2,\D_3,\D_4)$.

\subsection{Four dimensions} \label{4d6j}
The four dimensional calculation is very similar to the one in two dimensions.  For simplicity, we only consider scalar external operators. We use \eqr{lorinv} with $d=4$. We can exclude the integral over $\chi,\bar\chi\in(-\oo,0)$ because $\mathrm{dDisc}_u$ of a $t$-channel block vanishes, as discussed in the previous subsection. In four dimensions, the conformal block is given by \cite{Dolan:2000ut, Dolan:2003hv}
\begin{align}
G^{\De_i}_{\De,J}(x_i) &= T^{\De_i} \left(\frac{\chi \bar{\chi }}{\chi -\bar{\chi }} {{k}^{\frac{\De_i}{2}}_{\Delta +J}}(\chi ){{k}^{\frac{\De_i}{2}}_{\Delta +\tl J}}(\bar{\chi }) - (J\leftrightarrow \tl J)\right)~,
\end{align}
where $T^{\D_i}$ was given in \eqr{Tdef} and $k^{h_i}_{2h}(\chi)$ was given in (\ref{eq:sltwoblock}). The four-dimensional partial wave was given in terms of these blocks in Eq. (\ref{BSd}). For convenience, we have defined the ``spin shadow" affine Weyl reflection \cite{Kravchuk:2018htv},
\be
\tl J &\equiv -2-J.
\ee
From here, it is straightforward to proceed as in two dimensions. One difference is that in the four dimensional case, both the $s$ and $t$-channel partial waves contribute factors of $1/(\chi-\bar\chi)$. However, these factors cancel the $|\chi-\bar\chi|^2$ in the measure in (\ref{lorinv}), once again leading to a sum of factorized one-dimensional integrals.

The final result for the $6j$ symbol is
\begin{align}\label{6jd4}
\mathcal{J}_4 
&= K_{\tilde \Delta ',J'}^{{\Delta _1},{\Delta _4}} (\mathcal{B}_4)^{\De_i}_{[\De,J],[\De',J']} + 
K_{\De',J'}^{\De_3,\De_2} (\mathcal{B}_4)^{\De_i}_{[\De,J],[\tl \De',J']} \\
%(1\leftrightarrow 3, 2\leftrightarrow 4, \De\leftrightarrow \tl \De, \De'\leftrightarrow \tl \De') \\
(\mathcal{B}_4)^{\De_i}_{[\De,J],[\De',J']}&=
(-1)^J \a_{\De,J}\Bigg(
\Theta(\De'+\tl J') \Omega^{\frac{\De_i}{2}}_{\frac{\De+J}{2},\frac{\De'+J'}{2},\frac{\De_2+\De_3}{2}-1} \Omega^{\frac{\De_i}{2}}_{\frac{\tl \De+J}{2},\frac{\De'+\tl J'}{2},\frac{\De_2+\De_3}{2}-1}  \nn\\
&\qquad\qquad\qquad\qquad\qquad - (J' \leftrightarrow \tl J') - (\De\leftrightarrow \tl \De) + (J'\leftrightarrow \tl J', \De\leftrightarrow \tl \De) \Bigg),\label{eq:b4} %\nn\\
%&\quad + (1\leftrightarrow 3, 2\leftrightarrow 4, \De\leftrightarrow \tl \De, \De'\leftrightarrow \tl \De')\nn,
\end{align}
with
\begin{align}
\Theta(x) \equiv \frac{{4{\pi ^2}}}{{\Gamma (\frac{{{\Delta _3} + {\Delta _2} - x}}{2})\Gamma (1 - \frac{{{\Delta _3} + {\Delta _2} - x}}{2})\Gamma (\frac{{{\Delta _1} + {\Delta _4} - x}}{2})\Gamma (1 - \frac{{{\Delta _1} + {\Delta _4} - x}}{2})}}.
\end{align}
In the notation of \eqr{jdef}, the ordering in \eqr{6jd4} is $\mathcal{J}_4(\D,J;\D',J'|\D_1,\D_2,\D_3,\D_4)$.

This completes our evaluation of the $6j$ symbol. A simple check is that in the limit that one of the operator dimensions goes to zero, the corresponding partial wave degenerates into a product of two-point functions. For example, in four dimensions we have,
\begin{align}
\left\langle {\mathcal{O}_{1}}{\mathcal{O}_{2}}{\mathcal{O}_{3}}{\mathcal{O}_{4}} \right\rangle \to \frac{4{{\pi }^{2}}}{\Delta' }\frac{1}{{{\left| {{x}_{14}} \right|}^{2{{\Delta }_{1}}}}}\frac{1}{{{\left| {{x}_{23}} \right|}^{2{{\Delta }_{2}}}}}~,
\end{align}
where we have taken $J=J'=0$. In this limit, the $6j$ symbol is  given by relatively simple shadow integrals, 
\begin{align}
{{\cal J}_4} &\to \frac{{4{\pi ^2}}}{{\Delta '}}\int {\frac{{{d^4}{x_1}{d^4}{x_2}{d^4}{x_3}{d^4}{x_4}{d^4}{x_5}}}{{{\rm{vol}}({\rm{SO}}(5,1))}}} \frac{1}{{{{\left| {{x_{14}}} \right|}^{2{\Delta _1}}}}}\frac{1}{{{{\left| {{x_{23}}} \right|}^{2{\Delta _2}}}}}\left\langle {{{\tilde {\cal O}}_1}{{\tilde {\cal O}}_2}{{\tilde {\cal O}}_5}} \right\rangle \left\langle {{{\cal O}_5}{{\tilde {\cal O}}_3}{{\tilde {\cal O}}_4}} \right\rangle\nonumber\\
&= \frac{{4{\pi ^2}}}{{\Delta '}}S_{{{\tilde \Delta }_1}}^{{{\tilde \Delta }_2},\tilde \Delta }S_{{{\tilde \Delta }_2}}^{{{\tilde \Delta }_1},\Delta }\frac{1}{2^4{{\rm{vol}}({\rm{SO}}(3))}}~.
\end{align}

\subsection{Extracting CFT data from $6j$ symbols}\label{6jcft}

On what $s$-channel partial waves does the $t$-channel partial wave have support? The answer is given by the locations of poles of the $6j$ symbol. To see this, we recall that the output of the Lorentzian inversion formula \eqr{lorinv} is the OPE function $I_{\D,J}$ appearing in the four-point function written as in \eqr{partialwaveexpansion}, whose poles determine the locations of physical operators after contour deformation away from the principal series. The $6j$ symbol is just one instance of $I_{\D,J}$ for which $\la \O_1\O_2\O_3\O_4\ra$ is a single $t$-channel partial wave. 

Let us determine the locations of these poles. We keep fixed the quantum numbers $(\D_i,J_i)$ of the external operators and the $(\D',J')$ of the $t$-channel exchanged operator, and search for poles in the $s$-channel dimension, $\D$. Examining the expression for $\mathcal{J}_2$ for simplicity, we notice that the factor $\Omega^{\bar h_i}_{\tl{\bar h},\bar h',\bar h_2 + \bar h_3}$ contains poles of the form
\be\label{ompoles}
\frac{1}{\bar h_1+\bar h_2-\bar h+n} \qquad (n\in \mathbb{Z}_{\geq 0})~,
\ee
coming from the regime of the integral (\ref{eq:Omega}) near $\bar\chi=0$. These correspond to the case where $\O$ is a double-twist operator built out of $\O_1,\O_2$, with $\bar h=\bar h_1+\bar h_2+n$ and $h=\bar h+J$, where $J$ is arbitrary.\footnote{In our conventions, $\bar h = \frac{\De-J}{2}$, and $h=\frac{\De+J}{2}$.} By tetrahedral symmetry (\ref{eq:tetrahedrasixj}), we deduce that every vertex of the tetrahedron corresponding to a three-point structure $\<\cO_i\cO_j\cO_k\>$ contributes poles of the form
\be
\frac{1}{\bar h_i+\bar h_j - \bar h_k + n}~,\quad \frac{1}{\bar h_j+\bar h_k - \bar h_i + n}~,\quad\frac{1}{\bar h_k+\bar h_i - \bar h_j + n}.
\ee
There are also poles in $h$ at shadow locations, e.g. $\tl h_i = h_j+h_k+n$.

This is a general feature of $6j$ symbols: in all spacetime dimensions $d$, each vertex contributes double-twist poles
\be
\frac{1}{\frac{\De_i-J_i}{2} + \frac{\De_j-J_j}{2} - \frac{\De_k-J_k}{2} + n} \quad (n\in \mathbb{Z}_{\geq 0}).
\ee
When external dimensions are pairwise equal (e.g.\ $h_{13}=h_{24}=0$), the $6j$ symbol develops double poles. We can see these features directly from the $d$-dimensional Lorentzian inversion formula \eqr{lorinv} without having to compute the full $6j$ symbol, where we take the double commutator to be the dDisc of a $t$-channel partial wave. For simplicity, let us take all external operators to be identical, $\O_i\equiv\phi$. Focusing on the first line of \eqr{lorinv} (the second line will contribute identically up to a $(-1)^J$), we substitute the following:
\es{tchddisc}{&\frac{\<[\O_3,\O_2][\O_1,\O_4]\>}{T^{\De_i}} \\&=-2\dDisc_t\left[\left({\chi\chib\o (1-\chi)(1-\chib)}\right)^{\D_\phi}\left( K_{\tilde{\Delta'}, J'}^{\Delta_\phi, \Delta_\phi}\, G_{\Delta', J'}^{\Delta_\phi}(1-\chi,1-\chib) +(\D'\leftrightarrow \tilde\D')\right)\right]\\
&= -4\sin^2(\pi(h'-\D_\phi))\left({\chi\chib\o (1-\chi)(1-\chib)}\right)^{\D_\phi}\left( K_{\tilde{\Delta'}, J'}^{\Delta_\phi, \Delta_\phi}\, G_{\Delta', J'}^{\Delta_\phi}(1-\chi,1-\chib) +(\D'\leftrightarrow \tilde\D')\right)}
The relevant poles in $h$ come from the region of the integral near $\chi=0$. Taking the $\chi\ll1$ limit of the ``funny block'' in the measure yields
\e{}{G^{\tilde\D_i}_{J+d-1,\D-d+1}(\chi\ll1,\bar\chi) \sim \chi^{d-1-\left({\D-J\o2}\right)}k_{\D+J}(\chib)}
On the other hand, the $\chi\ll1$ limit of \eqr{tchddisc} takes the block near its radius of convergence, where its leading order behavior is
\e{}{G_{\Delta', J'}^{\Delta_\phi}(1-\chi,1-\chib)|_{\chi\ll1} \sim f_1(\chib)g_1(\chi)\log\chi +f_2(\chib)g_2(\chi)}
where the functions $f_i$ and $g_i$ are regular at $\chib=1$ and $\chi=0$, respectively. Collecting factors, the first line of \eqr{lorinv} becomes proportional to
%\alpha_{\D,J}(-1)^J
\e{350}{\int_0^1{d\chib\o \chib^2}\left({\chib\o 1-\chib}\right)^{\D_\phi}k_{\D+J}(\chib)\int_0^1 d\chi \chi^{-1-\left({\D-J\o2}\right)+\D_\phi}(f_1(\chib)g_1(\chi)\log\chi + f_2(\chib)g_2(\chi))}
The $\chi$ integral generates both single and double poles at twists $\D-J=2\D_\phi+2n$ for all $n\in\mathbb{Z}_{\geq 0}$ and any $J$. As explained in the introduction, these are the twists of the double-twist operators\foot{The $\log\chi$ term, and hence the anomalous dimensions, is absent when external operators do not have pairwise equal dimensions.}
\e{phiphi}{[\phi\phi]_{n,J} = \phi(\partial^2)^n\partial_{\mu_1}\ldots \partial_{\mu_J}\phi - (\text{traces})}
The parameter $n$ is correlated with the order in the power series expansions of $g_1(\chi)$ and $g_2(\chi)$ around $\chi=0$. 

We are now in a position to extract the CFT data that we are after, because the residues of the poles give the contribution to the double-twist OPE data: double poles encode their anomalous dimensions, and single poles encode their OPE coefficients with the external operators $\phi$. The residues are determined by the $\chib$ integral, which depends on the form of the conformal blocks via $f_i(\chib)$. In all even dimensions, the $f_i(\chib)$ may be derived from the explicit expressions for the blocks \cite{Dolan:2000ut, Dolan:2003hv}. In $d=2$ for instance,
\e{}{f_1(\chib) = k_{2\bar h'}(1-\chib).}
Plugging into \eqr{350}, the $\chib$ integral is just $\Omega^{h_\phi,h_\phi,h_\phi,h_\phi}_{h,h',2h_\phi}$, previously defined in \eqr{eq:Omega}. This is exactly what one finds by analyzing the singularities of the two-dimensional $6j$ symbol \eqr{6jd2}, as we did in \eqr{ompoles}. By keeping track of the various constant factors above and using the explicit form of the conformal blocks in any even $d$, one can derive all anomalous dimensions and OPE coefficients of the double-twist operators \eqr{phiphi}. 

In particular, let us consider the contribution to anomalous dimensions from a single $t$-channel conformal block.
These are coefficients of double-poles in the Lorentzian inversion of the block, divided by OPE coefficients of MFT. OPE coefficients of MFT are themselves residues of single poles of the Lorentzian inversion of the unit operator. For example, consider equal external operators $h_i=h_\f$, $\bar h_i=\bar h_\f$ in $d=2$.  We find the following elegant formula for the contribution of a $t$-channel operator with quantum numbers $(h',\bar h')$ to leading-twist ($n=0$) anomalous dimensions in the $s$-channel,
\be
\label{eq:resulttwod}
\g_{0,J}^{(2d)} &= \mathop{\mathrm{Res}}_{\De=4\bar h_\f+J} \frac{(\mathcal{B}_2)^{h_\f,\bar h_\f}_{[h,\bar h],[h',\bar h']}}{(\mathcal{B}_2)^{h_\f,\bar h_\f}_{[h,\bar h],[0,0]}} \nn\\
%&=\mathrm{Res}_{\De=4\bar h_\f+J} \frac{
%\sin^2(\pi(\bar h'-2\bar h_\f)) \, \Omega^{h_\f}_{h,h',2h_\f}\, \Omega^{\bar h_\f}_{\tl{\bar h},\bar h',2\bar h_\f}
%}
%{\sin^2(\pi(-2\bar h_\f)) \, \Omega^{h_\f}_{h,0,2h_\f}\, \Omega^{\bar h_\f}_{\tl{\bar h},0,2\bar h_\f}
%}\nn\\
&=
-\left.\frac{2\sin^2(\pi(\bar h'-2\bar h_\f))}{\sin^2(\pi(-2\bar h_\f))} \frac{\G(h)^2 \G(2\bar h')}{\G(2h)\G(\bar h')^2} \frac{\G(h-2h_\f+1)}{\G(h+2h_\f-1)} \frac{\Omega^{h_\f}_{h,h',2h_\f}}{\G(1-2h_\f)^2}
\right|_{\De=4\bar h_\f+J}.
\ee
Here, $\mathcal{B}_2$ represents the Lorentzian inversion of a single conformal block, and is given in (\ref{eq:inverseofblock}). Similarly, in $d=4$, for the contribution of an operator with quantum numbers $\De',J'$ we find
\be
\label{eq:resultfourd}
\gamma_{0,J}^{(4d)} &= \mathop{\mathrm{Res}}_{\De=2\De_\f+J}\frac{(\mathcal{B}_4)^{\De_\f}_{\De,J,\De',J'}}{(\mathcal{B}_4)^{\De_\f}_{\De,J,0,0}} \nn\\
&= \frac{2\Gamma(\De_\f)^2}{\G(\De_\f-\frac{\De'-J'}{2})^2} \frac{\G(\tfrac{\De+J}{2})^2}{\G(\De+J)} \frac{\G(\tfrac{\De+J}{2}-\De_\f+1)}{\G(\tfrac{\De+J}{2}+\De_\f-1)} \frac{1}{\G(1+\frac{\De'-J'}{2}-\De_\f)^2} \nn\\
&\quad \x \left.\left(
\frac{\G(\De'-J'-2)}{\G(\frac{\De'-J'-2}{2})^2}
\Omega^{\frac{\De_\f}{2}}_{\frac{\De+J}{2},\frac{\De'+J'}{2},\De_\f-1}
 - \frac{\G(\De'+J')}{\G(\frac{\De'+J'}{2})^2}
  \Omega^{\frac{\De_\f}{2}}_{\frac{\De+J}{2},\frac{\De'-J'-2}{2},\De_\f-1}
\right)\right|_{\De=2\De_\f+J},
\ee
where $\mathcal{B}_4$ is given in (\ref{eq:b4}). These results agree with large-spin perturbation theory \cite{Fitzpatrick:2012yx,Komargodski:2012ek,Alday:2016njk,Simmons-Duffin:2016wlq,Sleight:2018ryu} when expanded at large $J$.~\footnote{
Our results (\ref{eq:resulttwod}, \ref{eq:resultfourd}) disagree with those of \c{Cardona:2018dov} for an interesting reason. The calculations of \c{Cardona:2018dov} used an exact formula for an integral  \c{Caron-Huot:2017vep}
\be
\int_0^1 \frac{d\bar z}{\bar z^2} \kappa_{2h} k_{2h}(\bar z) \mathrm{dDisc}[\bar y^a] = \frac{1}{\G(-a)^2}\frac{\G(h)^2}{\G(2h-1)} \frac{\G(h-a-1)}{\G(h+a+1)},
\label{eq:examplebad}
\ee
where $\bar y=\frac{1-\bar z}{\bar z}$ and $\kappa_{2h}$ is given in \c{Caron-Huot:2017vep}. By expanding $t$-channel blocks in powers of $\bar y$, one can attempt to perform the Lorentzian inversion term-by-term using (\ref{eq:examplebad}) and then resum the result. This precisely reproduces the solution to large-spin perturbation theory originally derived in \c{Simmons-Duffin:2016wlq}. (The right-hand side of (\ref{eq:examplebad}) is $S_a(h)$, defined in \c{Simmons-Duffin:2016wlq}.) However, while the expansion in $\bar y$ works well near $\bar z = 1$, it is poorly behaved near $\bar z = 0$. This is reflected on the right-hand side of (\ref{eq:examplebad}) which has poles at $h=a+1-k$, $k\in \mathbb{Z}_{\geq 0}$. For larger $a$, these poles move farther to the right, resulting in a highly oscillatory function of $h$. Because the expansion in $\bar y$ is only convergent for $\bar z \in [1/2,1]$, it ultimately only gives an {\it asymptotic\/} large-spin expansion for OPE data (which is determined by $\bar z\sim 1$) --- it does not give correct result at finite spin (which depends on all $\bar z\in [0,1]$).

The Mellin-space version of expanding in $\bar y$ is to miss certain poles in Mellin variables that contribute when inverting a block. The inverse of a block with dimension $\De'$ should decay in the right-half $\De'$ plane, so that it is picked up when deforming the contour in the spectral integral over $\De'$.

If instead of inverting $\bar y^a$ and resumming, one directly inverts $k_{2h'}(1-\bar z)$ (or any function well-behaved near $\bar z=0$), then unphysical poles are not present. Indeed, unphysical poles cancel between the two ${}_4F_3$'s in the formula (\ref{omegaexp}) for $\Omega_{h,h',p}^{h_i}$.  We have checked that if we take our result (\ref{eq:resultfourd}) for $\g^{(4d)}_{0,J}$ and artificially remove one of the ${}_4F_3$'s from each of the $\Omega$'s, then we reproduce the formulas of \c{Cardona:2018dov}. This procedure of removing a ${}_4F_3$ also reproduces the formulas of \c{Sleight:2018epi,Sleight:2018ryu}, which resummed the asymptotic large-spin expansion but did not provide formulas at finite spin. Unphysical poles in $h$ of the form (\ref{eq:examplebad}) appear in all of \c{Cardona:2018dov,Sleight:2018epi,Sleight:2018ryu} and are discussed in \c{Sleight:2018ryu}.

These subtleties show why the Lorentzian inversion formula is a concrete improvement over the naive procedure of solving the all-orders large-spin expansion and trying to resum it manually.

We thank David Poland, Soner Albayrak, Charlotte Sleight, and Massimo Taronna for discussion on these points. \label{foot27}
} In addition to capturing the contribution to anomalous dimensions of a single $t$-channel block, these formulas can also be interpreted as the contribution of a $t$-channel Witten diagram for the exchange of a bulk field with quantum numbers $\De',J'$, since the $\mathrm{dDisc}$ of a Witten diagram is the same as the $\mathrm{dDisc}$ of a block \cite{Hijano:2015zsa}. (See Sec. \ref{tree}.) It is completely straightforward to obtain anomalous dimensions for higher $n=1,2,\dots$ by taking residues of other families of poles $\De=2\De_\f+2n+J$. By using the $6j$ symbol itself, $\mathcal{J}_d$, instead of the Lorentzian inversion of a block, $\mathcal{B}_d$, we can determine the contribution of a partial wave in the $t$-channel to anomalous dimensions in the $s$-channel. The resulting formulae match, for example, the anomalous dimensions derived in Eqs.~4.11 and 4.20 of \c{Giombi:2018vtc}.

%For leading twist operators $(n=0)$, one easily finds a match to the anomalous dimensions derived in Eqs.~4.11 and 4.20 of \c{Giombi:2018vtc}.\foot{More precisely, \c{Giombi:2018vtc} branch a $t$-channel partial wave into $s$-channel confomal blocks rather than partial waves, so the residues of the $6j$ symbol give a sum of that result and its shadow. In general, given \eqr{BSd}, it is simple to formally read off the Lorentzian inversion of a conformal block $G^{\D_i}_{\D',J'}(x_i)$, rather than a partial wave $\Psi_{\D',J'}$, from a $6j$ symbol: one just drops the shadow piece of the $6j$ symbol and sets $K^{\D_1,\D_4}_{\tilde \D',J'}$ to one. For instance, in the $d=4$ $6j$ symbol \eqr{6jd4}, the inversion of $G^{\D_i}_{\D',J'}$ would yield only the coefficient of $K_{\tilde\D',J'}^{\D_1,\D_4}$, without the last line.\label{foot26}} Results for subleading twists $\g_{n,\ell}$, and extension to general $d$, can be found in \c{Cardona:2018dov,Sleight:2018epi,Sleight:2018ryu}.

\subsection{$6j$ open questions}

We  computed general $6j$ symbols in $d=1,2$ and a large class of $6j$ symbols in 4d using the Lorentzian inversion formula. (All other $6j$ symbols in 4d can be obtained by applying weight-shifting operators to the case we computed \cite{Karateev:2017jgd}, or by applying the generalized Lorentzian inversion formula \cite{Kravchuk:2018htv}.) For other values of $d$, we do not know a simple way to proceed using this technique: in odd $d$, the explicit form of the partial waves is not known, while in even $d>4$, somewhat surprisingly, the Lorentzian inversion integral does not  factorize into one-dimensional integrals. It would obviously be useful to find a way to perform these integrals. %The uniformity of double-twist anomalous dimensions as a function of $d$ \c{Cardona:2018dov,Sleight:2018epi,Sleight:2018ryu} suggests that the $6j$ symbol may have simple $d$-dependence.

Our method of computing $6j$ symbols breaks most of the manifest tetrahedral symmetry of the starting point (\ref{6jDef}). Nevertheless, we have checked numerically that the resulting expressions exhibit tetrahedral symmetry. In all cases, the result can be written in terms of so-called ``balanced" ${}_4F_3$ hypergeometric functions, i.e.\ ${}_4F_3$'s where the sum of the arguments in the first row minus the arguments in the second row is $-1$. These functions enjoy nontrivial identities that are presumably responsible for tetrahedral symmetry.\footnote{We thank Petr Kravchuk for these observations.} The same functions appear in the $6j$ symbol for $\mathrm{SU}(2)$.

\section{AdS applications of $6j$ symbols} \label{AdS}

We now show how the $6j$ symbol plays a central role in the computation of AdS amplitudes. 
\ssec{Tree-level}\label{tree}
Consider a four-point function of identical scalars $\phi$ for simplicity. Restricting to tree-level in AdS, the $\phi\x\phi$ OPE contains the operators
\e{treeope}{\phi\x\phi \sim \sum_{\D',J'}\O_{\D',J'} + \sum_{n=0}^\i\sum_{J=0}^\i [\phi\phi]_{n,J}\qquad (\text{tree-level})}
where $\cO_{\D',J'}$ are single-trace conformal primaries. The AdS amplitude $\cA_{\rm tree} = \la\phi\phi\phi\phi\ra_{\rm tree}$ is a sum of exchange diagrams in the $s$-, $t$- and $u$-channels, with one such diagram for every bulk cubic coupling $\lambda_{\phi\phi\phi_{\D',J'}}$, where $\phi_{\D',J'}$ is dual to $\O_{\D',J'}$. There may also be $\phi^4$-type contact interactions. 

The information contained in $\cA_{\rm tree}$ is the set of OPE data in \eqr{treeope}. Given some set of single-traces $\lbrace \O_{\D',J'}\rbrace$, the double-trace OPE data in \eqr{treeope} -- i.e.\ the leading corrections to the MFT dimensions $\D_{n,J}=2\D_\phi+2n+J$ and three-point functions $\la\phi\phi[\phi\phi]_{n,J}\ra$ -- receive contributions fixed by crossing symmetry. In fact, the $6j$ symbol compactly packages essentially all of the double-trace OPE data. Writing $\cA_{\rm tree}$ in contour-integral form \eqr{partialwaveexpansion}, the main observation, to be explained below, is the following:
\vs
\vs
{{{\it The OPE function ${I_{\D,J}/ n_{\D,J}}$ is a sum of $6j$ symbols, one for each single-trace operator $\O_{\D',J'}$,}}}

{{{\it plus terms non-analytic in $J$.}}}
\vs
\vs
\noindent By ``$6j$ symbol'' we mean the inversion of the conformal block $G_{\D',J'}$ rather than the partial wave  $\Psi_{\D',J'}$ -- that is, $\mathcal{B}_d$, not $\mathcal{J}_d$. Note that for $\O_{\D',J'}$ exchange, the $6j$ symbol fully determines the OPE data only for spins $J>J'$, due to the Regge bound discussed below \eqr{eq:ddiscu}. In particular, this excludes the conformal block for $\O_{\D',J'}$ itself, a fact to which we return in Section \ref{adsq}. Likewise, $\cA_{\rm tree}$ may have contact contributions with $N_{\partial}$ derivatives, which contribute only to double-trace OPE data of spin $J\leq \lfloor {N_\p/ 2}\rfloor$; these are also not captured by the $6j$ symbol. In all, we may write 
\e{inj}{{I_{\D,J}\o n_{\D,J}} = \sum_{\D',J'} \lambda_{\phi\phi\O_{\D',J'}}^2\Big(\mathcal{B}_d(\D,J;\D',J'|\D_\phi,\D_\phi,\D_\phi,\D_\phi) + \text{($J\leq J'$ terms)}\Big) + \text{(contact terms)}}
In other words, the double-trace part of the OPE decomposition of the AdS exchange diagrams is, modulo low-spin non-analyticities, determined by the residues of the inverted blocks $\mathcal{B}_d$. 

It is straightforward to prove \eqr{inj}. The idea is that dDisc of the sum of exchange diagrams kills all but the single-trace conformal blocks, whose Lorentzian inversion is the definition of the $6j$ symbol. Consider the sum of exchange diagrams from a single $\O_{\D',J'}$,
\e{}{W_{\D',J'} = W_{\D',J'}^{(s)} + W_{\D',J'}^{(t)}+W_{\D',J'}^{(u)}}
The first statement is that dDisc of a crossed-channel exchange diagram vanishes: e.g.
\e{}{\dDisc_t[W_{\D',J'}^{(s)}] = \dDisc_t[W_{\D',J'}^{(u)}]=0}
This follows from the fact that a crossed-channel exchange diagram branches into direct-channel double-trace blocks only, and dDisc kills direct-channel double-trace blocks:
\e{ddiscz}{\dDisc_t[G_{\D',J'}^{(t)}] =2 \sin^2\left(\pi\left({\D'-J' \o 2} - \D_\phi\right)\right)G_{\D',J'}^{(t)}}
These zeroes are visible in our $6j$ symbols \eqr{6jd2} and \eqr{6jd4}, and are present in all $d$. Next, dDisc of a direct-channel exchange diagram is equivalent to dDisc of the single-trace block alone:
\e{}{\dDisc_t[W_{\D',J'}^{(t)}] = \dDisc_t[G_{\D',J'}^{(t)}]}
This follows because the exchange diagram branches into one single-trace block plus a sum of double-trace blocks, but the latter are killed by the zeroes \eqr{ddiscz}. Altogether, then
\e{dwgt}{\dDisc_t[W_{\D',J'}] = \dDisc_t[G_{\D',J'}^{(t)}]}
Plugging the result into the Lorentzian inversion formula \eqr{lorinv} by recalling the definitions \eqr{eq:ddisct} and \eqr{eq:ddiscu}, and adding the dDisc$_u$ pieces (which just multiplies \eqr{dwgt} by an overall $1+(-1)^J$), the LHS yields the $s$-channel OPE function ${I_{\D,J}/ n_{\D,J}}$ for the exchange diagrams, while the RHS yields the $6j$ symbol with the shadow contribution projected away. This concludes the proof.

An interesting aspect of \eqr{dwgt} is that conformal blocks are represented in AdS as geodesic Witten diagrams \c{Hijano:2015zsa}. Thus, the geometrization of the dDisc action \eqr{dwgt} is that the integration over bulk vertices becomes restricted to geodesics.

A holographic CFT with a weakly coupled, local bulk dual has a gap to single-trace higher-spin operators ($\D_{\rm gap}\rar\i$): all single-trace operators have $J\leq2$. Likewise, there are only three quartic interactions -- $\phi^4, (\p\phi)^4, (\p\phi)^2(\p^2\phi)^2$ -- that are consistent with the chaos bound \c{MSS}, all of which contribute only to $J'\leq 2$.\foot{Of these, the last two are expected to be suppressed by powers of $\D_{\rm gap}$ predicted by dimensional analysis, based on studies of higher-derivative contributions to cubic vertices and effective field theory reasoning \c{Caron-Huot:2017vep, cemz, Meltzer:2017rtf, Afkhami-Jeddi:2018own}. However, this has not been proven. } So for such theories, the $6j$ term in \eqr{inj} gives the complete answer for $J>2$. Such CFTs typically have a finite number of light single-trace operators (``sparseness''), so \eqr{inj} is a finite sum. 

To summarize, $6j$ symbols determine the full OPE content of AdS tree-level exchange diagrams $W_{\D',J'}$ for all $J>J'$: all double-trace data for $[\phi\phi]_{n,J>J'}$ are sums of residues of the $6j$ symbol, in the manner described in Sec.~\ref{6jcft}. (The $J\leq J'$ corrections as determined from crossing symmetry may be found in \c{Alday:2017gde}.) In \eqr{eq:resulttwod} and \eqr{eq:resultfourd}, we gave explicit formulas for the leading-twist anomalous dimensions in $d=2,4$. This is a concrete situation where $\dDisc[\cA]$ is simpler than $\cA$, but can nevertheless be used to construct (almost) the full amplitude.\foot{One may construct the amplitude using a dispersion relation of the form $\cA(z,\zb) = \int dz' d\zb' K(z,\zb;z',\zb') \dDisc[\cA(z',\zb')]$ for some kernel $K(z,\zb;z',\zb')$, which can be determined by plugging the inversion formula back into the original expression for the four-point function \c{deansimon}.} The preceding discussion applies without modification to the case of distinct external operators.  

\ssec{One-loop}\label{loop}

We now compute one-loop diagrams in AdS -- specifically, $n$-gons, which have $n$ external legs of arbitrary spin, connected by cubic vertices to an internal loop made of scalars, and show that they can be written as glued $6j$ symbols. The procedure used to prove this is simple:

\vs
{\bf 1)} Write all internal propagators in the split representation.
\vs
{\bf 2)} Do the AdS three-point integrals.
\vs
From our expressions, it will be apparent that structure of the one-loop $n$-gon AdS diagram is the same as that of the SYK bilinear $n$-point planar diagrams with no exchanged melons.

Let us first set up our notations.\foot{Various identities used below may be found in many works, e.g.\ \c{Penedones:2010ue,Costa:2014kfa}. We use the conventions of \c{Giombi:2018vtc}.} We work in Poincar\'e AdS, with bulk coordinates $y=(z,\vec x)$, where the radial coordinate is $z$. We use shorthand
\e{}{dy \equiv d^{d+1}y\sqrt{g(y)}~, ~~ dx \equiv d^dx~.}
We write scalar bulk-boundary propagators as $K_\D(x_1,y)$, bulk-bulk propagators as $G_\D(y_1,y_2)$, and harmonic functions as $\Om_\D(y_1,y_2)$. The latter may be defined by 
\e{}{\Om_\nu(y_1,y_2) = {i\nu\o2\pi}(G_\nu(y_1,y_2)-G_{-\nu}(y_1,y_2))~,}
where we have introduced the spectral parameter $\nu$,%\foot{}
\e{Dnu}{\D={d\o2}+i\nu~.}
We sometimes use $\nu$ as a subscript, with the understanding that it is related to $\D$ by \eqr{Dnu}, in order to make the shadow transformation more obvious. We normalize the bulk-boundary propagators as
\e{}{K_{\Delta}(x_1,y) = {\cal C}_{\Delta} \left(\frac{z}{z^2+(\vec{x}-\vec{x}_1)^2}\right)^{\Delta}\,,}
where
\e{}{\C_{\Delta}=\frac{\Gamma(\Delta)}{2\pi^{d/2}\Gamma(\Delta+1-d/2)}~.}
%\end{equation}
%
With this choice of normalization of the bulk-to-boundary propagator, the two-point function of the dual operator is normalized as
\e{onorm}{\langle \O (x_1)\O (x_2)\rangle = {{\cal C}_{\Delta}\o x_{12}^{2\Delta}}~.}
The analogous normalization constant for a spinning operator is $\C_{\D,J} = \C_\D\left({\D-1+J\o \D-1}\right)$. In all that follows, we set the bulk cubic couplings to one. 

\begin{figure}
\centering
\includegraphics[width=2.07in]{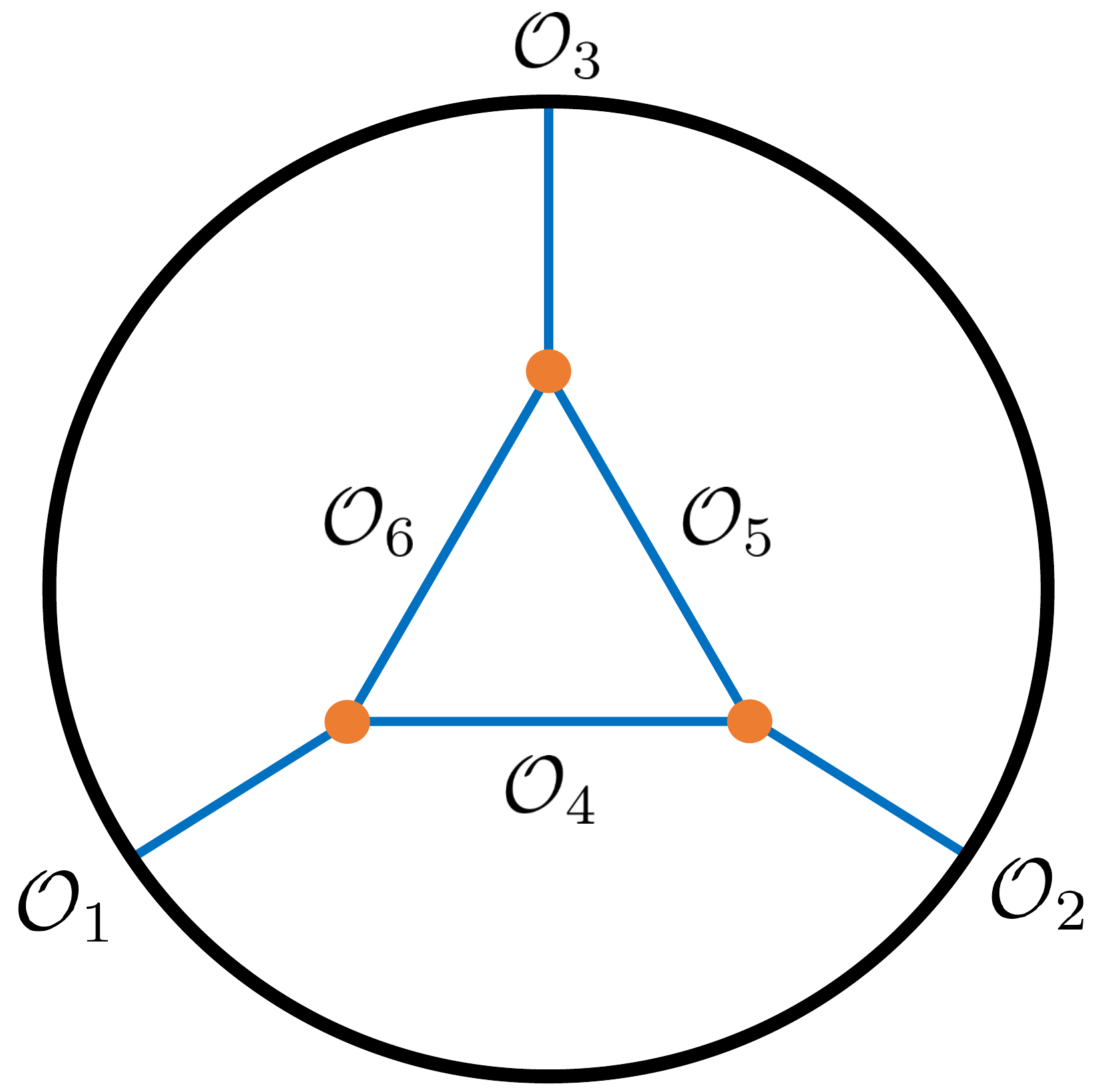}
\caption{The AdS three-point triangle, $\mathcal{A}^\1_{123}$. Orange points denote AdS integration.}
\label{figtri}
\end{figure}

\sssec{Three-point triangle}
We start with the one-loop, three-point triangle diagram in AdS. For simplicity, we take all operators to be scalars; the generalization to external spins will follow trivially. The external legs are labeled $\O_{1,2,3}$, and the internal legs are $\O_{4,5,6}$, with corresponding conformal dimensions, as in Fig. \ref{figtri}.
The diagram is computed as
\es{}{&\mathcal{A}^\1_{123}(x_i) = \int dy_1dy_2dy_3 K_{\nu_1}(x_1,y_1)K_{\nu_2}(x_2,y_2)K_{\nu_3}(x_3,y_3)G_{\nu_4}(y_1,y_2)G_{\nu_5}(y_2,y_3)G_{\nu_6}(y_3,y_1)~,}
where the three integrals run over all of AdS. Due to conformal symmetry,
\e{}{\mathcal{A}^\1_{123}(x_i) = \C^\1_{123} \la\O_1(x_1)\O_2(x_2)\O_3(x_3)\ra~.}

Let's first study the simpler object --- call it $A^\1_{123}$ --- where the propagators $G_\nu$'s are replaced by harmonic functions $\Om_\nu$:
\es{}{&A^\1_{123}(x_i) = \int dy_1dy_2dy_3 K_{\nu_1}(x_1,y_1)K_{\nu_2}(x_2,y_2)K_{\nu_3}(x_3,y_3)\Om_{\nu_4}(y_1,y_2)\Om_{\nu_5}(y_2,y_3)\Om_{\nu_6}(y_3,y_1)~.}
As we will recall in a moment, $\mathcal{A}^\1_{123}$ is simply a triple spectral integral over $A^\1_{123}$.\foot{For this reason, $A^\1_{123}$ was referred to as a pre-amplitude in \cite{Yuan:2018qva}. We point out that pre-amplitudes correspond to physical quantities: namely, they may be seen as linear combinations of amplitudes (i.e.\ CFT correlators) in which one employs either standard or alternative quantization for the fields propagating in the internal lines. This is a loop-level version of the interpretation of the conformal partial wave, represented in AdS as an exchange diagram with a harmonic function, as a difference of four-point functions in two CFTs related by a double-trace RG flow \cite{Hartman:2006dy,Giombi:2018vtc}. } Again, by conformal symmetry, 
\e{}{A^\1_{123}(x_i)  =  C^\1_{123}\la\O_1(x_1)\O_2(x_2)\O_3(x_3)\ra~.}
for some $C^\1_{123}$. To evaluate this, we use the split representation for the $\Om_{\nu_i}$'s,
\e{omsplit}{\Om_\nu(y_1,y_2) = {\nu^2\o\pi}\int dx \,K_\nu(x_1,y_1){K_{-\nu}}(x_1,y_2)~,}
to write $A^\1_{123}$ as three three-point functions stitched together along the boundary, as in Fig. \ref{figsplittri}:
\begin{figure}
\centering
\includegraphics[width=2in]{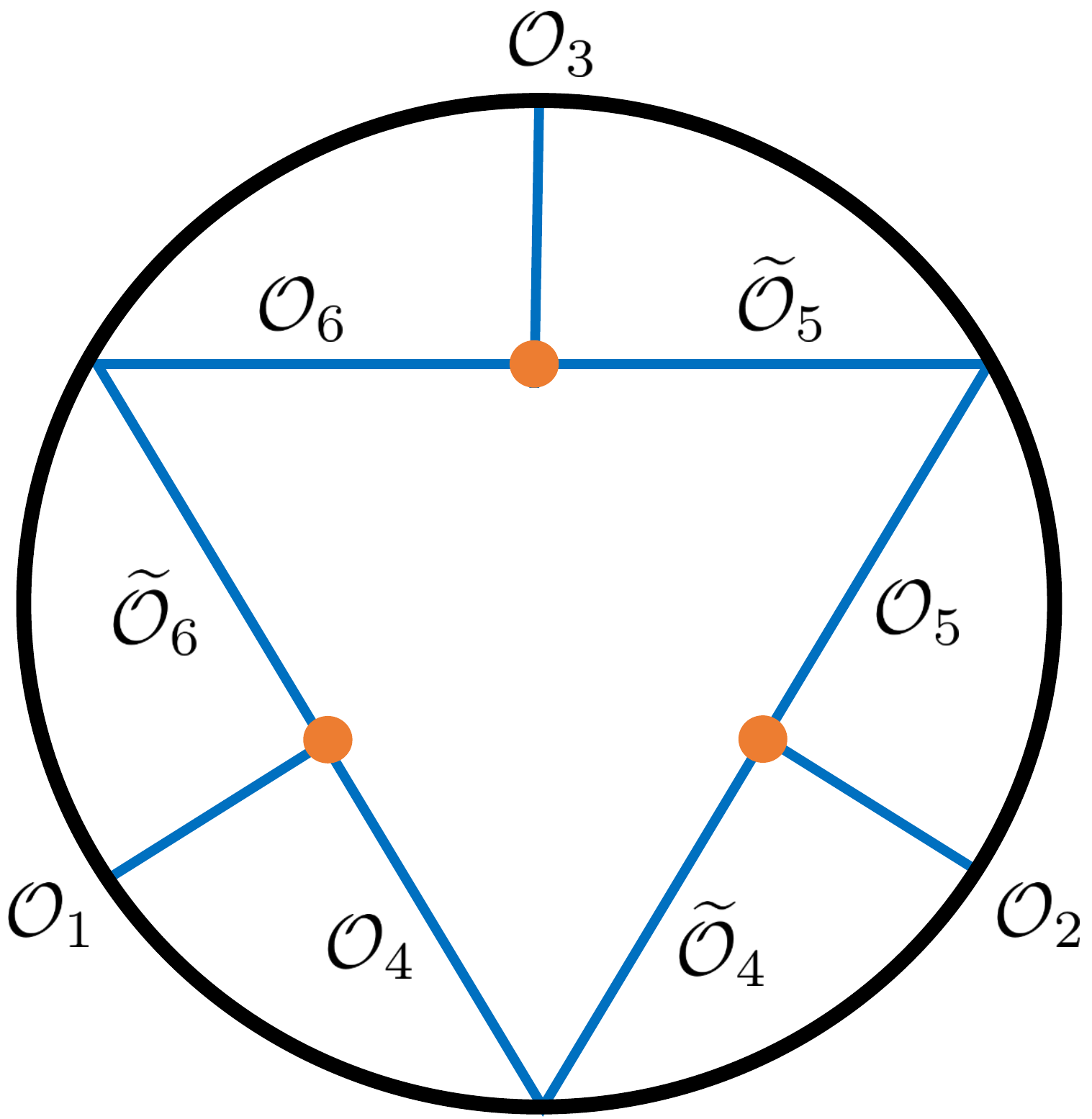}
\caption{The AdS harmonic three-point triangle, $A^\1_{123}$, after using the split representation on internal lines. Boundary vertices arising from the split representation are also integrated over.}
\label{figsplittri}
\end{figure} 
\es{}{A^\1_{123}(x_i) = \left(\prod_{i=4,5,6}{\nu_i^2\o\pi}\right)\int dx_4dx_5dx_6&\left(\int dy_1K_{\nu_1}(x_1,y_1)K_{-\nu_6}(x_6,y_1)K_{\nu_4}(x_4,y_1)\right)\\
\times&\left(\int dy_2 K_{\nu_2}(x_2,y_2)K_{-\nu_4}(x_4,y_2)K_{\nu_5}(x_5,y_2)\right)\\
\times&\left(\int dy_3 K_{\nu_2}(x_3,y_3)K_{-\nu_5}(x_5,y_3)K_{\nu_6}(x_6,y_2)
\right)~.}
Each object in parenthesis is itself a boundary three-point function, times a factor obtained by performing the AdS integral: 
\e{}{\int dy_1 K_{\nu_1}(x_1,y_1)K_{-\nu_6}(x_6,y_1) K_{\nu_4}(x_4,y_1)= b_{\nu_1,-\nu_6,\nu_4}\la \O_1(x_1)\Ot_{6}(x_6)\O_4(x_4)\ra}
where, in terms of the $\D$'s (e.g.\ \cite{Freedman:1998tz, Costa:2014kfa}),
\e{}{{b_{\D_1\D_2\D_3} = {\C_{\D_1}\C_{\D_2}\C_{\D_3,J_3}}{\pi^{d/2}\G({\D_1+\D_2+\D_3+J_3-d\o2})\G({\D_1+\D_2-\D_3+J_3\o2})\G({\D_2+\D_3-\D_1+J_3\o2})\G({\D_3+\D_1-\D_2+J_3\o2})\o 2^{1-J_3}\G(\D_1)\G(\D_2)\G(\D_3+J_3)}}~.}
(We have included the spin of $\O_3$ for later generalization to the spinning case.) The AdS integrals are now gone, and we have
\e{a1l}{A^\1_{123}(x_i) = f_{\rm AdS}(\D_i) \times I^{(2)}_{\lbrace \D_i \rbrace}(x_i)~,}
where $f_{\rm AdS}(\D_i)$ is a kinematic prefactor,
\e{fads3}{f_{\rm AdS}(\D_i) \equiv {\nu_4^2\nu_5^2\nu_6^2\o\pi^3}\, b_{\nu_1,-\nu_6,\nu_4}b_{\nu_2,-\nu_4,\nu_5}b_{\nu_3,-\nu_5,\nu_6}~,}
and~\foot{It is perhaps useful to write the bulk diagram slightly differently, using the bulk-boundary identity 
\be \nonumber
K_{-\nu}(x_1,y) =-2i\nu\int dx_2 K_{\nu}(x_2,y)\la \Ot_1(x_1)\Ot_2(x_2)\ra
\ee
where we normalize the shadow two-point function as in \eqr{onorm}. This allows us to trade shadow operators $\Ot$'s for $\O$'s in \eqr{I2loop}.} %
\e{I2loop}{I^{(2)}_{\lbrace \D_i \rbrace}(x_i) \equiv \int dx_4dx_5dx_6\la \O_1(x_1)\Ot_{6}(x_6)\O_{4}(x_4)\ra\la \O_2(x_2)\Ot_{4}(x_4)\O_{5}(x_5)\ra\la \O_3(x_3)\Ot_{5}(x_5)\O_{6}(x_6)\ra~.}
Contracting both sides of \eqr{a1l} with a shadow three-point structure, 
\e{}{C^\1_{123} = {1\o t_0}\,{(A^\1_{123}(x_i) ,\la\tilde\O_1\tilde\O_2\tilde\O_3\ra)}}
where $t_0= (\la \O_1\O_2\O_3\ra,\la \Ot_1\Ot_2\Ot_3\ra) = (2^d\vol(\SO(d-1)))^{-1}$ was computed in \eqr{t0}. The conformally covariant pairing is proportional to the $6j$ symbol computed earlier, 
\e{}{\mathcal{J}_d(\tilde\Delta_4,0;\D_3,0|\D_6,\tilde\Delta_1,\tilde\Delta_2,\tilde\Delta_5)\equiv{\left( I^{(2)}_{\lbrace \D_i\rbrace}(x_i),\la\tilde\O_1\tilde\O_2\tilde\O_3\ra\right) }~.}
Therefore, $C^\1_{123}$ equals a $d$-dimensional $6j$ symbol times a universal kinematic prefactor:
\e{com456}{C^\1_{123} ={f_{\rm AdS}(\D_i)\o t_0} \times\,\mathcal{J}_d(\tilde\Delta_4,0;\D_3,0|\D_6,\tilde\Delta_1,\tilde\Delta_2,\tilde\Delta_5)~.}
Note that if we had used the split representation on only two of the three internal legs, we would have obtained a convolution of two three-point functions and a conformal partial wave, the latter being represented as a bulk four-point exchange diagram with $\Om_\D$ exchange. This makes the relation to the overlap of two four-point conformal partial waves more transparent; the operators $\O_3$ and $\O_4$ are the internal operators when the diagram is viewed as an overlap of conformal partial waves, as is visible in Fig. \ref{figsplittri}.

Now we return to the full one-loop triangle diagram, $\mathcal{A}^\1_{123}(x_i)$, where the $\Om_{\D_i}$ are replaced by bulk-to-bulk propagators $G_{\D_i}$. The propagators admit a split representation 
\e{}{G_\D(y_1,y_2) = \int_{-\i}^\i {d\nu \o (\nu^2+(\D-{d\o2})^2)}\Om_\nu(y_1,y_2)~,}
which implies that $\mathcal{A}^\1_{123}$ is also a spectral integral over $A^\1_{123}$,
\e{a1aom}{\mathcal{A}^\1_{123}(x_i) =\prod_{i=4,5,6} \int_{-\i}^\i {d\nu_i \o (\nu_i^2+(\D_i-{d\o2})^2)} A^\1_{123}(x_i)~.}
Since $\D_{4,5,6}$ are only internal leg variables, the OPE coefficient $\C^\1_{123}$ is just a spectral integral over $C^\1_{123}$, 
\e{c1loop}{{\C^\1_{123} = \left(\prod_{i=4,5,6}\int_{-\i}^\i {d\nu_i\o (\nu_i^2 +(\D_i-{d\o2})^2)}\right)C^\1_{123} }~,}
where $C^\1_{123}$ was determined in \eqr{com456}. 

We have thus reached our final result: in any dimension $d$, the one-loop correction to $\la \O_1\O_2\O_3\ra$ is a spectral integral over the $6j$ symbol for the conformal group, times the universal kinematic factor $f_{\rm AdS}(\D_i)$. Both the $6j$ symbol and $f_{\rm AdS}(\D_i)$ are meromorphic functions, whose poles and zeroes may be read off from our explicit expressions and the discussions in Secs.~\ref{6jcft} and \ref{tree}; this concludes our computation. 

The generalization of the above to the case of external spinning operators $\mO_{1,2,3}$, still with internal scalars $\mO_{4,5,6}$, is immediate. The result follows from the same manipulations, which only really involve the internal scalar propagators. Now we have a sum over structures,
\e{}{A^\1_{123}(x_i)  =  \sum_a C^\1_{123,a}\la\O_1(x_1)\O_2(x_2)\O_3(x_3)\ra_a}
The result for $C^\1_{123,a}$ is given by essentially the same formula, now with a sum over structures:
\e{com456j}{\quad C^\1_{123,a} =f_{\rm AdS}(\D_i) \x(t_0^{-1})^{ab}\,\mathcal{J}_d(\tilde\Delta_4,0;\D_3,J_3|\D_6,\tilde\Delta_1,\tilde\Delta_2,\tilde\Delta_5)_b}
where $(t_0)^{ab}$ was defined in \eqr{t0ab}.

\begin{figure}
\centering
\includegraphics[width=2.2in]{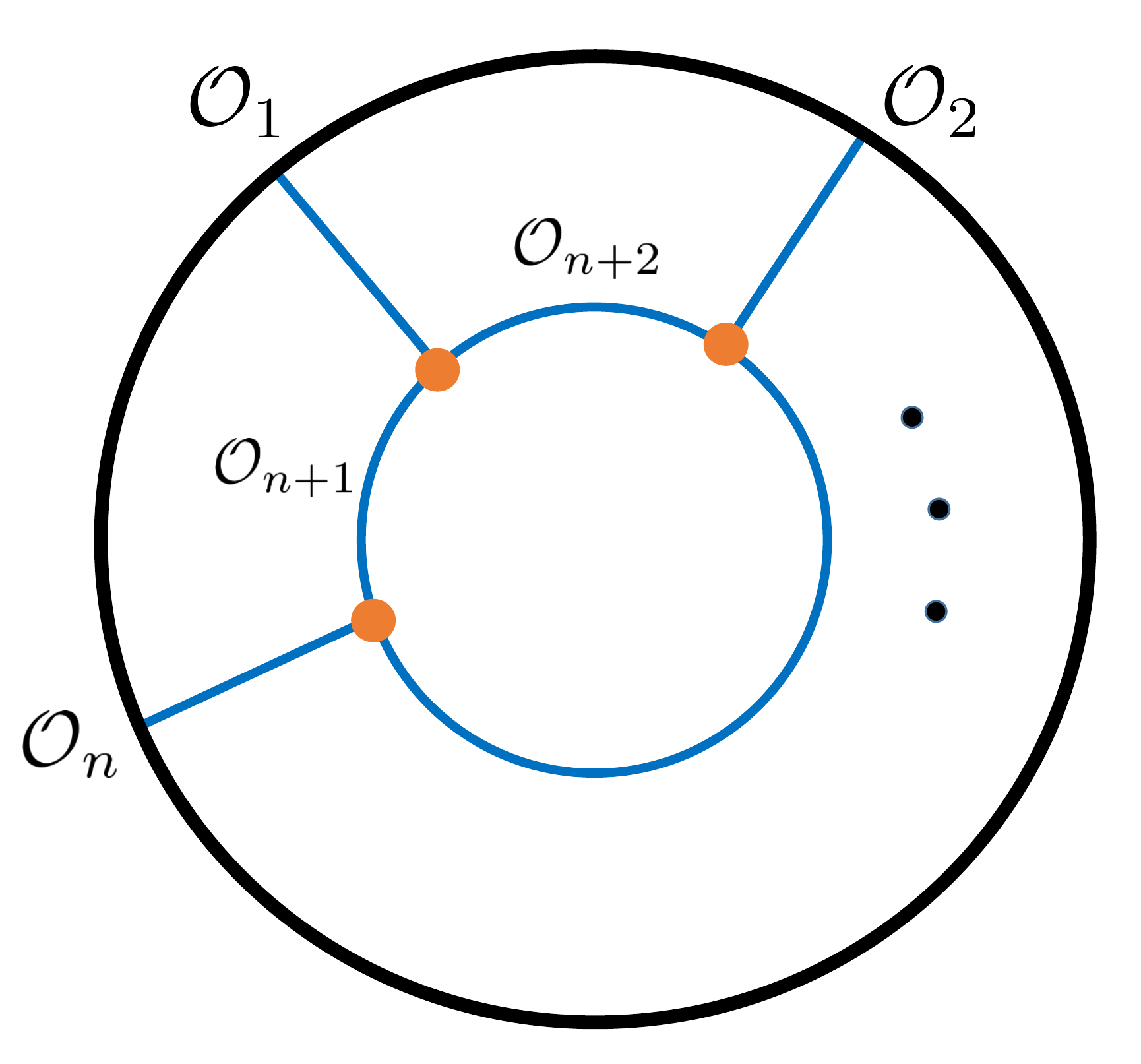}
\caption{The AdS $n$-gon.}
\label{figngon}
\end{figure}

\sssec{$n$-gons}
This story generalizes to more legs, with arbitrary external spins. We focus on $n$-gons, which are one-loop diagrams with $n$ external legs and only cubic vertices. We take the external legs to have labels $i=1,2,\ldots, n$, and internal legs with labels $i=n+1,n+2,\ldots, 2n$, as in Fig. \ref{figngon}.
Let us introduce the $n$-point pre-amplitude $A^\1_n(x_i)$, which is the $n$-gon with harmonic functions in the internal legs. Equivalently, it is the integrand of the $n$ spectral integrals of $\mathcal{A}^\1_n(x_i)$:
\e{A1n}{\mathcal{A}^\1_n(x_i) = \left(\prod_{i=n+1}^{2n}\int_{-\i}^\i {d\nu_i\o (\nu_i^2 +(\D_i-{d\o2})^2)}\right)A^\1_n(x_i)~,}
where $\nu_i$ are spectral parameters of the $n$ internal legs.

One can perform the same straightforward procedure for $A^\1_n(x_i)$ as was done in the previous section for the triangle diagram, $A^\1_3(x_i)$: one uses the split representation for the $\Omega_{\nu_i}$ and then does the AdS integrals. For the $4$-gon (box diagram) the result is a gluing of four three-point structures, 
\es{a4predist}{A^\1_{4}(x_i) = f_{{\rm AdS},4}(\D_i)\int dx_5dx_6dx_7dx_8\Big(&\la \O_1(x_1)\Ot_{6}(x_6)\O_{5}(x_5)\ra\la \O_2(x_2)\Ot_{7}(x_7)\O_{6}(x_6)\ra\\&\la \O_3(x_3)\Ot_{8}(x_8)\O_{7}(x_7)\ra\la \O_4(x_4)\Ot_{5}(x_5)\O_{8}(x_8)\ra\Big)~,}
where $f_{{\rm AdS},4}(\D_i)$ is the $n=4$ analog of  \eqr{fads3}. The $n$-gon generalization is evident. 

This integral, and  the integral in the  evaluation of the AdS triangle diagram (\ref{I2loop}), are of course familiar from our study of the SYK correlation function. 
Indeed, the integral that appeared in the evaluation of the AdS triangle diagram, Eq. (\ref{I2loop}), is identical to the one that appeared in  the planar bilinear three-point diagram in SYK, Eq. \eqr{I2v2}, while Eq. (\ref{a4predist}) is identical to the one that appeared in the evaluation of the planar bilinear four-point diagram in SYK with no exchanged melons, Eq.~\ref{4ptNoMelon}.~\footnote{The AdS diagrams were set up to be slightly more general, in that there are distinct internal operators instead of just one $(\phi)$. To generalize on the SYK side, one can generalize the diagrams to involve multiple species $\phi$, something like what occurs in the flavored SYK model \cite{GR1}.}

This connection continues to hold for any $n$: $A^\1_n$ equals a simple prefactor times the planar bilinear $n$-point function of SYK with no exchanged melons,
\e{ngonsyk}{A^\1_n(x_i) = {f_{{\rm AdS},n}(\D_i)}\times \text{(SYK bilinear $n$-point planar diagram with no exchanged melons)}\nn}
where the relevant SYK diagram for $n=3$ was shown in Fig.~\ref{tetraCut2}(a), for $n=4$ in Fig.~\ref{figsykpl}, and the generalization to higher $n$ is evident. In addition, the right hand side above denotes just the functional form of the integral that appears -- i.e. without factors of $b,\cJ, c_{\D,J}$ or shadow factors from amputation -- in the same sense observed above in the connection between the AdS triangle diagram and the SYK bilinear three-point planar diagram. The prefactor above is given by the product of AdS kinematic factors arising from each use of the split representation:
\e{}{f_{{\rm AdS},n}(\D_i) = \prod_{i=n+1}^{2n}{\nu_i^2\o \pi}b_{\nu_{i+1-n},-\nu_{i},\nu_{i+1}}~, ~\text{where}~\nu_{2n+j} \equiv \nu_{n+j}~.}

\sssec{Example: Box diagram ($n=4$) and OPE structure}

Let us give a further treatment of the particularly interesting case $n=4$, the box diagram, with internal scalars $\O_{5,6,7,8}$ and external operators $\mO_{1,2,3,4}$ with spins $J_{1,2,3,4}$. We follow the labeling convention of Fig. \ref{figngon}. 

The pre-amplitude for the box diagram was given by Eq. (\ref{a4predist}). It would be useful to have an expression that is in terms of conformal blocks. In fact, since (\ref{a4predist}) already appeared in the context of the SYK model (in the form of Eq.~\ref{4ptNoMelon}), we can use the results found there, which showed that Eq.~\ref{4ptNoMelon} is equal to the conformal block expansion Eq.~\ref{4Ans2}. We need only to generalize Eq.~\ref{4Ans2}, to account for the distinct internal operators $\O_{5,6,7,8}$. This involves using $\rho^{\rm MFT}(\D,J)$ for the disconnected correlator $\la \O_5\O_5\ra\la\O_7\O_7\ra$, which is, following the same steps that led to \eqr{rhomft},
\e{}{\rho^{\rm MFT}(\D,J) = {t_0 \o n_{\D,J}} S_{\tl \Delta_{7}}^{\tl \Delta_{5}, [\Delta, J]}\, S_{\tl \Delta_{5}}^{\Delta_{7}, [\Delta, J]}~.}
Thus, the conformal block expansion of $A^\1_4$ is,
\es{a4pre}{A_{4}^\1(x_i) &= \sum_{J=0}^{\infty}\int_{\frac{d}{2}- i\infty}^{\frac{d}{2} + i \infty} \frac{d\Delta}{2\pi i}\,\rho^\1_{4;a,b}(\D,J) \, G_{\Delta, J}^{\Delta_i, J_i; a,b }(x_i)~.}
where
\e{rho4pre}{\rho^\1_{4;a,b}(\D,J) \equiv f_{{\rm AdS},4}(\D_i)\x K_{[\tilde{\Delta}, J]}^{\Delta_{5}, \Delta_{7}}\rho^{\rm MFT}(\D,J)\,\mathcal{I}^{(2)}_{\Delta_1, J_1, \Delta_2, J_2, \Delta, J; a} \,\mathcal{I}^{(2)}_{\Delta, J, \Delta_3, J_3, \Delta_4, J_4; b}~. }
Recalling that $\mathcal{I}^{(2)}_{\D_1,J_1,\D_2,J_2,\D,J;a}$ and $\mathcal{I}^{(2)}_{\D,J,\D_3,J_3,\D_4,J_4;a}$ are just $6j$ symbols, we have
\es{I26jbox}{\mathcal{I}^{(2)}_{\Delta_1, J_1, \Delta_2, J_2, \Delta, J; a} &= (t_0^{-1})^{ab} \mathcal{J}_{d}(\D_6,0;\D,J|\tilde\D_1,\tilde\D_5,\D_7,\tilde\D_2)_b~,\\
\mathcal{I}^{(2)}_{\D,J,\D_3,J_3,\D_4,J_4;a} &= (t_0^{-1})^{ab} \mathcal{J}_{d}(\D_8,0;\D,J|\D_5,\tilde\D_4,\tilde\D_3,\tilde\D_7)_b~.}
This is our final expression for the conformal block decomposition of the box pre-amplitude $A^\1_4$, with external operators $\O_{1,2,3,4}$ of arbitrary spins and arbitrary internal scalars $\O_{5,6,7,8}$, in the $s$-channel $12 \rar 34$. One may thus view the construction of the box diagram as two $6j$ symbols ``glued together'' by the conformal block. 

In general, diagrams involving glued three-point functions can be simplified into products of $6j$ symbols by repeatedly applying crossing transformations (\ref{eq:crossingkernel}). For example, in the case of a box diagram, we can write schematically
\be
\vcenter{\hbox{\begin{tikzpicture}
\draw (-1.5,1.5) -- (-1,1);
\draw (1.5,1.5) -- (1,1);
\draw (-1.5,-1.5) -- (-1,-1);
\draw (1.5,-1.5) -- (1,-1);
\draw (1,1) -- (1,-1) -- (-1,-1) -- (-1,1) -- (1,1);
\draw[fill=black] (1,1) circle (.3ex);
\draw[fill=black] (-1,1) circle (.3ex);
\draw[fill=black] (1,-1) circle (.3ex);
\draw[fill=black] (-1,-1) circle (.3ex);
\end{tikzpicture}}}
&=
\sum_J \int d\De \left\{\phantom{\frac 1 2}\cdots\phantom{\frac 1 2}\right\}
\vcenter{\hbox{\begin{tikzpicture}
\draw (-1.5,1.5) -- (-1,1);
\draw (2,0.5) -- (1.5,0);
\draw (-1.5,-1.5) -- (-1,-1);
\draw (2,-0.5) -- (1.5,0);
\draw (-1,1) -- (0,0) -- (-1,-1) -- (-1,1);
\draw (0,0) -- (1.5,0);
\draw[fill=black] (1.5,0) circle (.3ex);
\draw[fill=black] (-1,1) circle (.3ex);
\draw[fill=black] (0,0) circle (.3ex);
\draw[fill=black] (-1,-1) circle (.3ex);
\node[below] at (0.8,0) {$\De,J$};
\end{tikzpicture}}} \nn\\
&=
\sum_J \int d\De \left\{\phantom{\frac 1 2}\cdots\phantom{\frac 1 2}\right\}\left\{\phantom{\frac 1 2}\cdots\phantom{\frac 1 2}\right\}
\vcenter{\hbox{\begin{tikzpicture}
\draw (-1,0.5) -- (-0.5,0);
\draw (-1,-0.5) -- (-0.5,0);
\draw (1,0.5) -- (0.5,0);
\draw (1,-0.5) -- (0.5,0);
\draw (-0.5,0) -- (0.5,0);
\draw[fill=black] (0.5,0) circle (.3ex);
\draw[fill=black] (-0.5,0) circle (.3ex);
\node[below] at (0,0) {$\De,J$};
\end{tikzpicture}}}.
\label{eq:boxmanipulation}
\ee
In the notation above, the dots represent conformal three-point functions, and lines between them represent conformally-invariant integrals over common points $\int d^d x\, \cO(x) \tl \cO(x)$.
The $\{\cdots\}$ represent $6j$ symbols, and we suppress their arguments for brevity.
In the first line, we have interpreted the right-half of the box diagram as a partial wave and applied a crossing transformation, expressing it as an integral of a $6j$ symbol times a partial wave in the other channel. In the second line, we use the fact that a triangle diagram is itself a $6j$ symbol to obtain the integral of a partial wave times a product of two $6j$ symbols. This is what the formulas (\ref{a4pre}) and (\ref{rho4pre}) express in more detail for the case of the box pre-amplitude in AdS, which we previously showed to be a gluing of four three-point functions.\foot{In this case, the right half of the box in \eqr{eq:boxmanipulation} is literally the AdS geometrization of a partial wave (up to a constant prefactor).} Note that when some of the lines represent identical operators, the box diagram can be automatically crossing-symmetric. In this case, the expression on the last line of (\ref{eq:boxmanipulation}) is a crossing symmetric combination of partial waves. This solution to crossing symmetry was written down in \cite{Gadde:2017sjg}. One can apply similar manipulations to (\ref{eq:boxmanipulation}) to express $n$-gon diagrams in terms of integrals of multi-point partial waves times products of $6j$ symbols, and similarly for higher-loop amplitudes.

The full amplitude $\mathcal{A}^\1_4$, and hence its conformal block decomposition, follows from spectral integration of \eqr{a4pre}. One new feature of the conformal block decomposition of the one-loop diagrams, relative to tree-level, is the exchange of the double-trace operators $[\cO_5\cO_7]_{n,\ell}$. In the bulk, these come from the ``AdS unitarity cut'' that crosses the two internal lines \cite{Fitzpatrick:2011dm,Aharony:2016dwx}. Consequently, the full amplitude $\mathcal{A}^\1_4$ must have simple poles at twists
\e{57pole}{\tau=\D_5+\D_7+2n \qquad (n\in\mathbb{Z}_{\geq 0})~.}
(There are also other poles, already present at tree-level, due to $[\cO_1\cO_2]_{n,\ell}$ and $[\cO_3\cO_4]_{n,\ell}$ exchanges.)
Bearing in mind that $A^\1_4$ is a pre-amplitude, we can ask whether $\rho^\1_{4;a,b}(\D,J) $ has the right poles. The poles \eqr{57pole} should be visible already in the pre-amplitude, because they do not come from the region of integration where the bulk points collide (unlike the  $[\cO_1\cO_2]_{n,\ell}$ and $[\cO_3\cO_4]_{n,\ell}$ poles).\foot{A useful analogy at tree-level is the following. A four-point exchange Witten diagram where the internal line is a harmonic function, not a propagator, is dual to a conformal partial wave for single-trace exchange. In the direct-channel decomposition, there are obviously no double-traces. In the bulk, this happens because the diagram is a difference of two exchange diagrams for bulk-bulk propagators with different quantizations, and the contact region of integration, which gives rise to the double-trace exchanges in the dual CFT, cancels in the difference.} Examining  $\rho^\1_{4;a,b}(\D,J) $, we find these poles in $S_{\tilde\D_5}^{\D_7,[\D,J]}$:
\e{}{S_{\tilde\D_5}^{\D_7,[\D_5+\D_7+2n+J+\eps,J]}\approx {(-1)^{n+1}\o\eps}\frac{2 \pi ^{d/2} \, \Gamma \left(\frac{d}{2}-\D_5\right) \Gamma (\D_5+J+n)}{ n! \Gamma (\D_5) \Gamma \left({d\o 2}-\D_5-n\right) \Gamma \left(\frac{d}{2}+J+n\right)}~.}
Indeed, these are the only poles of $\rho^\1_{4;a,b}(\D,J) $ at $\D=\D_5+\D_7+2n+J$. This provides a check on our computation.\footnote{Note that the operators $[\cO_5\cO_7]_{n,\ell}$ are singlet bilinears from the SYK perspective of this diagram. }

\subsection{AdS open questions}\label{adsq}

In $d=2$, one can also define a $6j$ symbol for the Virasoro algebra. This has, in fact, been computed explicitly as a contour integral in \c{Ponsot:2000mt,Ponsot:2003ju}. In a semiclassical limit, the Virasoro $6j$ symbol may be computed as the volume of a tetrahedron in hyperbolic space \c{MY}. One might ask whether there is an analogous geometric picture in Euclidean AdS for the higher-dimensional $6j$ symbols for the conformal group, perhaps involving the gluing of two geodesic Witten diagrams \c{Hijano:2015zsa}. 

In our tree-level discussion, recall the absence of the single-trace OPE data in the Lorentzian inversion of exchange Witten diagrams $W_{\D',J'}$: since the inversion formula is only valid down to spins greater than the Regge spin of the object being inverted, it is not possible to recover the $\O_{\D',J'}$ exchange in CFT from Lorentzian inversion of $W_{\D',J'}$, despite the obvious fact that $W_{\D',J'}$ includes the $\O_{\D',J'}$ conformal block. Nor does the $6j$ symbol, whose only poles are at double-trace twists, have a pole at twist $\D'-J'$. This reflects the fact that light single-trace operators in holographic CFTs seem not to belong to families analytic in spin: in CFTs with parametrically large higher spin gap, $\D_{\rm gap}\gg1$, all higher-spin members parametrically decouple.\foot{A stark version of this issue concerns a theory of ``pure'' Einstein gravity, i.e.\ general relativity with no other parametrically light elementary fields, dual to a large $N$ CFT whose only light operators are the unit operator and the stress tensor. In this case, the question is how to recover the stress tensor contribution from the Lorentzian inversion formula.} This suggests the possibility of relating the spectra of low-energy elementary fields and high-energy states in consistent theories of AdS quantum gravity, by combining the demand for single-trace poles from Lorentzian inversion with the requirement of consistent Regge behavior.

We have computed the four-point box diagram in AdS for internal scalars and external spinning operators. It would be worthwhile to do a thorough analysis of our result, e.g.\ extracting the dual CFT OPE data for all double-trace operators. It also would be valuable to derive our result for the box diagram in different ways: for instance, using a crossing symmetry-based derivation a l\'a \c{Aharony:2016dwx} or the Lorentzian inversion formula, or in Mellin space. Further extension of the crossing technique in \c{Alday:2017xua,Alday:2017vkk,Aharony:2018npf} has essentially developed the necessary ingredients. On the other hand, our derivation provides an interpretation of the one-loop amplitudes as a gluing of $6j$ symbols. It would be interesting to develop a set of ``Feynman rules'' for this gluing with which one generates higher-loop amplitudes. 

A simple case of our result is the pre-amplitude of the box diagram in $\phi^3$ theory, or other pure-scalar theories in AdS. Together with previous results for the four-point bubble and triangle diagrams \c{Penedones:2010ue,Fitzpatrick:2011dm,Aharony:2016dwx}, we now have a catalog of all one-loop, four-point amplitudes of scalar theories in AdS. In flat space, four-point boxes, three-point triangles, two-point bubbles and tadpoles form a basis for all one-loop amplitudes \c{Passarino:1978jh}; moreover, the one-loop scalar amplitudes appear after imposing unitarity cuts of higher-loop amplitudes. Are there sharp analogous statements in AdS?

\section*{Acknowledgments}
We thank Bartek Czech, Abhijit Gadde, David Gross, Denis Karateev, Petr Kravchuk, Murat Kologlu, Shota Komatsu, Grisha Korchemsky, Douglas Stanford, Herman Verlinde, and Ellis Ye Yuan for discussions. VR thanks H.~Verlinde for valuable early discussions on the possible connection between SYK and the 6j symbol. We also thank the December 2017 Southern California Strings Seminar at UCLA, where this work was initiated, and the Bootstrap 2018 workshop at Caltech, where this work was nearly completed, for hospitality. EP thanks the Yukawa Institute for Theoretical Physics and National Taiwan University for hospitality. VR thanks the Nordita program ``Correlation functions in solvable models'' for hospitality. JL is supported in part by the Institute for Quantum Information and Matter (IQIM), an NSF Physics Frontiers Center (NSF Grant PHY-1125565) with support from the Gordon and Betty Moore Foundation (GBMF-2644), and by the Walter Burke Institute for Theoretical Physics. DSD and EP are supported by Simons Foundation grant 488657, and by the Walter Burke Institute for Theoretical Physics. VR is supported  by NSF grant PHY-1125915 and PHY-1606531.  
\appendix   

\section{Further properties of the four-point function}\label{appa}
In this Appendix we discuss some properties of the four-point function of fundamentals in the $d$-dimensional SYK model. 

\sssec*{\it Non-singlet operator dimensions from large spin asymptotics}
As is by now well-known, the large spin spectrum of any CFT contains double-twist operators of asymptotically large spin \cite{Fitzpatrick:2012yx, Komargodski:2012ek}. For any two conformal primaries $\phi_1$ and $\phi_2$, there exist ``double-twist'' primaries $[\phi_1\phi_2]_{n,J}$, of schematic form
\e{}{[\phi_1\phi_2]_{n,J} = \phi_1\p^{2n}\p_{\mu_1}\ldots\p_{\mu_J}\phi_2~-~\text{(traces)}~,}
with twists $\tau=\D-J$ asymptotically approaching $\D_{1}+\D_{2}+2n$ at large spin $J\gg1$, where $n\in \mathbb{Z}_{\geq 0}$. At large $N$, these are double-trace operators, that exist for all $J$.

In Sec.~\ref{LadderSum} we computed the four-point function $\langle \phi_i \phi_i \phi_j \phi_j\rangle$,  expanding it in terms of conformal blocks of the singlet operators $[\phi_i\phi_i]_{n,J}$, whose dimensions are given by solutions to $k(\Delta, J) = 1$, where $k(\Delta, J)$ are the eigenvalues of the kernel, (\ref{kernelE}). 
Here we will look at the kernel  at large spin, and  extract the anomalous dimensions $\g_{n,J}$ of $[\phi_i\phi_i]_{n,J}$ in a $1/J$ expansion. Applying the lightcone bootstrap to this correlator, computed above in the singlet channel $\phi_i\phi_i \rightarrow \phi_j\phi_j$, allows us to read off the dimensions of low-twist non-singlet operators in the crossed channel $\phi_i\phi_j \rightarrow \phi_i\phi_j$. 

We evaluate the kernel $k(\D,J)$ at $\D={2d\o q}+J+\gamma_{0,0}$, expand in $J\gg1$ and $\g_{0,0}\ll1$, and set $k(\D,J)=1$. This yields the anomalous dimension $\g_{0,0}$ for the leading scalar double-twist operator, $[\phi_i\phi_i]_{0,0}$:
\e{a2}{\g_{0,0} \approx \frac{2 (q-1) \Gamma \left(\frac{d (q-2)}{2 q}\right) \Gamma \left(\frac{d (q-1)}{q}\right)}{\Gamma \left(\frac{d (2-q)}{2 q}\right) \Gamma \left(\frac{d}{q}\right) \Gamma \left(\frac{d (q-2)}{q}\right)}\,J^{-{d(q-2)\o q}}~.}
Note that this is negative for all unitary $q$, and becomes zero precisely at the unitarity bound $q\leq \frac{2 d}{d-2}$. We now match this to the lightcone bootstrap formula \cite{Fitzpatrick:2012yx, Komargodski:2012ek},
\e{}{\g_{0,0} \approx -c_{\tau_*} J^{-\tau_*}~,}
where $\tau_*$ is the twist of $\O_{ij}$, the lowest-twist operator in the $\phi_i\times \phi_j$ OPE. The coefficient is 
\e{}{c_{\tau_*} = {2^{1-J_*}\G\left( \tau_*+2J_*\right)\G(\D_\f)^2\o \G\left({\tau_*\o2}+J_*\right)^2\G\left(\D_\f-{\tau_*\o2}\right)^2}~{f_{\phi_i\phi_j\O_{ij}}^2\o C^2_{\phi\phi}C_{\O\O}}~,}
where $f_{\phi_i\phi_j\O_{ij}}$ is the OPE coefficient, and with norms defined as $\la \mathcal{O}(x)\mathcal{O}(0)\ra = C_{\mathcal{O}\mathcal{O}}/|x|^{2\D_i}$. By matching to \eqr{a2}, we read off the dimension of the tensor operator $\O_{ij}$,
\e{dimO}{\D(\O_{ij}) = {d\o q}(q-2)~.}
This simple result --- with vanishing order one anomalous dimension --- may be motivated heuristically\foot{We thank Douglas Stanford for this observation.} by cutting the ladder diagram (Fig. \ref{ladder}) ``along the middle'', representing the $\phi_i\phi_j \rar \phi_i\phi_j$ channel. The cut crosses $(q-2)$ lines, each of which represents a $\phi$ field with $\D=d/q$.  The fact that $\Delta(\O_{ij})$ is given by the free value implies that in the large $N$ melonic limit, the two-point function of this class of operators receives no renormalization due to ladder diagrams.
 Similar behavior was observed for bilinears of the form $\psi^{abc}\partial_{\tau}^{n} \psi^{ade}$  in fermionic $q=4$ tensor models in $d=1$\cite{Bulycheva:2017ilt}. 
Highly systematic approaches to large spin perturbation theory have been developed in many works, e.g.\ \c{Alday:2016njk,Simmons-Duffin:2016wlq,Caron-Huot:2017vep}, which could be applied to the present case.

\sssec*{\it Central charge}
The contribution of the stress tensor to the conformal block expansion of a four point function in the limit $\chi\ll \bar \chi \ll 1$ is \cite{Osborn:1993cr}
\begin{align}
\left\langle {\phi ({x_1})\phi ({x_2})\phi ({x_3})\phi ({x_4})} \right\rangle  \supset \frac{{d \Delta _\phi ^2}}{{d - 1}}\frac{{{C_{{\text{free}}}}}}{{{C_T}}}\frac{1}{4}{(\chi\bar \chi)^{(d - 2)/2}}{{\bar \chi}^2}~,
\end{align}
where $C_\mathrm{free}$ is the central charge for the free boson in $d$ dimensions.
In $d$-dimensional SYK, the coefficient of these powers in $\chi,\bar\chi$ is
\begin{align}
- \mathop{\mathrm{Res}}_{\De=d} {\frac{{{\rho ^{{\rm{MFT}}}}(\Delta ,2)K_{\tilde \Delta ,2}^{{\Delta _\phi },{\Delta _\phi }}}}{{1 - {k}(\Delta,2)}}}~.
\end{align}
Thus we have 
\begin{align}
C_T(d,q) &\equiv \frac{{{C_T}}}{{C_{{\text{free}}}}} =  - \frac{1}{4}\frac{{d \Delta _\phi ^2}}{{d - 1}}\left(\mathop{\mathrm{Res}}_{\De=d} {\frac{{{\rho ^{{\rm{MFT}}}}(\Delta ,2)K_{\tilde \Delta ,2}^{{\Delta _\phi },{\Delta _\phi }}}}{{1 - {k}(\Delta,2)}}}\right)^{-1}~ \nn\\
  &= -\frac{\Gamma ( 1+d-\tfrac{d}{q})\Gamma ( \tfrac{d+q}{q} )\sin ( \tfrac{d\pi (q-2)}{2q} )}{2\pi q\Gamma (\tfrac{d}{2})\Gamma (\tfrac{4+d}{2})} \nn\\
  &\quad \x \left(d(q-2)\left(\pi\cot ( \tfrac{d\pi (q-2)}{2q} )+H( \tfrac{d(q-1)}{q} )-H( \tfrac{d}{q} )\right) -2q\right),
\end{align}
where 
\e{}{H(x) = \psi(x+1)+\gamma}
is the analytic continuation of the harmonic number to non-integer argument and $\gamma$ is the Euler-Mascheroni constant. Setting $d=2$, we recover the result from \cite{Murugan:2017eto}~,
\begin{align}
{{C}_{T}}(d=2,q)=\frac{{{(q-2)}^{3}}}{{{q}^{3}}}.
\end{align}

For $q\leq \frac{2 d}{d-2}$ such that an infrared fixed point exists, the central charge is always positive, and obeys $dC_T(d,q)/dq>0$, i.e.\ a ``$C_T$-theorem'' in the space of SYK-like models due to the possibility of flowing from larger to smaller $q$. While the central charge shows no pathology,  the bosonic $d$-dimensional SYK models are certainly unphysical, due to the Lagrangian being unbounded from below, as noted previously. This manifests itself in, for instance,  the presence of bilinear operators of complex dimension \cite{Giombi:2017dtl}.

\section{One dimensional $6j$ symbol} \label{app1d6j}
In this appendix we compute the $6j$ symbols in one dimension. An operator in one dimension has no spin, and its dimension is labeled by $h$. In this notation, the $6j$ symbols (\ref{6j}) are 
\be  \label{6j1d}
\mathcal{J}_1=\int {\frac{{d{x_1} \cdots d{x_4}}}{{{\rm{vol}}({\rm{SO}}(2,1))}}} \Psi _{\tilde h}^{{{\tilde h}_1},{{\tilde h}_2},{{\tilde h}_3},{{\tilde h}_4}}({x_1},{x_2},{x_3},{x_4})\,\Psi _{h'}^{{h_3},{h_2},{h_1},{h_4}}({x_3},{x_2},{x_1},{x_4})~,
\ee
where, as before, $\tilde{h} \equiv 1- h$ is the dimension of the shadow operator. 
The partial wave in one dimension is a sum of a conformal block -- a hypergeometric function of one cross ratio -- and the shadow block. The above integral is thus similar to an integral we encountered earlier, see Eq. (\ref{eq:Omega}); however, the integration range above is over the entire $x_i$ axis. As a result, to evaluate the integral in this manner, we would need to split it into different integration regions; a straightforward but tedious exercise. 
A faster route  is to use the form  of the $6j$ symbol as an integral of four three-point functions (\ref{6jDef}), which we write out as, 
\be
 \mJ_1 =\int {\frac{{d{x _1}d{x _2}d{x _a}}}{{{\rm{vol}}({\rm{SO}}(2,1))}}} \langle {{\cal O}_{{{\tilde h}_1}}}({x _1}){{\cal O}_{{{\tilde h}_2}}}({x _2}){{\cal O}_{\tilde h}}({x _a})\rangle I_{{h_i}}^{h,h'}({x _a},{x _1},{x _2})~,
 \ee
 where we defined,
\begin{align}
&I_{{h_i}}^{h,h'}({x _a},{x _1},{x _2}) = \int d {x _3}d{x _4}d{x _b}\langle {{\cal O}_h}({x _a}){{\cal O}_{1 - {h_3}}}({x _3}){{\cal O}_{1 - {h_4}}}({x _4})\rangle \nonumber\\
&\times\langle {{\cal O}_{{h_3}}}({x _3}){{\cal O}_{{h_2}}}({x _2}){{\cal O}_{h'}}({x _b})\rangle \langle {{\cal O}_{\tilde h'}}({x _b}){{\cal O}_{{h_1}}}({x _1}){{\cal O}_{{h_4}}}({x _4})\rangle ~.
\end{align}
 By conformal invariance, $ I_{h_i}^{h, h'}$ will take the form of a three-point function, 
 \be
 I_{h_i}^{h, h'}(x_a, x_1, x_2)
 = \mI_{h_i}^{h, h'} \langle \mO_{h}(x_a)  \mO_{h_1}(x_1) \mO_{h_2}(x_2)\rangle~,
 \ee
and the $6j$ symbol becomes, $\mJ_1 = t_0\,  \mI_{h_i}^{h, h'}$. 
Let us now evaluate $I_{h_i}^{h, h'}$. Written out, we have, 
\be
I_{{h_i}}^{h,h'} = \int d {x _3}d{x _4}d{x _b}\frac{{|{x _{34}}{|^{h + {h_3} + {h_4} - 2}}|{x _{3b}}{|^{{h_{23}} - h'}}|{x _{4b}}{|^{{h_{14}} - \tilde h'}}}}{{|{x _{a3}}{|^{h - {h_{34}}}}|{x _{a4}}{|^{h + {h_{34}}}}|{x _{23}}{|^{{h_2} + {h_3} - h'}}|{x _{2b}}{|^{h' + {h_{23}}}}|{x _{1b}}{|^{\tilde h' + {h_{14}}}}|{x _{14}}{|^{{h_1} + {h_4} - \tilde h'}}}}~,
 \ee
 where we use notation $x_{i j} \equiv x_i - x_j$ and $h_{i j} \equiv h_i - h_j$. 
We do a change of variables, 
 \be
 A= \frac{x_{b1} x_{a2}}{x_{b2} x_{a1}}~, \ \ \ B= \frac{x_{1a} x_{b4}}{x_{1b}x_{a4}}~, \ \ \ \ C = \frac{x_{2 b}x_{a 3}}{x_{2 a} x_{b3}}~,
 \ee
 which gives, 
 \begin{align}
  &\mI_{h_i}^{h, h'} =\int d AdBdC\times\nonumber\\
  &\frac{{|1 - ABC{|^{h + {h_3} + {h_4} - 2}}}}{{|A{|^{ - h' + {h_1} + {h_4}}}|1 - A{|^{1 + h - {h_1} - {h_2}}}|B{|^{1 - h' - {h_1} + {h_4}}}|1 - B{|^{ - 1 + h' + {h_1} + {h_4}}}|C{|^{h - {h_3} + {h_4}}}|1 - C{|^{{h_2} + {h_3} - h'}}}}~.
  \end{align}
  We evaluate this integral using the method in \cite{GR4}, where a similar integral appeared in the evaluation of the SYK three-point function of bilinears. 
  The result is, 
\be
\mJ_1/t_0 = \mI_{h_i}^{h,h'} =\(\gamma\, \mF_{h, h'}^{h_1, h_2, h_3, h_4} +\eta \, \mF_{h, h'}^{h_1, h_2, h_3, \tilde{h}_4}\)+ (1\leftrightarrow 3, 2\leftrightarrow 4, h\leftrightarrow \tl h, h'\leftrightarrow \tl h')~.
\ee  
where, 
\be
\mF_{h, h'}^{h_1, h_2, h_3, h_4} =
{}_4F_3\left( {\begin{array}{*{20}{c}}
{1\!+\!h'\!-\!h_1\! -\! h_4,\,  h'\!+\!h_1\! -\! h_4,\, 2\!-\!h_3\! -\!h_4 \!-\! h,\, 1\!+ \!h_3\! -\! h_4\! -\! h}\\
{2\!-\!2h_4,\, 2\!+\!h'\!-\! h_2\!-\!h_4\! -\! h,\, 1\!+\! h'\! +\! h_2\! -\! h_4\! -\! h}
\end{array};1} \right)~.
\ee
and
\be
\gamma = {\cal B}_{ - h' + {h_1} + {h_4}}^{1 + h - {h_1} - {h_2}}\,{\cal B}_{1 - h' - {h_1} + {h_4}}^{ - 1 + h' + {h_1} + {h_4}}\,{\cal B}_{h - {h_3} + {h_4}}^{{h_2} + {h_3} - h'}~,\ \ \ \ \eta={\cal B}_{1 - h' + {h_1} - {h_4}}^{1 + h - {h_1} - {h_2}}\,{\cal B}_{2{h_4}}^{2 - h - {h_3} - {h_4}}\,{\cal B}_{1 - {h_3} - {h_4} + h}^{{h_2} + {h_3} - h'}~,
\ee
where we defined, 
\be
\mathcal{B}_{a_1}^{ a_2} = \pi^{\frac{1}{2}} \frac{\Gamma(\frac{1- a_1}{2})}{\Gamma(\frac{a_1}{2})} \frac{\Gamma(\frac{1- a_2}{2})}{\Gamma(\frac{a_2}{2})} \frac{\Gamma(\frac{a_1 + a_2 - 1}{2})}{\Gamma(\frac{2 - a_1- a_2}{2})}~.% = S_{\frac{a_1 + a_2}{2}}^{\frac{a_1}{2}, \frac{a_2}{2}}
\ee

\section{Contact diagram in higher dimensions} \label{contact}

\begin{figure}[t]
\centering
\includegraphics[width=1.8in]{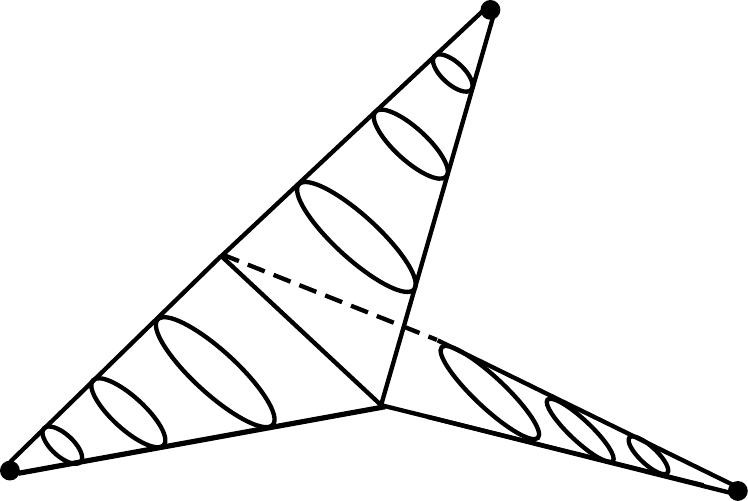}
\caption{The contribution of the ``contact'' diagrams to the SYK bilinear three-point function. Even though this diagram is nonplanar, it appears at the same order in $1/N$ as the planar diagram, shown previously in Fig.~\ref{tetraCut2} (a).} \label{FigContact}
\end{figure} 
In the main body of the paper we focused on the contribution of the planar Feynman diagrams to the SYK three-point function of bilinears. There is, however, an additional contribution to the three-point function,  shown in Fig.~\ref{FigContact}, which was named the contact diagram in \cite{GR2}. In one dimension, the result for the contact diagram was found to be a generalized hypergeometric function, ${}_3 F_2$, which simplified to a ratio of gamma functions \cite{GR4}. 

In this appendix we evaluate the contact diagram in two and four dimensions. Our strategy will be similar to the one  used in the evaluation of the $6j$ symbols: we apply the Lorentzian inversion formula, factorizing the integral into a product of two one-dimensional integrals, and find a result that is expressed as a product of two ${}_3 F_2$ functions. 

The contribution of the contact diagrams is, 
\begin{multline}
 \langle \mO_{\Delta_1, J_1}(x_1)\mO_{\Delta_2, J_2}(x_2) \mO_{\Delta_3, J_3}(x_3)\rangle_1 \\
 = (q-1)(q-2) \cJ^2 \int d^d x_a d^d x_b\, G(\tau_{ab})^{q-3}\, \prod_{i=1}^3  \langle \mO_{\Delta_i, J_i}(x_i) \phi(x_a) \phi(x_b)\rangle_{\mathrm{phys} } ~.
\end{multline}
Separating the coefficients from the functional form of the arguments, this becomes
\be
\langle \mO_{\Delta_1, J_1}(x_1)\mO_{\Delta_2, J_2}(x_2) \mO_{\Delta_3, J_3}(x_3)\rangle_1  = c_{\Delta_1, J_1}\, c_{\Delta_2, J_2}\,  c_{\Delta_3, J_3}\, (q-1)(q-2) {\cJ^2} b^q  \,  I_{\Delta_i, J_i}^{(1)} (x_i)~,
\ee
where
\be
I_{\Delta_i, J_i}^{(1)} (x_i) =\int d^d x_a d^d x_b\,\frac{1}{|x_{ab}|^{2( d- 3 \Delta_{\phi})}} \prod_{i=1}^3  \langle \mO_{\Delta_i, J_i}(x_i) \phi(x_a) \phi(x_b)\rangle~.
\ee
By conformal invariance, the result after integration will take the form of a conformal three-point function. As in the discussion of the planar diagrams in Sec.~\ref{6pt}, we extract the coefficient of the integral by contracting with a three-point function of shadow operators, see Eqs.~(\ref{246}) and (\ref{247}), to get
\begin{multline}
\mathcal{I}_{\Delta_i, J_i,a}^{(1)} = (t_0^{-1})_{ab}  \int \frac{d^d x_1 d^d x_2 d^d x_3\, d^d x_a d^d x_b }{\text{vol}(\text{SO}(d+1,1))}\frac{1}{|x_{ab}|^{2( d- 3 \Delta_{\phi})}} 
\prod_{i=1}^3  \langle \mO_{\Delta_i, J_i}(x_i) \phi(x_a) \phi(x_b)\rangle \\
\langle \tl \mO_{\Delta_1, J_1}(x_1)  \tl \mO_{\Delta_2, J_2}(x_2) \tl \mO_{\Delta_3, J_3}(x_3)\rangle^b~.
\end{multline}
We now recognize that, for instance, the integral over (say) $x_1$ yields a conformal partial wave. 
\be
\mathcal{I}_{\Delta_i, J_i}^{(1)} =\frac{1}{t_0} \int \frac{d^d x_2 d^d x_3\, d^d x_a d^d x_b }{\text{vol}(\text{SO}(d+1,1))}
\,  \Psi_{\tilde \Delta_1, J_1}^{\Delta_{\phi}, \Delta_{\phi},\tilde \Delta_2,\tilde \Delta_3}(x_a, x_b, x_2, x_3) \frac{\prod_{i=2}^3  \langle \mO_{\Delta_i, 0}(x_i) \phi(x_a) \phi(x_b)\rangle}{|x_{ab}|^{2( d- 3 \Delta_{\phi})}} 
\ee
where, for simplicity, we set the spins $J_2 =J_3=0$. This integral is now  of the type we are familiar with: an inner product between a four-point function and a partial wave.\footnote{The integrand is actually independent of $\Delta_{\phi}$, as is evident by writing it out explicitly.}

It is convenient at this stage to change notation. We will evaluate
\begin{align}
{{\cal U}_d} = 
 \int {\frac{{{d^d}{x_1}{d^d}{x_2}{d^d}{x_3}{d^d}{x_4}}}{{{\rm{vol}}({\rm{SO}}(d + 1,1))}}} \left\langle {{\mathcal{O}_1}{\mathcal{O}_2}{\mathcal{O}_3}{\mathcal{O}_4}} \right\rangle \Psi _{\tilde \Delta ,J}^{{{\tilde \Delta }_i}}({x_i})~,
\end{align}
where 
\begin{align}
\left\langle {{{\cal O}_1}{{\cal O}_2}{{\cal O}_3}{{\cal O}_4}} \right\rangle  = \frac{1}{{{{\left| {{x_{12}}} \right|}^{2{\Delta _\kappa } - {\Delta _3} - {\Delta _4}}}}}\frac{1}{{{{\left| {{x_{23}}} \right|}^{{\Delta _3}}}}}\frac{1}{{{{\left| {{x_{13}}} \right|}^{{\Delta _3}}}}}\frac{1}{{{{\left| {{x_{24}}} \right|}^{{\Delta _4}}}}}\frac{1}{{{{\left| {{x_{14}}} \right|}^{{\Delta _4}}}}}~.
\end{align}
and $\Delta_1=\Delta_2=\Delta_\kappa$. We get back $t_0\, \mathcal{I}_{\Delta_i, J_i}^{(1)} $ by sending  $\Delta_{\kappa} \rightarrow \tilde \Delta_{\phi}$, $\Delta_3 \rightarrow \Delta_2$, $\Delta_4\rightarrow \Delta_3$, and $\Delta, J \rightarrow \Delta_1, J_1$.

\subsection{Two dimensions}
We begin in two dimensions. In the notation of $h_i, \bar{h}_i$, the integral ${\cal U}_2$ is
\be
{{\cal U}_2} = \int {\frac{{{d^2}{z_1} d^2z_2d^2z_3 {d^2}{z_4}}}{{{\rm{vol}}({\rm{SO}}(3,1))}}} \left\langle {{\mathcal{O}_1}({z_1})\O_2(z_2)\O_3(z_3)  {\mathcal{O}_4}({z_4})} \right\rangle \Psi _{\tilde h,\tilde {\bar{h}}}^{{{\tilde h}_i},\tilde{\bar{h}}_i}({z_i})~,
\ee
where, 
\be
\left\langle {{{\cal O}_1}{{\cal O}_2}{{\cal O}_3}{{\cal O}_4}} \right\rangle  = \frac{1}{{{{\left[ {{z_{12}}} \right]}^{2{h_\kappa } - {h_3} - {h_4}}}}}\frac{1}{{{{\left[ {{z_{32}}} \right]}^{{h_3}}}}}\frac{1}{{{{\left[ {{z_{14}}} \right]}^{{h_4}}}}}\frac{1}{{{{\left[ {{z_{13}}} \right]}^{{h_3}}}}}\frac{1}{{{{\left[ {{z_{42}}} \right]}^{{h_4}}}}}~,
\ee
and  $h_1=h_2=h_{\kappa}$ and $\bar{h}_1=\bar{h}_2=\bar{h}_{\kappa}$. 
We will evaluate this integral by applying  the Lorentzian inversion formula, (\ref{eq:caronhuottwod}). The four-point function with external factors stripped off, as a function of the cross-ratio, is,
\begin{align}
&g(\chi ,\bar \chi )= \frac{{\left\langle {{{\cal O}_1}{{\cal O}_2}{{\cal O}_3}{{\cal O}_4}} \right\rangle }}{{\left| {{T_s}} \right|}} = {\left| \chi  \right|^{{h_3} + {h_4}}}{\left| {\bar \chi } \right|^{{{\bar h}_3} + {{\bar h}_4}}}{\left| {1 - \chi } \right|^{ - {h_3}}}{\left| {1 - \bar \chi } \right|^{ - {{\bar h}_3}}}~,
\end{align}
From this we  compute the double commutator, to get, 
\be
\frac{{\left\langle {\left[ {{{\cal O}_3},{{\cal O}_2}} \right]\left[ {{{\cal O}_1},{{\cal O}_4}} \right]} \right\rangle }}{{\left| {{T_s}} \right|}}= \frac{{\left\langle {\left[ {{{\cal O}_4},{{\cal O}_2}} \right]\left[ {{{\cal O}_1},{{\cal O}_3}} \right]} \right\rangle }}{{\left| {{T_s}} \right|}} = \frac{{ - 4{\pi ^2}}}{{\Gamma ({{\bar h}_3})\Gamma (1 - {{\bar h}_3})\Gamma ({{\bar h}_4})\Gamma (1 - {{\bar h}_4})}}g(\chi ,\bar \chi )~.
\ee

The two regions of integration, $R_1$ and $R_2$, will contribute equally when $j_H$ is even and will cancel when $j_H$ is odd, where $j_H$ was defined previously, below (\ref{eq:jH}). The result is,
\begin{align}
  & {\cal{U}}_{2}=(1+{{(-1)}^{{{j}_{H}}}})\frac{{{\pi }^{2}}{{\Gamma }^{2}}(h)\Gamma ({{{\bar{h}}}_{43}}+1-\bar{h})\Gamma ({{{\bar{h}}}_{34}}+1-\bar{h})}{4\Gamma (2h)\Gamma (2-2\bar{h})\Gamma ({{{\bar{h}}}_{3}})\Gamma (1-{{{\bar{h}}}_{3}})\Gamma ({{{\bar{h}}}_{4}})\Gamma (1-{{{\bar{h}}}_{4}})}  \nonumber\\
 & \int_{0}^{1}{d\chi }\, {{\chi }^{{{h}_{3}}+{{h}_{4}}+h-2}}{{(1-\chi )}^{-{{h}_{3}}}}{{\times }_{2}}{{F}_{1}}(h+{{h}_{43}},h,2h;\chi )  \nonumber\\
 & \int_{0}^{1}{d\bar{\chi }}\, {{{\bar{\chi }}}^{{{{\bar{h}}}_{3}}+{{{\bar{h}}}_{4}}-\bar{h}-1}}{{(1-\bar{\chi })}^{-{{{\bar{h}}}_{3}}}}{{\times }_{2}}{{F}_{1}}(1-\bar{h},1-\bar{h}+{{{\bar{h}}}_{43}},2-2\bar{h};\bar{\chi })~.
\end{align}
The integrals  that appear are just  the definition of ${}_3 F_2$, so we arrive at
\begin{align}
&{\mathcal{U}_2} = \frac{\pi ^2}{4}({1+( - 1)^{{j_H}}} )  \frac{{{\Gamma ^2}(h)\Gamma ({h_3} + {h_4} + h - 1)\Gamma (1 - {h_3})}}{{\Gamma (2h)\Gamma ({h_4} + h)}}\nonumber\\
&\times{}_3{F_2}(h + {h_{43}},h,{h_3} + {h_4} + h - 1;2h,{h_4} + h;1)\nonumber\\
& \times \frac{{\Gamma ({{\bar h}_{43}} + 1 - \bar h)\Gamma ({{\bar h}_{34}} + 1 - \bar h)\Gamma ({{\bar h}_3} + {{\bar h}_4} - \bar h)}}{{\Gamma (2 - 2\bar h)\Gamma ({{\bar h}_3})\Gamma ({{\bar h}_4})\Gamma (1 - {{\bar h}_4})\Gamma ({{\bar h}_4} - \bar h + 1)}}\nonumber\\
&\times{}_3{F_2}(1 - \bar h,1 - \bar h + {{\bar h}_{43}},{{\bar h}_3} + {{\bar h}_4} - \bar h;2 - 2\bar h,{{\bar h}_4} - \bar h + 1;1)~.
\end{align}

\subsection{Four dimensions}
The  computation in four dimensions  is  similar to the one in two dimensions. The result is, 
\begin{align}
{{\cal U}_4} = \frac{A_{\Delta ,J}^{{\Delta _i}}}{2^4}(\Lambda _{\Delta  + J}^{(1),{\Delta _i}}\Lambda _{\tilde \Delta  + J}^{(2),{\Delta _i}} - \Lambda _{\tilde \Delta  + J}^{(1),{\Delta _i}}\Lambda _{\Delta  + J}^{(2),{\Delta _i}})~,
\end{align}
where
\begin{align} 
&A_{\Delta ,J}^{{\Delta _i}} = \frac{{ - 8{\pi ^2}{\alpha _{\Delta ,J}}\left( {{{( - 1)}^J} + 1} \right)}}{{\Gamma ({\Delta _3}/2)\Gamma (1 - {\Delta _3}/2)\Gamma ({\Delta _4}/2)\Gamma (1 - {\Delta _4}/2)}}~,\nonumber\\
  & \Lambda _\Delta ^{(1),{\Delta _i}} = \frac{{\Gamma (\frac{{{\Delta _3} + {\Delta _4} + \Delta }}{2} - 1)\Gamma (1 - \frac{{{\Delta _3}}}{2})}}{{\Gamma (\frac{{{\Delta _4} + \Delta }}{2})}}{ \times _3}{F_2}(\frac{\Delta }{2},\frac{{\Delta  - {\Delta _{34}}}}{2},\frac{{{\Delta _3} + {\Delta _4} + \Delta }}{2} - 1;\Delta ,\frac{{{\Delta _4} + \Delta }}{2};1)~,\nonumber\\
 & \Lambda _\Delta ^{(2),{\Delta _i}} = \frac{{\Gamma (\frac{{{\Delta _3} + {\Delta _4} + \Delta }}{2} - 2)\Gamma (1 - \frac{{{\Delta _3}}}{2})}}{{\Gamma (\frac{{{\Delta _4} + \Delta }}{2} - 1)}}{ \times _3}{F_2}(\frac{\Delta }{2},\frac{{\Delta  - {\Delta _{34}}}}{2},\frac{{{\Delta _3} + {\Delta _4} + \Delta }}{2} - 2;\Delta ,\frac{{{\Delta _4} + \Delta }}{2} - 1;1)~. \nonumber
\end{align}
where $\alpha_{\D,J}$ was defined earlier, in \eqr{adj}. 

\section{Shadow transforms of three-point functions}\label{shadow}

In this section, we give explicit computations of some shadow factors obtained by shadow transforming scalar-scalar-spin-$J$ three point functions.
These are generalizations of the famous star-triangle relation \cite{DEramo:1971hnd}. Shadow factors can also be computed efficiently using weight-shifting operators \cite{Karateev:2017jgd,MFTfuture}. Here we give elementary computations using embedding-space integrals. We follow the notation of \cite{Costa:2011mg,SimmonsDuffin:2012uy}.

\subsection{Shadow transforming the scalar}
The shadow transform of an operator was defined earlier in Eq.~\ref{ShadowT}. 
Let's compute the shadow transform of the three-point function $\<\f_1(x_1)\f_2(x_2) \cO^{\mu_1\cdots\mu_J}(x_3)\>$ with respect to $x_1$. In the embedding space, the three-point structure is given by
\begin{align}
\<\f_1(X_1)\f_2(X_2) \cO(X_3,Z_3)\> &= \frac{V_{3,12}^J}{X_{12}^{\frac{\De_1+\De_2-\De_3}{2}} X_{23}^{\frac{\De_2+\De_3-\De_1}{2}} X_{13}^{\frac{\De_1+\De_3-\De_2}{2}}}~,
\end{align}
where
\be
X_{ij} &=-2X_i\.X_j, \nn\\
V_{3,12} &= -2\frac{(Z_3\.X_1)(X_2\.X_3) - (Z_3\.X_2)(X_1\.X_3)}{X_{12}^{1/2} X_{23}^{1/2} X_{13}^{1/2}} = \frac{X_{14}}{X_{12}^{1/2} X_{23}^{1/2} X_{13}^{1/2}}~, \nn\\
X_4 &\equiv Z_3(X_2\.X_3) - X_3(Z_3\.X_2)~.
\ee
Note that 
\be
\label{eq:extraorthogonality}
X_4^2 = 0~,\quad X_4\.X_2 = 0~,\quad X_4\.X_3 = 0~.
\ee
Thus, the shadow integral we'd like to perform is
\be
\<\mathbf{S}[\f_1](X_0) \f_2(X_2) \cO(X_3,Z_3)\> &=
\frac{1}{X_{23}^{\frac{\De_2+\De_3-\De_1+J}{2}}}\int D^d X_1 \frac{1}{X_{01}^{d-\De_1}}\frac{X_{14}^J}{X_{12}^{\frac{\De_1+\De_2-\De_3+J}{2}} X_{13}^{\frac{\De_1+\De_3-\De_2+J}{2}}}~.
\label{eq:ourshadowintegral}
\ee
Let us temporarily think of $J$ as continuous and treat this as a conformal four-point integral. It will be simpler than the usual four-point integral because of the orthogonality conditions (\ref{eq:extraorthogonality}).

We will need the following result for a simple conformal integral \cite{SimmonsDuffin:2012uy}
\be
\label{eq:basicconformalintegral}
\int D^d X \frac{1}{(-2 X\.Y)^d} &= \frac{\pi^{\frac d 2}\G(\frac d 2)}{\G(d)} \frac{1}{(-Y^2)^{d/2}}~,
\ee
and also the following formula for combining factors using Feynman/Schwinger parameters
\be
\label{eq:feynmanschwinger}
\frac{1}{\prod_i A_i^{a_i}} &= \frac{\G(\sum_i a_i)}{\prod_i \G(a_i)} \int_0^\oo \prod_{i=2}^n \frac{d\alpha_i}{\alpha_i} \a_i^{a_i} \frac{1}{(A_1+ \sum_{i=2}^n \a_i A_i)^{\sum_i a_i}}~.
\ee
As a special case of the latter formula, we have
\be
\label{eq:singleparameter}
\int \frac{d\a}{\a} \a^b \frac{1}{(A+\a B)^c} &= \frac{\G(c-b)\G(b)}{\G(c)} \frac{1}{A^{c-b} B^b}~.
\ee
Let us define
\be
a_1 = d-\De_1~,\quad a_2 = \frac{\De_1+\De_2-\De_3+J}{2}~,\quad a_3 = \frac{\De_1+\De_3-\De_2+J}{2}~,\quad a_4 = -J~,
\ee
which satisfy $a_1+a_2+a_3+a_4=d$.
We start by combining the factors in (\ref{eq:ourshadowintegral}) using (\ref{eq:feynmanschwinger}), and then applying (\ref{eq:basicconformalintegral})
\be
&\int D^d X_1 \frac{1}{X_{01}^{a_1}X_{12}^{a_2} X_{13}^{a_3} X_{14}^{a_4}}\nn\\
&=
\frac{\G(d)}{\G(a_1)\G(a_2)\G(a_3)\G(a_4)} \int \frac{d\a}{\a}\a^{a_2} \frac{d\beta}{\beta} \beta^{a_3} \frac{d\g}{\g} \g^{a_4} \int D^d X_1 \frac{1}{(-2X_1\.(X_0 + \a X_2 + \beta X_3 + \g X_4))^d}\nn\\
&=
\frac{\G(d)}{\G(a_1)\G(a_2)\G(a_3)\G(a_4)}\frac{\pi^{\frac d 2}\G(\frac d 2)}{\G(d)}\int \frac{d\a}{\a}\a^{a_2} \frac{d\beta}{\beta} \beta^{a_3} \frac{d\g}{\g} \g^{a_4}  \frac{1}{(\a X_{02} + \beta X_{03} + \a \beta X_{23} + \g X_{04})^{d/2}}~.\nn\\
\ee
The factor in the denominator is relatively simple because of the orthogonality conditions (\ref{eq:extraorthogonality}).
Now we repeatedly use (\ref{eq:singleparameter}), first for $\g$, then $\beta$, then $\a$, giving
\be
&= \frac{\pi^{\frac d 2} \G(\frac d 2 - a_3 - a_4) \G(\frac d 2 - a_2 - a_4) \G(\frac d 2 - a_1)}{\G(a_1)\G(a_2)\G(a_3)} \frac{1}{X_{02}^{\frac d 2 - a_3 - a_4} X_{03}^{\frac d 2 - a_2 - a_4} X_{23}^{\frac d 2 - a_1} X_{04}^{a_4}}~.
\ee
Plugging everything back in, we find
\be
\<\mathbf{S}[\f_1](X_0) \f_2(X_2) \cO(X_3,Z_3)\> &= S^{\De_2,[\De_3,J]}_{\De_1} \<\tl \f_1(X_0) \f_2(X_2)\cO(X_3,Z_3)\> \nn\\
S^{\De_2,[\De_3,J]}_{\De_1} &= 
\frac{\pi^{\frac d 2}\G(\De_1 - \frac d 2)\G(\frac{\tl \De_1 + \De_2 - \De_3+J}{2}) \G(\frac {\tl \De_1 + \De_3 - \De_2+J}{2}) }{\G(d-\De_1)\G(\frac{\De_1+\De_2-\De_3+J}{2}) \G(\frac{\De_1+\De_3 - \De_2+J}{2})}~.
\ee

\subsection{Shadow transforming the spinning operator}

Let us also compute the shadow transform with respect to the spinning operator,
\be
\mathbf{S}[\cO]_{m_1\cdots m_J}(X) &= \int D^d Y \frac{\prod_i -2(\eta^{m_i n_i}(X\.Y)-Y^m X^n)}{(-2X\.Y)^{d-\De+J}}\cO_{n_1\cdots n_J}(Y)~,
\ee
or in index-free notation
\be
\mathbf{S}[\cO](X,Z) &= \int D^d Y \frac{1}{(-2X\.Y)^{d-\De+J}}\cO(Y,2X(Z\.Y)-2Z(X\.Y))~,
\ee
Thus, we would like to compute
\be
&\int D^d X_3 \frac{1}{(-2X_0\.X_3)^{d-\De_3+J}}\<\cO(X_3,2X_0(Z_0\.X_3)-2Z_0(X_0\.X_3))\f_1(X_1)\f_2(X_2)\>\nn\\
&=\int D^d X_3 \frac{1}{(-2X_0\.X_3)^{d-\De_3+J}} \frac{(-2(2X_0(Z_0\.X_3)-2Z_0(X_0\.X_3))\.(X_1(X_2\.X_3)-X_2(X_1\.X_3)))^J}{
X_{12}^{\frac{\De_1+\De_2-\De_3+J}{2}} X_{23}^{\frac{\De_2+\De_3-\De_1+J}{2}} X_{13}^{\frac{\De_1+\De_3-\De_2+J}{2}}
}~.
\ee
The factor in the numerator is given by
\be
-2(2X_0(Z_0\.X_3)-2Z_0(X_0\.X_3))\.(X_1(X_2\.X_3)-X_2(X_1\.X_3)) &= X_{23}X_{31'}-X_{13}X_{32'}~,
\ee
where
\be
X_1' &\equiv (X_1\.Z_0) X_0 - (X_1\.X_0) Z_0~,\nn\\
X_2' &\equiv (X_2\.Z_0) X_0 - (X_2\.X_0) Z_0~.
\ee
We have defined $X_1'$ and $X_2'$ in this way because many dot products involving them vanish, using $X_0^2 = Z_0\.X_0=Z_0^2=0$.  The only nonzero dot products involving them are
\be
X_{21'} &= -X_{12'} = -2((X_2\.X_0)(X_1\.Z_0)-(X_2\.Z_0)(X_1\.X_0)) \nn\\
&= V_{0,12} X_{12}^{\frac 1 2} X_{20}^{\frac 1 2} X_{10}^{\frac 1 2}~.
\ee
In terms of these quantities, our integral becomes
\be
&
\int D^d X_3 \frac{1}{X_{03}^{d-\De_3+J}} \frac{(X_{23}X_{31'}-X_{13}X_{32'})^J}{
X_{12}^{\frac{\De_1+\De_2-\De_3+J}{2}} X_{23}^{\frac{\De_2+\De_3-\De_1+J}{2}} X_{13}^{\frac{\De_1+\De_3-\De_2+J}{2}}}
\nn\\
&=
\frac{1}{X_{12}^{\frac{\De_1+\De_2-\De_3+J}{2}}}\sum_{n=0}^J{J \choose n}(-1)^{J-n}\int D^d X_3 \frac{1}{X_{03}^{d-\De_3+J}} \frac{X_{31'}^nX_{32'}^{J-n}}{
 X_{23}^{\frac{\De_2+\De_3-\De_1+J}{2}-n} X_{13}^{\frac{\De_1+\De_3-\De_2+J}{2}+n-J}}~.
\ee
We can evaluate the integral over $X_3$ using the same techniques as before: combining factors using (\ref{eq:feynmanschwinger}), applying the basic conformal integral (\ref{eq:basicconformalintegral}), and then repeatedly using (\ref{eq:singleparameter}) to split the factors apart. The result is
\be
\frac{\pi^{\frac d 2}\G(a_1+a_2-\frac d 2)\G(\frac d 2 - a_1 - b_1)\G(\frac d 2 - a_2 - b_2)}{\G(a_0)\G(a_1)\G(a_2)}
 X_{10}^{a_2+b_2-\frac d 2} X_{20}^{a_1+b_1-\frac d 2} X_{12}^{-a_1-a_2+\frac d 2} X_{12'}^{-b_2} X_{21'}^{-b_1}~.
\ee
Plugging this in to the above and performing the sum over $n$, we obtain
\be\label{spinshadow}
\<\f_1(X_1)\f_2(X_2)\mathbf{S}[\cO](X_0,Z_0)\> &= S^{\De_1,\De_2}_{[\De_3,J]} \<\f_1(X_1)\f_2(X_2)\tl \cO(X_0,Z_0)\> \nn\\
S^{\De_1,\De_2}_{[\De_3,J]} &=
\frac{\pi^{\frac d 2} \G(\De_3-\frac d 2)\G(\De_3+J-1)\G(\frac{\tl\De_3+\De_1-\De_2+J}{2})\G(\frac{\tl\De_3+\De_2-\De_1+J}{2})}{\G(\De_3-1) \G(d-\De_3+J) \G(\frac{\De_3+\De_1-\De_2+J}{2})\G(\frac{\De_3+\De_2-\De_1+J}{2})}~.
\ee
As a check, when $J=0$, we have $S^{\De_1,\De_2}_{[\De_3,0]} = S^{\De_1,[\De_2,0]}_{\De_3}$.

\bibliographystyle{utphys}
\bibliography{Higherbib}

\end{document}